\documentclass[prd,aps,groupedaddress,showpacs,nobibnotes]{revtex4}
\usepackage{epsf,epsfig,graphicx}

\newcounter{muni}

\begin{document}
\hbadness=10000 \pagenumbering{arabic}

\title{Glauber gluons in spectator amplitudes for $B\to\pi M$ decays}

\author{Hsiang-nan Li$^{1}$}
\email{hnli@phys.sinica.edu.tw}
\author{Satoshi Mishima$^2$}
\email{Satoshi.Mishima@roma1.infn.it}

\affiliation{$^{1}$Institute of Physics, Academia Sinica, Taipei,
Taiwan 115, Republic of China,}

\affiliation{$^{1}$Department of Physics, Tsing-Hua University,
Hsinchu, Taiwan 300, Republic of China,}

\affiliation{$^{1}$Department of Physics, National Cheng-Kung
University, Tainan, Taiwan 701, Republic of China}

\affiliation{$^{2}$Dipartimento di Fisica, Universit\`a
di Roma ``La Sapienza'', I-00185 Roma, Italy}

\affiliation{$^{2}$SISSA, Via Bonomea 265, I-34136 Trieste, Italy}

\begin{abstract}

We extract the Glauber divergences from the spectator amplitudes for
two-body hadronic decays $B\to M_1M_2$ in the $k_T$ factorization
theorem, where $M_2$ denotes the meson emitted at the
weak vertex. Employing the eikonal approximation, the divergences
are factorized into the corresponding Glauber phase factors
associated with the $M_1$ and $M_2$ mesons.
It is observed that the latter factor
enhances the spectator contribution to the color-suppressed tree
amplitude by modifying the interference pattern between the two
involved leading-order diagrams. The first factor rotates the
enhanced spectator contribution by a phase, and changes its
interference with other tree diagrams. The above Glauber effects are
compared with the mechanism in elastic rescattering among various
$M_1M_2$ final states, which has been widely investigated in the
literature. We postulate that only the Glauber effect associated
with a pion is significant, due to its special role as a $q\bar q$
bound state and as a pseudo Nambu-Goldstone boson simultaneously.
Treating the Glauber phases as additional inputs in the
perturbative QCD (PQCD) approach, we find a good fit to all the
$B\to\pi\pi$, $\pi\rho$, $\pi\omega$, and $\pi K$ data, and
resolve the long-standing $\pi\pi$ and $\pi K$ puzzles. The
nontrivial success of this modified PQCD formalism is elaborated.

\end{abstract}


\maketitle

\section{INTRODUCTION}

The known $B\to\pi\pi$ and $B\to\pi K$ puzzles have stimulated a lot
of discussions in the literature: the measured $B^0\to\pi^0\pi^0$
branching ratio \cite{Amhis:2012bh} is several times larger than the naive
expectation, and the measured direct CP asymmetry in the
$B^\pm\to\pi^0 K^\pm$ decays dramatically differs from the
$B^0\to\pi^\mp K^\pm$ one. It has been pointed out that these
puzzles are sensitive to the least-understood color-suppressed tree
amplitudes $C$ \cite{Charng2,Pham:2009ti,CC09}. Other similar
discrepancies were also observed: the $B^0\to\pi^0\rho^0$ branching
ratios from the perturbative QCD (PQCD) and QCD factorization (QCDF)
approaches, being sensitive to $C$, are lower than the data
\cite{LY02,RXL,BN}. However, the estimate of $C$ from PQCD is well
consistent with the measured $B^0\to\rho^0\rho^0$ branching ratio
\cite{LM06}. Proposals resorting to new physics
\cite{HLMN,S08,FJPZ,Baek:2008vr,Bhattacharyya:2008vj,Mohanta:2008ce,
Chang:2009wt,Baek:2009hv,Khalil:2009zf,Huitu:2009st,Cho:2011ta,
Imbeault:2008ge,Endo:2012gi}
mainly resolved the $\pi K$ puzzle without addressing the peculiar
feature of $C$ in the $\pi^0\pi^0$, $\pi^0\rho^0$, and
$\rho^0\rho^0$ modes, while those to QCD effects are usually
strongly constrained by the $\rho\rho$ data \cite{BRY06}. The recent
resolution of the $B\to\pi K$ puzzle by means of the so-called Pauli
blocking mechanism seems to be lack of a solid theoretical support
\cite{Lipkin:2011ds}. It manifests the difficulty of this subject.

Motivated by the above puzzles, we have carefully investigated the
subleading contributions to the amplitudes $C$ and their impact on
the $B\to\pi\pi$, $\pi K$ decays in the PQCD approach based on the
$k_T$ factorization theorem \cite{KLS,LUY}. For example, the
next-to-leading-order (NLO) contributions from the vertex
corrections, the quark loops and the magnetic penguin have been
calculated \cite{LMS05,X14}. Nevertheless, once a mechanism identified
for $C$ respects the conventional factorization theorem, it is
unlikely to be a resolution due to the $B\to\rho\rho$ constraint
mentioned above \cite{LM06}. This is the reason why the above NLO
corrections could not resolve the puzzles completely, though the
consistency between the PQCD predictions and the data was improved.
For a similar reason, higher-order corrections evaluated in QCDF
\cite{BY05}, which obey the collinear factorization, cannot resolve
the $B\to\pi\pi$ puzzle either. In a recent work \cite{LM11}, we
have analyzed high-order corrections to the spectator diagrams in
the $k_T$ factorization theorem, and found a new type of infrared
divergences, called the Glauber gluons \cite{CQ06}. The all-order
summation of the Glauber gluons leads to a phase factor, which
modifies the interference between the spectator diagrams for $C$. We
postulated that only the Glauber factors associated with a pion give
significant effects, due to its special role as a $q\bar q$ bound
state and as a pseudo Nambu-Goldstone (NG) boson simultaneously
\cite{NS08}. It was then demonstrated that the Glauber effect,
enhancing the magnitude of $C$, partially resolved the $B\to\pi\pi$
and $B\to\pi K$ puzzles. Our prediction for the $B^0\to\pi^0\pi^0$
branching ratio around $1.0\times 10^{-6}$ \cite{LM11} turns out to
be consistent with the recent Belle data
$(0.90 \pm 0.12 \pm 0.10) \times 10^{-6}$ \cite{update}.

The above progress implies that the Glauber gluons in the $k_T$
factorization theorem deserve a thorough study. In this paper we
shall examine whether the Glauber divergences in the spectator
diagrams for the $B\to M_1M_2$ decay, where $M_2$ denotes the
meson emitted at the weak vertex, have been extracted
completely, and whether the same Glauber effect improves the
consistency of the PQCD predictions with other data involving the
pion, such as the $B\to\pi\rho,\pi\omega$ data. It will be shown
that there exist the Glauber divergences associated with the $M_1$
meson, in additional to those
associated with the $M_2$ meson \cite{LM11}. The all-order
organization of the Glauber divergences follows the standard
procedures, relying on the eikonal approximation for soft gluons.
The resultant Glauber factor $\exp(-iS_{e1})$ from $M_1$ is the same
for the two leading-order (LO) spectator diagrams. The Glauber
factor from $M_2$ carries opposite phases, namely, $\exp(iS_{e2})$
for one diagram, and $\exp(-iS_{e2})$ for another. Therefore, they
have different impacts on the amplitude $C$: the latter enhances the
spectator contribution to $C$ by modifying the interference pattern
between the two LO diagrams as mentioned before. The former rotates
the enhanced spectator contribution by a phase, and changes its
interference with other tree diagrams.
The correspondence will be made explicit between
the Glauber factors and the mechanism in elastic rescattering among
various $M_1M_2$ final states, including the singlet exchange and
the charge exchange, which have been widely explored in the
literature \cite{CHY,Chua08}.

The Glauber factors $\exp(-iS_{e1})$ (as $M_1=\pi$), and $\exp(\pm
iS_{e2})$ (as $M_2=\pi$) are introduced into the PQCD factorization
formulas for the spectator diagrams in the $B\to\pi\pi$, $\pi\rho$,
$\pi\omega$, and $\pi K$ modes (totally 13 modes), and the phases
$S_{e1}$ and $S_{e2}$ are treated as additional inputs. It turns out
that the equal value $S_{e1}=S_{e2}\approx-\pi/2$ leads to
a good fit to all the $B\to\pi M$ data. It will be observed that
the Glauber effects render the NLO PQCD predictions for the
$B^0\to\pi^+\pi^-$, $B^+\to\pi^+\pi^0$, and $B^0\to\pi^0\pi^0$
branching ratios agree well with the data. In particular, the
rotation of the spectator amplitude by $\exp(-iS_{e1})$ is crucial
for enhancing the ratio of the $B^+\to\pi^+\pi^0$ branching ratio over
the $B^0\to\pi^+\pi^-$ one: this ratio depends on
both the color-allowed tree amplitude $T$ and the
color-suppressed tree amplitude $C$, so the relative phase between
them matters. It is a nontrivial success that all
the $B\to\pi\pi$, $\pi\rho$, and $\pi K$ puzzles mentioned before
are resolved at the same time by introducing two Glauber phases.

In Sec.~II we construct the standard meson wave functions for the
$B\to M_1M_2$ decays in the $k_T$ factorization theorem, and analyze
the residual infrared divergences caused by the Glauber gluons in
the NLO spectator diagrams. The Glauber gluons associated with the
$M_1$ and $M_2$ mesons are then factorized into the Glauber factors
$\exp(-iS_{e1})$ and $\exp(\pm iS_{e2})$, respectively. In Sec.~III
we investigate the numerical impacts of the Glauber factors on the
$B\to\pi\pi$, $\pi\rho$, $\pi\omega$, and $\pi K$ decays by
presenting NLO PQCD predictions as contour plots in the
$S_{e1}$-$S_{e2}$ plane. The agreement between the predictions and
the data for the branching ratios and direct CP asymmetries as
$S_{e1}=S_{e2}\approx-\pi/2$ is highlighted. Section IV contains the
conclusion. The existence of the Glauber divergences is illustrated
in the Appendix by means of the Feynman parametrization of
loop integrands.

\section{FACTORIZATION OF GLAUBER GLUONS}

It was pointed out \cite{CQ06} that the $k_T$ factorization theorem
holds for simple processes like deeply inelastic scattering,
but residual infrared divergences from the Glauber region may appear
in complicated QCD processes like high-$p_T$ hadron hadroproduction.
To factorize the collinear gluons associated with, say, one of the
initial-state hadrons, one eikonalizes the particle lines to which
the collinear gluons attach. Those eikonal lines from other hadrons
should cancel in order to maintain the universality of the
considered parton distribution function. However, the required
cancellation is not exact in the $k_T$ factorization, leading to
imaginary infrared logarithms, though it is in the collinear
factorization. It has been demonstrated that the residual
divergences can be factorized into a Glauber factor for low-$p_T$
hadron hadroproduction: the contour of a collinear gluon momentum
can be deformed away from the Glauber region at low $p_T$, such that
the usual eikonalization still holds \cite{CL09}. The above investigation
was then extended to two-body hadronic $B$ meson decays $B\to
M_1M_2$, and the residual infrared divergences in a spectator
amplitude associated with the $M_2$ meson were found, and factorized
into the same Glauber factor \cite{LM11}. Note that the $k_T$
factorization for a factorizable emission amplitude, i.e., a $B$
meson transition form factor, has been proved in \cite{NL03}.

\begin{figure}[t]
\begin{center}
\begin{tabular}{cc}
\includegraphics[height=3.5cm]{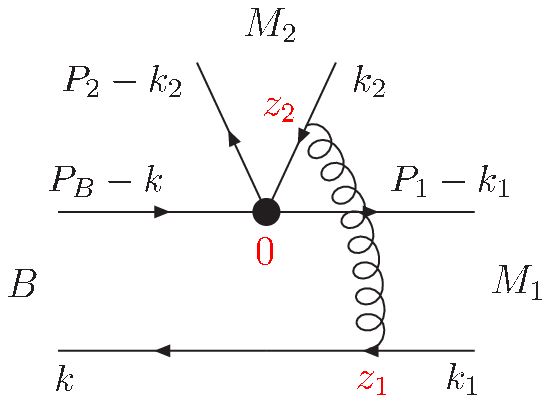}
\hspace{1mm} &
\includegraphics[height=3.5cm]{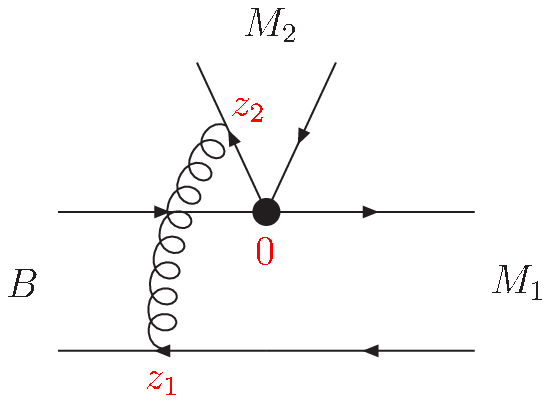}
\\
\hspace{-2mm}(a) & (b)
\end{tabular}
\caption{LO diagrams for a spectator amplitude.} \label{fig1}
\end{center}
\end{figure}

In this section we shall perform a thorough study of the infrared
divergences in the spectator diagrams at one-loop level of the $k_T$
factorization, following reasoning different from that in \cite{LM11}.
Both the infrared divergences, which are absorbed
into the standard meson wave functions, and the residual infrared
divergences from the Glauber gluons associated with the mesons $M_1$
and $M_2$ will be extracted. Since we have postulated that only the
Glauber effect from the pion is significant, it is not necessary to
discuss the Glauber divergences associated with the $B$ meson.
In principle, the Glauber gluons also exist in spectator penguin
diagrams and in nonfactorizable annihilation diagrams, in which
the hard gluon is emitted by the $b$ quark or by the spectator quark
in the $B$ meson. As explained
in \cite{LM11}, these diagrams are larger at LO, so they are more
stable against subleading corrections. The Glauber effect is
expected to be more significant in the spectator tree amplitudes,
because of their tininess at LO.

Consider the $B(P_B)\to M_1(P_1)M_2(P_2)$ decay, where $P_B$, $P_1$,
and $P_2$ represent the momenta of the $B$, $M_1$, and $M_2$ mesons,
respectively. For convenience, we choose $P_B=(P_B^+,P_B^-,{\bf
0}_T)$ with $P_B^+=P_B^-=m_B/\sqrt{2}$, $m_B$ being the $B$ meson
mass, and $P_1$ ($P_2$) in the plus (minus) direction. The parton
four-momenta $k$, $k_1$, and $k_2$ are labelled in
Fig.~\ref{fig1}(a). After performing loop integrations, we keep
$k^-=xP_B^-$, $k_1^+=x_1P_1^+$, $k_2^-=x_2P_2^-$, and transverse
components $k_T$, that appear in the hard kernel for the $b$-quark decay.
The order of magnitude $x_2\sim 0.5$, $x_1\sim 0.3$, $x\sim 0.1$,
$m_B\sim 5$ GeV, and $k_T\lesssim 1$ GeV implies the hierarchy among
the scales involved in exclusive $B$ meson decays in the small-$x$ region
\cite{LSW12}
\begin{eqnarray}
m_B^2,\, x_2m_B^2 \gg x_1 m_B^2 \gg x m_B^2 \gg  x x_1 m_B^2, \,
k_{T}^2, \label{hie}
\end{eqnarray}
which will serve as a basis for higher-order analysis below.

\begin{figure}[t]
\begin{center}
\begin{tabular}{ccc}
\includegraphics[height=2.5cm]{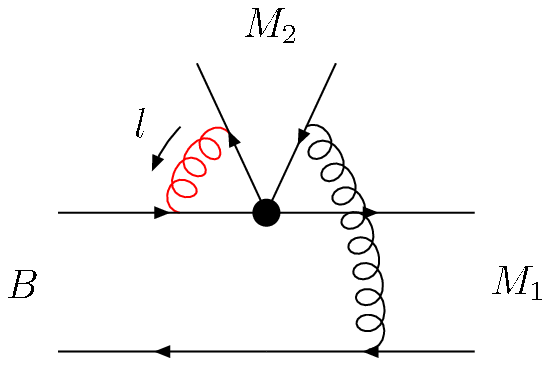}\hspace{0.3cm} &
\includegraphics[height=2.5cm]{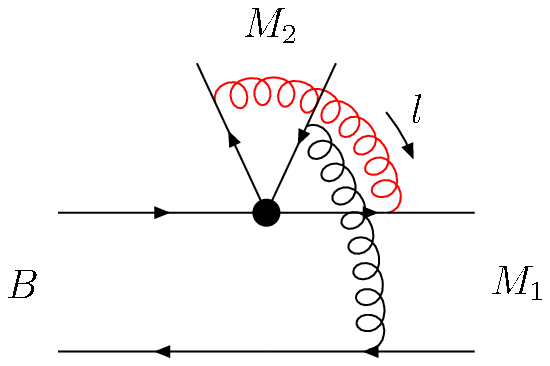}\hspace{0.3cm} &
\includegraphics[height=2.5cm]{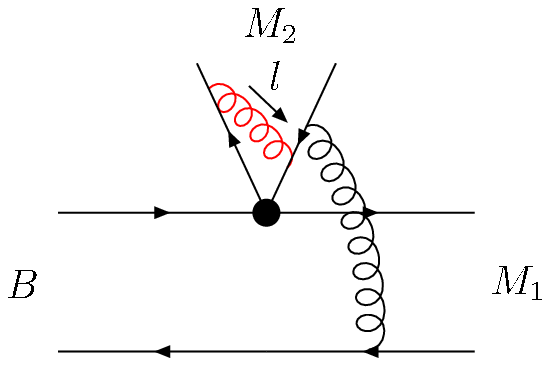}\hspace{0.3cm}  \\
\hspace{-5mm} (a) &\hspace{-5mm} (b) &\hspace{-5mm} (c) \\
\\
\includegraphics[height=2.5cm]{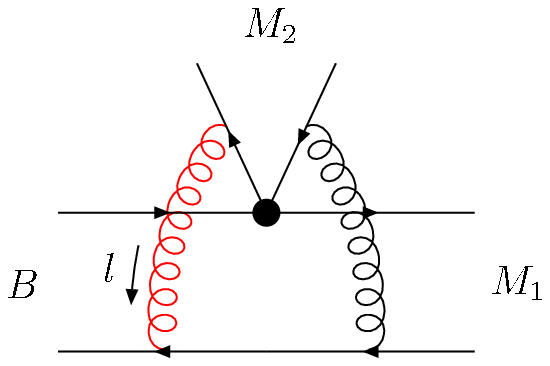}\hspace{0.3cm}&
\includegraphics[height=2.5cm]{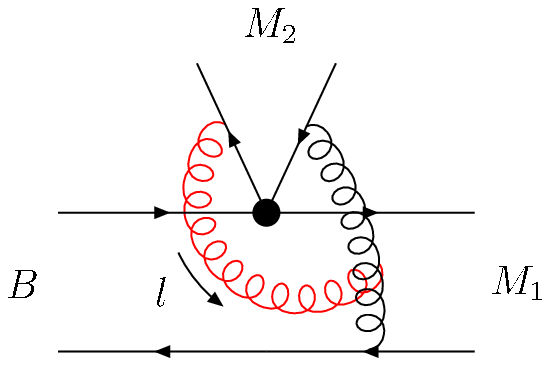}\hspace{0.3cm} &
\includegraphics[height=2.5cm]{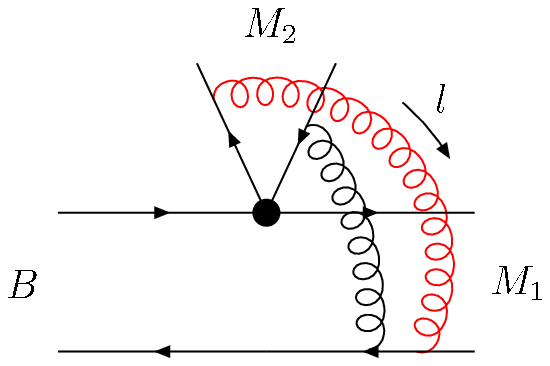}\hspace{0.3cm}\\
\hspace{-5mm} (d) &\hspace{-5mm} (e) &\hspace{-5mm} (f)
\end{tabular}
\caption{NLO diagrams for Fig.~\ref{fig1}(a)
that are relevant to the factorization of the $M_2$ meson wave
function. Figures~\ref{fig2}(d)-\ref{fig2}(f)
contribute to the Glauber divergences.} \label{fig2}
\end{center}
\end{figure}

We first identify the Glauber gluons associated with the LO spectator tree
diagram in Fig.~\ref{fig1}(a), originating from the operator $O_2$ \cite{REVIEW}.
Start with the set of NLO diagrams with a radiative gluon of momentum
$l$ being emitted by the valence quark of $M_2$, as displayed in Fig.~\ref{fig2}.
Due to the soft cancellation between the gluons radiated by the valence quark
and by the valence anti-quark of $M_2$ \cite{LT98}, only the collinear
region with $l$ being collimated to $P_2$ is relevant here, and the $k_T$
dependence of parton propagators in the $B$ and $M_1$ mesons is negligible.
The propagators of theses partons attached by the collinear
gluons can then be approximated by the eikonal propagators
$1/(l^-\pm i\epsilon)$. For a loop diagram to generate an imaginary
Glauber logarithm, a necessary (but not sufficient) condition is
that the interval of $l^-$ covers the origin $l^-=0$. The
corresponding integral then contains an imaginary piece,
\begin{eqnarray}
{\rm Im} \int_{-a}^{b} dl^-\frac{1}{l^- + i\epsilon}
=-\pi \int_{-a}^{b} dl^- \delta(l^-)=-\pi,\label{int}
\end{eqnarray}
under the principal-value prescription.

It has been shown that Figs.~\ref{fig2}(a)-\ref{fig2}(c) do not contain
Glauber divergences, and contribute to the $M_2$ meson wave function \cite{LM11}.
Take the vertex correction in Fig.~\ref{fig2}(a) as an example.
The integrand is proportional to
\begin{eqnarray}
& &\frac{1}{2(P_2^--k_2^-+l^-)l^+-|{\bf k}_{2T}-{\bf l}_T|^2+i\epsilon}
\frac{1}{2l^-l^+-l_T^2+i\epsilon}\nonumber\\
& &\times\frac{1}{2(P_B^--k^-)l^++2(P_B^+-k^++l^+)l^-
+i\epsilon},\label{pb}
\end{eqnarray}
where $l$ denotes the loop momentum, and the transverse-momentum-dependent
terms of the virtual $b$ quark propagator have been neglected in the
heavy-quark limit. The contour integration over $l^+$ indicates that the
loop integral does not vanish only for $l^- <0$: in this range there are
poles located in the different half complex planes of $l^+$. Picking up the
pole $l^+\approx 0-i\epsilon$ (see the power counting in Eq.~(\ref{hie}))
associated with the valence quark propagator in $M_2$, namely, the first factor of
Eq.~(\ref{pb}), the $b$ quark propagator reduces to the eikonal propagator proportional
to $1/(l^-+i\epsilon)$. In the range $l^- <0$ this propagator does not generate a
Glauber divergence according to Eq.~(\ref{int}). Because it is factorized in color
flow by itself with the color factor $C_F$, Fig.~\ref{fig2}(a)
leads to a Wilson line running from minus infinity to the origin,
i.e, the weak vertex, which appears in the definition of the $M_2$
meson wave function. Similarly, the vertex correction in Fig.~\ref{fig2}(b)
is free of a Glauber divergence. Figure~\ref{fig2}(c), with the collinear
gluon attaching to the virtual quark line, does not produce an infrared
Glauber divergence: the virtual quark line remains highly
off-shell by $O(x_1m_B^2)$ before and after the attachment of the
collinear gluon according to Eq.~(\ref{hie}), so no Glauber
divergence is generated in this diagram.

As observed in \cite{LM11}, Figs.~\ref{fig2}(d)-\ref{fig2}(f) produce
residual Glauber divergences under the hierarchical relation in
Eq.~(\ref{hie}), which demand introduction of an additional nonperturbative
input. The integrand for Fig.~\ref{fig2}(d) contains the five denominators
\begin{eqnarray}
[(P_2-k_2+l)^2+i\epsilon][(k+l)^2+i\epsilon][(k-k_1+l)^2+i\epsilon]
(l^2+i\epsilon)[(k_2-k+k_1-l)^2+i\epsilon].\label{2e}
\end{eqnarray}
Non-vanishing contributions come from the ranges $0< l^- < k_2^-$,
$-k^- < l^- <0$, and $-(P_2^--k_2^-)<l^-<-k^-$, where
the poles of $l^+$ are given by
\begin{eqnarray}
& &l^+\approx \frac{|{\bf l}_T-{\bf
k}_{2T}|^2}{2(l^-+P_2^--k_2^-)}-i\epsilon(-i\epsilon,\;-i\epsilon),\label{po1}\\
& &l^+=-k^++\frac{|{\bf l}_T+{\bf k}_T|^2}{2(l^-+k^-)}
-i\epsilon(-i\epsilon,\;+i\epsilon),\label{po2}\\
& &l^+=k_1^++\frac{|{\bf l}_T-{\bf k}_{1T}+{\bf
k}_T|^2}{2(l^-+k^-)}-i\epsilon(-i\epsilon,\;+i\epsilon),\label{po3}\\
& &l^+=\frac{l_T^2}{2l^-}-i\epsilon (+i\epsilon,\; +i\epsilon),\label{po4}\\
& &l^+=k_1^++\frac{|{\bf l}_T-{\bf k}_{2T}-{\bf
k}_{1T}+{\bf k}_T|^2}{2(l^--k_2^-)}
+i\epsilon(+i\epsilon,\;+i\epsilon),\label{po5}
\end{eqnarray}
respectively. We pick up the first pole
$l^+\sim O(\Lambda_{\rm QCD}^2/m_B)-i\epsilon$,
which corresponds to the collinear gluon associated with the valence
quark of $M_2$. It is seen that the allowed range for this pole,
$-(P_2^--k_2^-)<l^-<k_2^-$, covers the origin $l^-=0$,
leading to a Glauber divergence from the eikonalized spectator
propagator $1/(k+l)^2$ and the on-shell radiative gluon. 
The other poles, such as those in Eqs.~(\ref{po2})
and (\ref{po3}) in the range $-k^- < l^- <0$, should be included.
However, it is easy to confirm that they are irrelevant to the
analysis of the Glauber divergences. An alternative demonstration
of the existence of the Glauber divergence in Fig.~\ref{fig2}(d)
by means of the Feynman parametrization of the corresponding
loop integrand is presented in Appendix A.

For Fig.~\ref{fig2}(e), the Ward identity is applied to the virtual
gluon propagators,
\begin{eqnarray}
\frac{1}{[(k-k_1)^2+i\epsilon][(k-k_1+l)^2+i\epsilon]}=
\left[\frac{1}{(k-k_1)^2+i\epsilon}
-\frac{1}{(k-k_1+l)^2+i\epsilon}\right]\frac{1}{l^2+2(k-k_1)\cdot
l+i\epsilon}. \label{split1}
\end{eqnarray}
Here we have chosen the sign of the $i\epsilon$
term in the factor outside the square brackets, such that this
factor reduces to the eikonal propagator $1/(-l^-+i\epsilon)$, after
picking up the pole $l^+\approx 0-i\epsilon$. With this
choice the first term in the above splitting can be combined with
Figs.~\ref{fig2}(b) and \ref{fig2}(c), and contribute to the $M_2$
meson wave function with the piece of Wilson lines from a coordinate
$z_2$ to plus infinity \cite{Li01}, where $z_2$ has been labelled in
Fig.~\ref{fig1}. As explicitly shown in Appendix A, the first term does not
involve a Glauber divergence, so it does not break the universality of
the $M_2$ meson wave function.
The second piece in Eq.~(\ref{split1}) with the color factor $N_c/2$,
$N_c$ being the number of colors,
contains the original Glauber divergence of Fig.~\ref{fig2}(e).
The eikonal approximation for the spectator propagator
$1/[(k_1-l)^2+i\epsilon]$ in Fig.~\ref{fig2}(f) also gives
$1/(-l^-+i\epsilon)$ but with the color factor $-1/(2N_c)$ \cite{LT98}. The sum of
the second piece in Eq.~(\ref{split1}) and Fig.~\ref{fig2}(f)
then leads to the Glauber divergence with the color factor
$N_c/2-1/(2N_c)=C_F$.

\begin{figure}[t]
\begin{center}
\begin{tabular}{ccc}
\includegraphics[height=2.5cm]{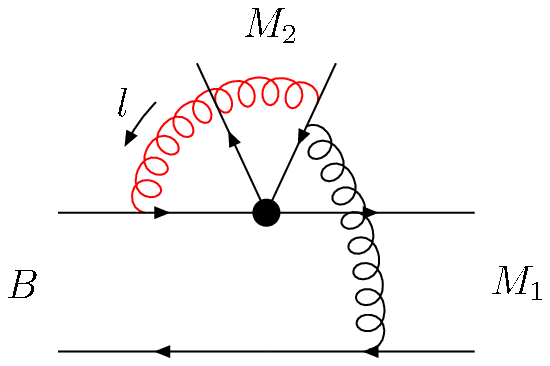}\hspace{0.3cm} &
\includegraphics[height=2.5cm]{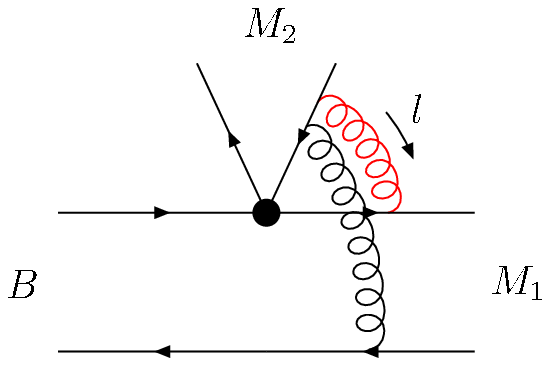}\hspace{0.3cm} &
\includegraphics[height=2.5cm]{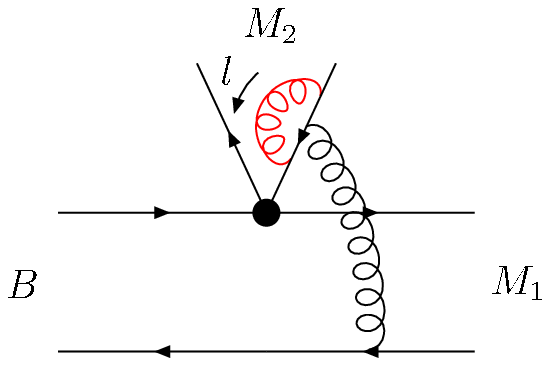}\hspace{0.3cm} \\
\hspace{-5mm} (a) &\hspace{-5mm} (b) &\hspace{-5mm} (c) \\
\\
\includegraphics[height=2.5cm]{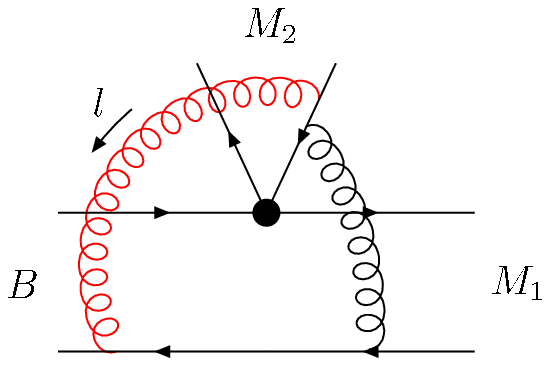}\hspace{0.3cm} &
\includegraphics[height=2.5cm]{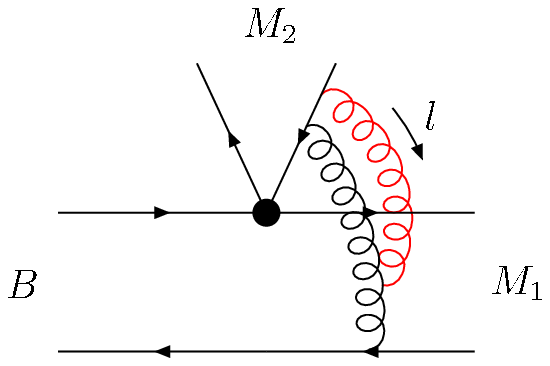}\hspace{0.3cm} &
\includegraphics[height=2.5cm]{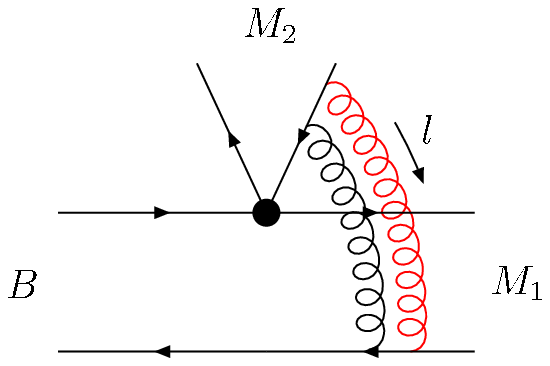}\hspace{0.3cm} \\
\hspace{-5mm} (d) &\hspace{-5mm} (e) &\hspace{-5mm} (f)
\end{tabular}
\caption{More NLO diagrams for Fig.~\ref{fig1}(a).}\label{fig1am}
\end{center}
\end{figure}

We examine the effects from Fig.~\ref{fig1am}, which is
similar to Fig.~\ref{fig2} but with the collinear gluon being
emitted by the valence anti-quark of $M_2$.
Figures~\ref{fig1am}(a)-\ref{fig1am}(c)
do not generate Glauber divergences, and also contribute to the
$M_2$ meson wave function. For example, the attachment to the $b$
quark in Fig.~\ref{fig1am}(a) gives rise to the eikonal propagator
$1/(l^-+i\epsilon)$ as in Eq.~(\ref{pb}), namely, the first piece of
Wilson lines, which runs from minus infinity to the origin.
Figure~\ref{fig1am}(d) contains the four denominators
\begin{eqnarray}
[(k_2+l)^2+i\epsilon][(k+l)^2+i\epsilon]
[(k-k_1+l)^2+i\epsilon](l^2+i\epsilon),\label{4a}
\end{eqnarray}
whose corresponding $l^+$ poles are the same as in
Eqs.~(\ref{po1})-(\ref{po4}). Therefore, the allowed
range of $l^-$ reduces to $-(P_2^--k_2^-)<l^-<0$ without
the pole in Eq.~(\ref{po5}), and this diagram does not
contain a Glauber divergence. This observation has been also
confirmed in Appendix A by means of the Feynman parametrization of
the corresponding loop integrand. Figures~\ref{fig2}(d) and
\ref{fig1am}(d) have the same amplitudes in the soft region with
$l\sim O(\Lambda_{\rm QCD})$ except a sign difference,
which is attributed to the emissions of the collinear gluon by the
valence quark and by the valence anti-quark in $M_2$.
Because of this soft cancellation, the contour of $l^-$
in Fig.~\ref{fig2}(d) can be deformed away the $O(\Lambda_{\rm
QCD})$ region, and the eikonalization of the spectator
$1/[(k+l)^2+i\epsilon]$ into $1/(l^-+i\epsilon)$ is justified \cite{LM11}.
That is, Fig.~\ref{fig1am}(d) provides soft subtraction
for Fig.~\ref{fig2}(d), but does not remove its Glauber divergence.
The soft cancellation also occurs between Figs.~\ref{fig2}(e) and
\ref{fig1am}(e), and between Figs.~\ref{fig2}(f) and
\ref{fig1am}(f).

The NLO residual infrared divergences in Figs.~\ref{fig2}(d)-\ref{fig2}(f)
are then extracted from the Glauber region,
\begin{eqnarray}
& &gC_F\int \frac{d^4l}{(2\pi)^4}tr\bigg[...\frac{-i(\not
\! k_2-\not\! k+\not\! k_1-\not l)}{(k_2-k+k_1- l)^2+i\epsilon}
(-ig\gamma_\beta)\gamma_5\not\!\! P_2(-ig\gamma^-)\frac{i(\not\!\! P_2-\not\!
k_2+\not l)}{(P_2-k_2+l)^2+i\epsilon}\bigg]\nonumber\\
& &\times\frac{-i}{(k-k_1+l)^2+i\epsilon}
\frac{-i}{l^2+i\epsilon}2\pi i \delta(l^-),\label{il}
\end{eqnarray}
where the $...$ denotes the rest of the
integrand, and $\gamma_5\!\not\!\!P_2$ comes from the twist-2 structure
of the $M_2$ meson wave function.
The $l^+$ poles in Eq.~(\ref{il}) are given by
Eqs.~(\ref{po1}), (\ref{po3}), and (\ref{po5})
with $l^-=0$ from the valence quark propagator, the virtual gluon
propagator, and the virtual quark propagator, respectively. Only the
pole in Eq.~(\ref{po1}) is of $O(\Lambda_{\rm QCD}^2/m_B)$. As long
as $k_1^+$ is of or greater than $O(\Lambda_{\rm QCD})$, we can deform
the contour of $l^+$, such that $l^+$ remains $O(\Lambda_{\rm QCD})$,
and the hierarchy
\begin{eqnarray}
(P_2^--k_2^-)l^+ \sim O(m_B\Lambda_{\rm QCD})\gg |{\bf l}_T-{\bf k}_{2T}|^2
\sim O(\Lambda_{\rm QCD}^2),
\end{eqnarray}
holds. The valence quark carrying the momentum $P_2-k_2+l$ in
Eq.~(\ref{il}) can then be eikonalized into $1/(l^++i\epsilon)$.

Equation~(\ref{il}) is factorized into
\begin{eqnarray}
& &g^2C_F\int
\frac{d^4l}{(2\pi)^4}tr\bigg[...\frac{-i(\not \!
k_2-\not\! k+\not\! k_1-\not l)}{(k_2-k+k_1- l)^2+i\epsilon}
(-ig\gamma_\beta)\gamma_5\not\!\! P_2
\bigg]\frac{-i}{(k-k_1+l)^2+i\epsilon}\nonumber\\ &
&\times\frac{1}{l^++i\epsilon}\frac{-i}{l^2+i\epsilon}2\pi i
\delta(l^-).\label{il2}
\end{eqnarray}
The above factorization of the Glauber gluon follows exactly the reasoning 
applied to the low-$p_T$ hadron hadroproduction in \cite{CL09}.
We close the contour in the lower half plane of $l^+$, and pick up
only the pole $l^+\approx 0-i\epsilon$ from the eikonal propagator
$1/(l^++i\epsilon)$, which corresponds to
an on-shell valence quark propagator. Another pole that
corresponds to the on-shell right gluon 
contributes to the Glauber divergence associated with
Fig.~\ref{fig1}(b) \cite{LM11}.
We then derive explicitly the imaginary logarithm,
\begin{eqnarray}
i\frac{\alpha_s}{\pi}C_F
\int\frac{d^2l_T}{l_T^2}{\cal
M}_a^{(0)}({\bf l}_T), \label{vi2}
\end{eqnarray}
where ${\cal M}_a^{(0)}$ denotes the LO spectator amplitude from
Fig.~\ref{fig1}(a). The gluon propagator proportional to $1/l_T^2$
indicates that the infrared divergence we have identified arises
from the Glauber region.

\begin{figure}[t]
\begin{center}
\begin{tabular}{ccc}
\includegraphics[height=2.5cm]{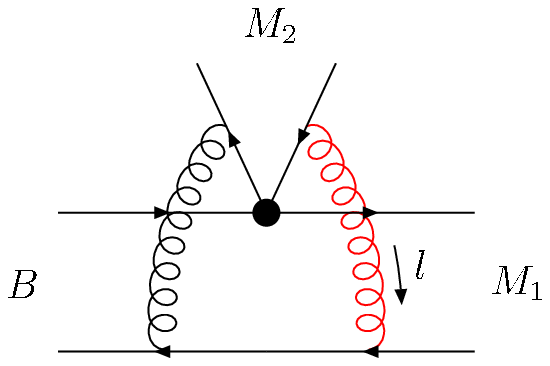}\hspace{0.3cm} &
\includegraphics[height=2.5cm]{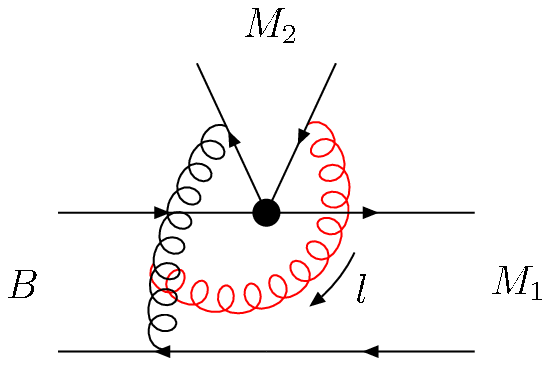}\hspace{0.3cm} &
\includegraphics[height=2.5cm]{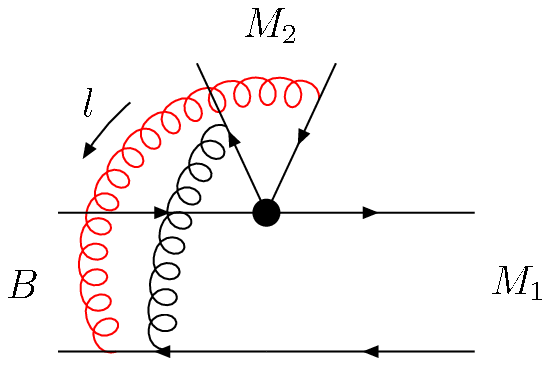}\hspace{0.3cm}
\\
\hspace{-5mm} (a) &\hspace{-5mm} (b) &\hspace{-5mm} (c) \\
\end{tabular}
\caption{NLO diagrams for Fig.~\ref{fig1}(b)
that contribute to the Glauber divergences associated with the $M_2$ meson.
}\label{figbg}
\end{center}
\end{figure}

Below we investigate the Glauber divergences appearing in the NLO
corrections to Fig.~\ref{fig1}(b), which are associated with the $M_2$ meson.
The relevant diagrams contain the attachments of the collinear gluons
emitted by the valence anti-quark of $M_2$ as depicted in Fig.~\ref{figbg}.
For the attachment to the virtual gluon
in Fig.~\ref{figbg}(b), we adopt the splitting
\begin{eqnarray}
\frac{1}{[(k-k_1)^2+i\epsilon][(k-k_1+l)^2+i\epsilon]}
=\left[\frac{1}{(k-k_1)^2+i\epsilon}
-\frac{1}{(k-k_1+l)^2+i\epsilon}\right]\frac{1}{l^2+2(k-k_1)\cdot
l-i\epsilon},\label{split3}
\end{eqnarray}
where the second term on the right-hand side contains the Glauber
divergence in the original NLO diagram. The first term then contributes to the
definition of the $M_2$ meson wave function. The similar analysis implies that
the diagrams in Fig.~\ref{figbg} contain the Glauber divergences,
\begin{eqnarray}
-i\frac{\alpha_s}{\pi}C_F
\int\frac{d^2l_T}{l_T^2}{\cal
M}_b^{(0)}({\bf l}_T), \label{vi3}
\end{eqnarray}
where ${\cal M}_b^{(0)}$ denotes the LO spectator amplitude from
Fig.~\ref{fig1}(b). The additional minus sign compared to Eq.~(\ref{vi2})
is attributed to the collinear gluon emission by the valence anti-quark
of $M_2$.

\begin{figure}[t]
\begin{center}
\begin{tabular}{cc}
\includegraphics[height=2.5cm]{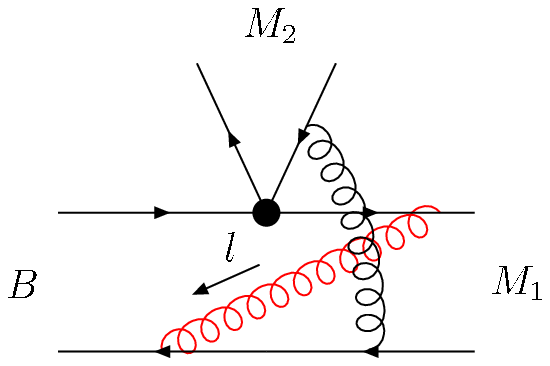}\hspace{1.0cm}  &
\includegraphics[height=2.5cm]{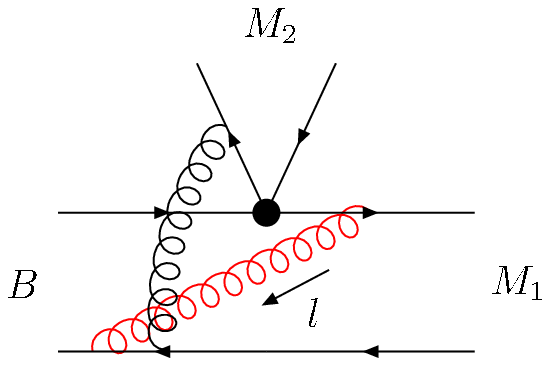}\hspace{1.0cm} \\
(a)\hspace{1.0cm} & \hspace{-9mm}(b) \hspace{1.0cm}\\
\\
\includegraphics[height=2.5cm]{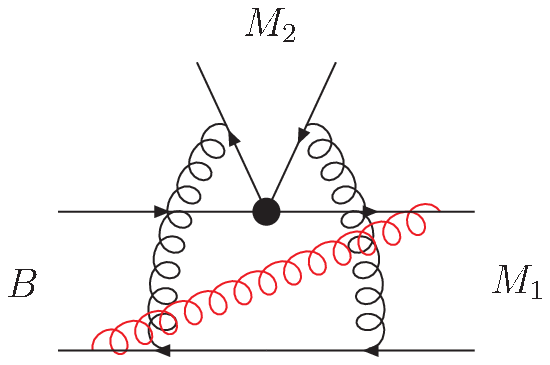}\hspace{1.0cm}  &
\includegraphics[height=2.5cm]{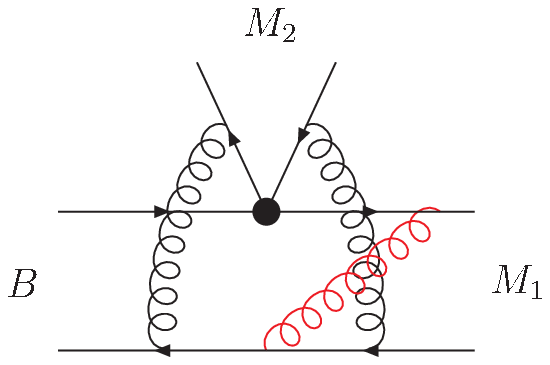}\hspace{1.0cm} \\
(c)\hspace{1.0cm} & (d)\hspace{1.0cm}
\end{tabular}
\caption{(a)-(c) Higher-order corrections to Fig.~\ref{fig1} that
contain the Glauber divergences associated with the $M_1$ meson.
(d) Higher-order correction to Fig.~\ref{fig1} that does not
contain the Glauber divergence associated with the $M_1$ meson.}\label{figa5}
\end{center}
\end{figure}

It has been shown that the residual infrared divergences appear
between the $M_2$ meson and the $B\to M_1$ transition \cite{LM11}.
It is natural to ask whether there exist more residual infrared divergences
in the spectator amplitude of the $B\to M_1M_2$ decay. We shall verify
that it is the case: additional Glauber divergences associated with the $M_1$
meson are induced by the inclusion of the Glauber divergences associated with
the $M_2$ meson. Consider all possible attachments of the collinear gluons
emitted by the valence quark of $M_1$ to particle lines in Fig.~\ref{fig1}(a),
among which the diagram in Fig.~\ref{figa5}(a) contains a Glauber divergence
as implied by the pole analysis of the following five denominators
\begin{eqnarray}
[(P_1-k_1+l)^2+i\epsilon][(k+l)^2+i\epsilon][(k-k_1+l)^2+i\epsilon]
(l^2+i\epsilon)[(k_2-k+k_1-l)^2+i\epsilon].\label{4f}
\end{eqnarray}
Non-vanishing contributions come from the ranges $0< l^+ < k_1^+$,
$-k^+ < l^+ <0$, and $-(P_1^+-k_1^+)<l^+<-k^+$, where
the poles of $l^-$ are given by
\begin{eqnarray}
& &l^-\approx \frac{|{\bf l}_T-{\bf
k}_{1T}|^2}{2(l^++P_1^+-k_1^+)}-i\epsilon(-i\epsilon,\;-i\epsilon),\label{pa1}\\
& &l^-=-k^-+\frac{|{\bf l}_T+{\bf k}_T|^2}{2(l^++k^+)}
-i\epsilon(-i\epsilon,\;+i\epsilon),\label{pa2}\\
& &l^-=\frac{l_T^2}{2l^+}-i\epsilon (+i\epsilon,\; +i\epsilon),\label{pa3}\\
& &l^-=-k^-+\frac{|{\bf l}_T-{\bf k}_{1T}+{\bf
k}_T|^2}{2(l^+-k_1^+)}+i\epsilon(+i\epsilon,\;+i\epsilon),\label{pa4}\\
& &l^-=k_2^-+\frac{|{\bf l}_T-{\bf k}_{2T}-{\bf
k}_{1T}+{\bf k}_T|^2}{2(l^+-k_1^+)}
+i\epsilon(+i\epsilon,\;+i\epsilon),\label{pa5}
\end{eqnarray}
respectively. We pick up the first pole
$l^-\sim O(\Lambda_{\rm QCD}^2/m_B)-i\epsilon$,
which corresponds to the collinear gluon associated with the valence
quark of $M_1$. It is seen that the allowed range for this pole,
$-(P_1^+-k_1^+)<l^+<k_1^+$, covers the origin $l^+=0$,
leading to a Glauber divergence from the eikonalized spectator
propagator $1/(k+l)^2$ and the on-shell radiative gluon.

The residual Glauber divergence in Fig.~\ref{figa5}(a) yields
the NLO spectator amplitude
\begin{eqnarray}
& &g\frac{-1}{2N_c}\int \frac{d^4l}{(2\pi)^4}tr\bigg[...(-ig\gamma^+)
\frac{i(\not\!P_1-\not\!k_1+\not
l)}{(P_1-k_1+l)^2+i\epsilon}\gamma_\mu(1-\gamma_5)\frac{-i(\not
\! k_2-\not\! k+\not\! k_1-\not l)}{(k_2-k+k_1- l)^2+i\epsilon}
(-ig\gamma_\beta)\gamma_5\not\!\! P_2\bigg]\nonumber\\
& &\times\frac{-i}{(k-k_1+l)^2+i\epsilon}
\frac{-i}{l^2+i\epsilon}\pi i\delta(l^+).\label{il3}
\end{eqnarray}
The $l^-$ poles in the above expression are given by Eqs.~(\ref{pa1}) and (\ref{pa5})
with $l^+=0$ from the valence quark propagator in $M_1$ and the virtual quark
propagator, respectively. The pole in Eq.~(\ref{pa1}) is of $O(\Lambda_{\rm
QCD}^2/m_B)$, and the pole in Eq.~(\ref{pa5}) is of $O(m_B)$, so we can deform
the contour of $l^-$, such that $l^-$ remains at least
$O(\Lambda_{\rm QCD})$, and the hierarchy
\begin{eqnarray}
(P_1^+-k_1^+)l^- \sim O(m_B\Lambda_{\rm QCD})\gg |{\bf l}_T-{\bf
k}_{1T}|^2 \sim O(\Lambda_{\rm QCD}^2),
\end{eqnarray}
holds. The valence quark carrying the momentum $P_1-k_1+l$ is thus
eikonalized into $1/(l^-+i\epsilon)$.
Equation~(\ref{il3}) is factorized into
\begin{eqnarray}
& &g^2 \frac{-1}{2N_c}\int \frac{d^4l}{(2\pi)^4}tr\bigg[...
\gamma_\mu(1-\gamma_5)\frac{-i(\not
\! k_2-\not\! k+\not\! k_1-\not l)}{(k_2-k+k_1- l)^2+i\epsilon}
(-ig\gamma_\beta)\gamma_5\not\!\! P_2\bigg]\frac{-i}{(k-k_1+l)^2+i\epsilon}\nonumber\\
& &\times \frac{1}{l^-+i\epsilon}
\frac{-i}{l^2+i\epsilon}\pi i\delta(l^+),\nonumber\\
&\approx&-i\frac{1}{2N_c}\frac{\alpha_s}{2\pi}\int\frac{d^2l_T}{l_T^2}{\cal
M}_a^{(0)}({\bf l}_T),\label{il6}
\end{eqnarray}
where we have closed the contour in the lower half plane of $l^-$ over the pole
$l^-\approx 0-i\epsilon$ from the eikonal propagator $1/(l^-+i\epsilon)$.

Figure~\ref{figa5}(b) gives the Glauber divergence the same as
in Eq.~(\ref{il6}), since the collinear gluon is also emitted by the valence
quark of $M_1$ and attaches to the spectator of the $B$ meson:
\begin{eqnarray}
-i\frac{1}{2N_c}\frac{\alpha_s}{2\pi}\int\frac{d^2l_T}{l_T^2}{\cal
M}_b^{(0)}({\bf l}_T).\label{il7}
\end{eqnarray}
The Glauber divergences in Eqs.~(\ref{il6}) and (\ref{il7}) can also
be verified by means of the Feynman parametrization of the loop integrands,
as shown in Appendix A. Because of the destruction between the LO
amplitudes ${\cal M}_a^{(0)}$ and ${\cal M}_b^{(0)}$, these Glauber divergences
cancel each other. The same cancellation also occurs between the pair of diagrams
with the collinear gluons attaching to the virtual gluons in Figs.~\ref{fig1}(a)
and \ref{fig1}(b). It then implies that there is no more Glauber divergence
at NLO in the $B\to M_1M_2$ decay, except those associated with the $M_2$ meson.
The other collinear gluon emissions from the valence quark and from the spectator of
$M_1$ contribute only to the construction of the $M_1$ meson wave function,
which contains the Wilson lines running from the origin to infinity, and then from
the infinity to the coordinate $z_1$ labelled in Fig.~\ref{fig1}(a).
The cancellation of the soft divergences, similar to that between
Figs.~\ref{fig2} and \ref{fig1am}, also occurs between the above
two sets of diagrams.

Nevertheless, the Glauber divergences associated
with the $M_1$ meson exist at next-to-next-to-leading order.
Once the Glauber gluons associated with the $M_2$ meson
are included, the interference between the two spectator amplitudes
${\cal M}_a$ and ${\cal M}_b$ becomes constructive, and the cancellation
between Eqs.~(\ref{il6}) and (\ref{il7}) does not take place anymore.
A corresponding diagram is displayed
in Fig.~\ref{figa5}(c), in which the two vertical gluon lines contribute to
the Glauber divergences for Figs.~\ref{fig1}(a) and \ref{fig1}(b), and the
third gluon emitted by $M_1$ gives a common Glauber divergence.
The color factor for Fig.~\ref{figa5}(c) is given by
\begin{eqnarray}
tr\left(T^cT^aT^bT^cT^bT^a\right)=\frac{1}{2}tr(T^aT^b)tr(T^bT^a)
-\frac{1}{2N_c}tr\left(T^aT^bT^bT^a\right),
\end{eqnarray}
where $T^a$, $T^b$, and $T^c$ are associated with the left vertical gluon,
the right vertical gluon, and the third gluon, respectively. The first
term in the above expression corresponds to a color flow from
the four-fermion operator $O_1$. Since we focus on the spectator amplitude
from $O_2$ in this work, this contribution
will be dropped. The second term corresponds to the color
flow of the original spectator amplitude, implying that the color factors for the
Glauber divergences associated with the $M_2$ and $M_1$ mesons remain
as in Eqs.~(\ref{vi2}), (\ref{vi3}), (\ref{il6}), and (\ref{il7}).

It can be shown that the attachments of the third gluon to other lines,
for example, to the spectator line between the two vertical gluons in
Fig.~\ref{figa5}(d),
do not produce Glauber divergences. The reason is explained below.
We route the loop momentum of the third gluon through the left-handed
vertical gluon. When this left-handed vertical gluon is hard (the right-handed
vertical gluon is soft), the third gluon contributes only to the $M_1$ meson
wave function: the diagram can be regarded as a two-particle reducible
correction to the $M_1$ meson wave function with the right-handed vertical
soft gluon coupling the $M_2$ meson and the $B$-$M_1$ system. That is, it does not
contribute to the Glauber divergence, which breaks the factorization.
When the left-handed
vertical gluon is soft (the right-handed vertical gluon is hard), the valence
quark of the $M_2$ meson remains on-shell and collimated to
the $M_2$ meson. In this case its momentum is independent of $k_1$, and it does
not constrain the contour in the $l^-$ plane. When both vertical gluons are hard,
Fig.~\ref{figa5}(d) contributes to the NLO hard kernel, which goes beyond the
accuracy of the present calculation.

A remark is in order. It has been shown that the Glauber divergence
exists in Fig.~\ref{fig2}(f), where the radiative gluon of momentum $l$
attaches partons in the $M_1$ and $M_2$ mesons. A simple way to tell
whether this Glauber divergence is associate with the $M_1$ or $M_2$ meson
is to investigate the pole structures. Replacing the spectator propagator
by $\delta(l^-)$ as done in Eq.~(\ref{il}), we check the pole positions
in the complex $l^+$ plane, and find that the $l^+$ contour for
Fig.~\ref{fig2}(f) is constrained by the valence quark propagator and
the valence anti-quark propagator of $M_2$. On the contrary, replacing
the valence quark propagator of $M_2$ by $\delta(l^+)$, we see that the
$l^-$ contour is not constrained. The above different pole structures
between $l^+$ and $l^-$ implies that the observed Glauber divergence
should be associated with the $M_2$ meson.
We then complete the investigation of the Glauber divergences in
the spectator amplitudes for the two-body hadronic $B$ meson decays.
The exponentiation of the NLO results in Eqs.~(\ref{vi2}) and (\ref{vi3})
\cite{LM11}, and in Eqs.~(\ref{il6}) and (\ref{il7}) leads to the
parametrization
\begin{eqnarray}
M_a^G &=& \exp(-iS_{e1})\exp(iS_{e2}){\cal M}_a^{(0)},\nonumber\\
M_b^G &=& \exp(-iS_{e1})\exp(-iS_{e2}){\cal M}_b^{(0)},\label{iab}
\end{eqnarray}
where the signs have followed the indication of the NLO results.
It is obvious that the destruction between $M_a^G$ and $M_b^G$ retains,
as the Glauber factors associated with the $M_2$ meson are turned off, i.e.,
$S_{e2}=0$. Strictly speaking, Eq.~(\ref{iab}), derived with the dependence on the
Glauber gluon transverse momentum being neglected, holds only approximately.
We shall treat the Glauber phases $S_{e1}$ and $S_{e2}$ as free
parameters in the numerical analysis later.
A definition for the Glauber factor in terms of a matrix element of
four Wilson lines has been constructed in \cite{CL09}.

At last, we point out the connection between the Glauber gluon
exchanges and the elastic scattering in two-body hadronic $B$ meson decays.
The analysis of \cite{CHY,Chua08} started with the amplitudes
evaluated in the QCDF approach, and final-state interaction effects
were included via the elastic rescattering. Take the rescattering only
between the $B^0\to\pi^+\pi^-$ and $B^0\to\pi^0\pi^0$ modes as an example,
\begin{equation}
\label{qs}
   \left( \begin{array}{c}
    \pi^+\pi^- \\ \pi^0\pi^0
   \end{array} \right)
   = S^{1/2}_{\rm res}
   \left( \begin{array}{c}
    \pi^+\pi^- \\ \pi^0\pi^0
   \end{array} \right)_{\rm QCDF},
\end{equation}
with the matrix $S^{1/2}_{\rm res}\equiv (1+i{\cal T})^{1/2}$
parameterizing the rescattering effects. The matrix $\cal T$
is written as
\begin{eqnarray}
\cal T=\left( \begin{array}{cc}
    r_0+2r_a+r_t &(2r_a-r_e+r_t)/\sqrt{2}\\
    (2r_a-r_e+r_t)/\sqrt{2} &r_0+(2r_a+r_e+r_t)/2
   \end{array} \right),
\end{eqnarray}
where the parameters $r_0$, $r_e$ $r_a$, and $r_t$ denote the
mechanism from the singlet exchange, the charge
exchange, the annihilation, and the total annihilation,
respectively. The best fit to the $B\to PP$ data gave the following
combined parameters defined in Eq.~(15) of \cite{Chua08}
\begin{eqnarray}
1+i(r_0+r_a)=0.94+0.58\,i,\;\;\;\; i(r_e-r_a)=0.06-0.58\,i,\;\;\;\;
i(r_a+r_t)=-0.12-0.09\,i,\label{resn}
\end{eqnarray}
which seem to indicate that the annihilation and the total annihilation
are less important, and $r_0$ and $r_e$ are roughly of the same order
of magnitude.

Compared to the above formalism, the standard NLO PQCD decay amplitudes
correspond to the inputs on the right-hand side of
Eq.~(\ref{qs}), and the Glauber
gluon exchanges correspond to the matrix $\cal T$. The
Glauber gluons do not generate the annihilation $r_a$ and $r_t$, an
observation consistent with the numerical outcomes in Eq.~(\ref{resn}).
We elaborate that the amplitude in Eq.~(\ref{vi2})
contributes to $r_0$, and that in Eq.~(\ref{vi3}) contributes to
$r_e$. Insert the identity for the color matrices
\begin{eqnarray}
I_{ij}I_{lk}=\frac{1}{N_c}I_{lj}I_{ik}+2
(T^c)_{lj}(T^c)_{ik},\label{col}
\end{eqnarray}
into $M_b^G$, with $I_{ij}$ ($I_{lk}$) being the unity matrix associated
with the meson $M_1$ ($M_2$). The second term in the decomposition,
associated with a meson in the color-octet state, will not be considered here.
The matrix $I_{lj}$ in the first term implies that the valence quark in
$M_1$ and the valence anti-quark in $M_2$ form a color-singlet state. The
matrix $I_{ik}$ implies that the valence anti-quark in
$M_1$ and the valence quark in $M_2$ form a color-singlet state. It is
easy to see that the resultant topology corresponds to the color-allowed
tree amplitude $T$. Therefore, $M_b^G$ can be regarded as
a contribution from the $B^0\to\pi^+\pi^-$ intermediate state,
dominated by the amplitude $T$, to the $B^0\to\pi^0\pi^0$ decay, dominated
by $C$, through the mechanism of charge exchange. The above color
rearrangement does not apply to the amplitude $M_a^G$, since
the color trace of $I_{lj}$ and the color matrix associated with
the hard gluon vertex vanishes. Hence, $M_a^G$
represents the contribution from the $B^0\to\pi^0\pi^0$
intermediate state to itself through the singlet exchange.
Certainly, the Glauber effect and the elastic rescattering are essentially
different. For instance, the former is crucial only in the pion-involved
decays, while the latter contributes to all relevant modes under the
$SU(3)$ flavor symmetry.

\section{NUMERICAL ANALYSIS}

As postulated in \cite{LM11}, the Glauber effect from the
multi-parton states is more significant in the pion than in other
mesons. This postulation can be understood by means of the
simultaneous role of the pion as a $q\bar q$ bound state and as a
NG boson \cite{NS08}: the valence quark and
anti-quark of the pion are separated by a short distance in order to
reduce the confinement potential energy, while the multi-parton
states of the pion spread over a huge space-time in order to meet
the role of a massless NG boson. That is, the multi-parton states
distribute more widely than the $q\bar q$ state does in the pion
compared to other mesons. This explains the strong Glauber effect
from the pion, which will be examined in this section.
The standard PQCD factorization formulas for the $B\to\pi\pi$ and $\pi K$ decays
are referred to \cite{LMS05}, while those for the $B\to\pi\rho$ and
$\pi\omega$ decays \cite{LY02,RXL} can be obtained by taking into account the
differences between $B\to PP$ and $PV$ modes as illustrated in
\cite{LM062}\footnote{Below Eq.~(6) in \cite{LM062},
the distribution amplitude $(-\phi_s^M)$
has to be corrected to $(+\phi_M^s)$.}.

Following Eq.~(\ref{iab}), we multiply the $b$-quark
spectator amplitudes in NLO PQCD, both tree and penguin,
by $\exp(iS_{e2})$ ($\exp(-iS_{e2})$) with the hard gluon being emitted by the
valence anti-quark (quark) in $M_2$, if $M_2$ denotes a pion. We also multiply
the above spectator amplitudes by $\exp(-iS_{e1})$, if $M_1$ denotes a pion.
As mentioned in \cite{LMS05}, the color-suppressed
tree amplitude in the $B\to\pi\pi$ decays is
small at LO due to the small Wilson coefficient $a_2$ for the
factorizable contribution and to the cancellation between
Figs.~\ref{fig1}(a) and \ref{fig1}(b) for the spectator
contribution. The presence of the Glauber factor $\exp(\pm iS_{e2})$
converts the destructive interference in Fig.~\ref{fig1} into a
constructive one, resulting in strong enhancement. The Glauber
factor $\exp(-iS_{e1})$ further rotates the enhanced spectator
amplitude, and modifies its interference with other emission
amplitudes. This effect will adjust the relative phase between the
color-allowed and color-suppressed tree amplitudes, such that all the
three $B\to\pi\pi$ branching ratios are accommodated at the same
time.

The choices of the distribution amplitudes for the
$B$ meson, pseudo-scalar mesons and vector mesons are the same as in
\cite{LM062}, but with the updated values of the meson decay
constants: $f_{B}= 191$ MeV, $f_{\pi}= 130$ MeV, $f_{K}= 156$ MeV,
$f_{\rho} = 216$ MeV, $f_{\rho}^T= 165$ MeV, $f_{\omega} = 187$ MeV,
and $f_{\omega}^T= 151$ MeV \cite{PDG,Ball06}.
We also update the meson masses $m_B = 5.28$ GeV, $m_\pi=0.137$ GeV,
$m_K=0.495$ GeV, $m_\rho=0.77$ GeV, and $m_\omega=0.783$ GeV, the
quark masses
$m_q=6.5$ MeV, $m_s=140$ MeV, $m_c=1.5$ GeV, and $m_b=4.8$ GeV,
which appear in the quark-loop and magnetic-penguin
amplitudes, the chiral scales $m_{0\pi}= 1.6$ GeV and $m_{0K} = 1.8$ GeV,
the $B$ meson lifetimes
$\tau_{B^0}=1.519\times 10^{-12}$ sec
and $\tau_{B^\pm}= 1.641\times 10^{-12}$ sec,
and the CKM matrix elements
$V_{ud}=0.97427$,
$V_{us}=0.22534$,
$|V_{ub}|=3.51\times 10^{-3}$,
$V_{cd} = - 0.22520$,
$V_{cs} = 0.97344$, and
$V_{cb} = 0.0412$, and
the weak phases $\phi_1 = 21.5^\circ$ and $\phi_3 = 70^\circ$
\cite{Amhis:2012bh,PDG}, while the other parameters are taken to
be the same as in \cite{LM062}. We employ the NLO Wilson
coefficients for the emission amplitudes, and the LO ones for the
the annihilation amplitudes, since the NLO corrections to the weak
vertices in the latter are not yet available. The resultant $B\to
\pi, K, \rho, \omega$ transition form factors are then given by
\begin{eqnarray}
F_0^{B\pi}(0) = 0.28,\ \ \ \ \ \
F_0^{BK}(0) = 0.39,\ \ \ \ \ \
A_0^{B\rho}(0) = 0.29,\ \ \ \ \ \
A_0^{B\omega}(0) = 0.27,
\end{eqnarray}
at maximal recoil, close to those obtained in \cite{KLS02}.

\begin{figure*}[t]
\begin{center}
\includegraphics[height=4.5cm]{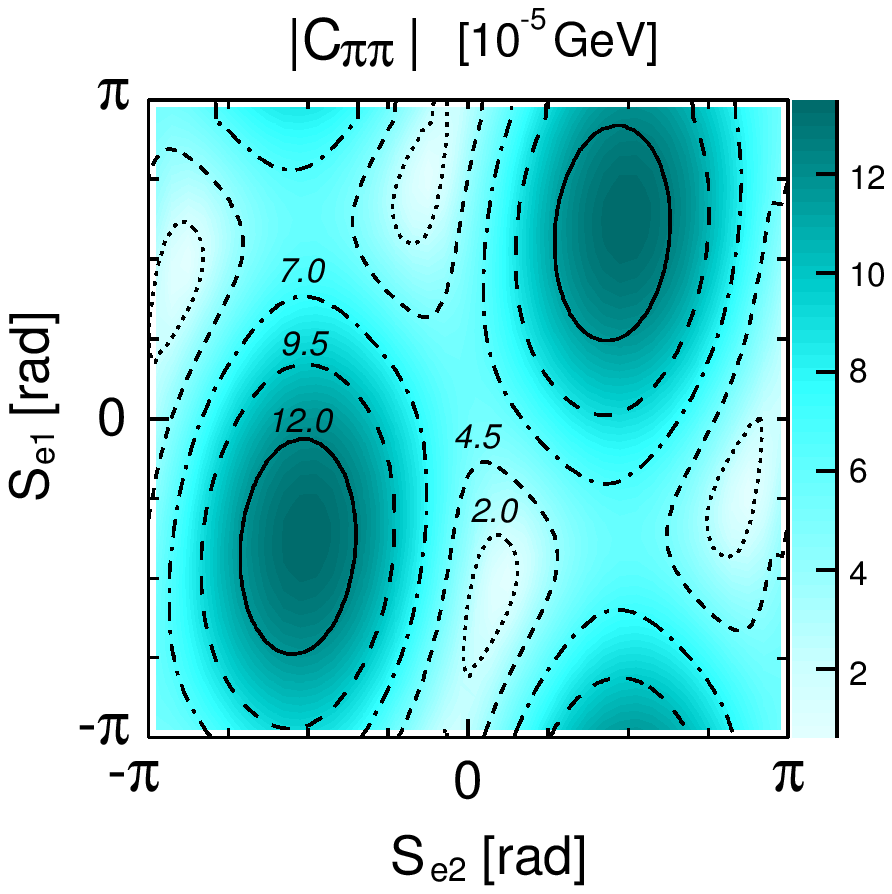}
\includegraphics[height=4.5cm]{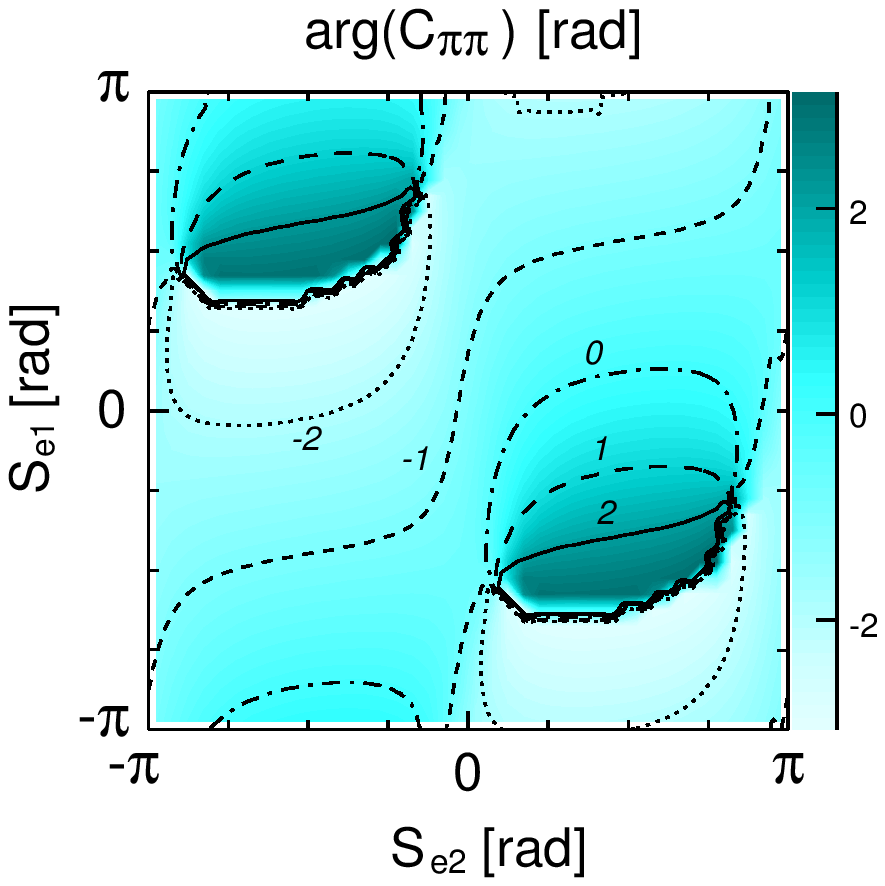}
\includegraphics[height=4.5cm]{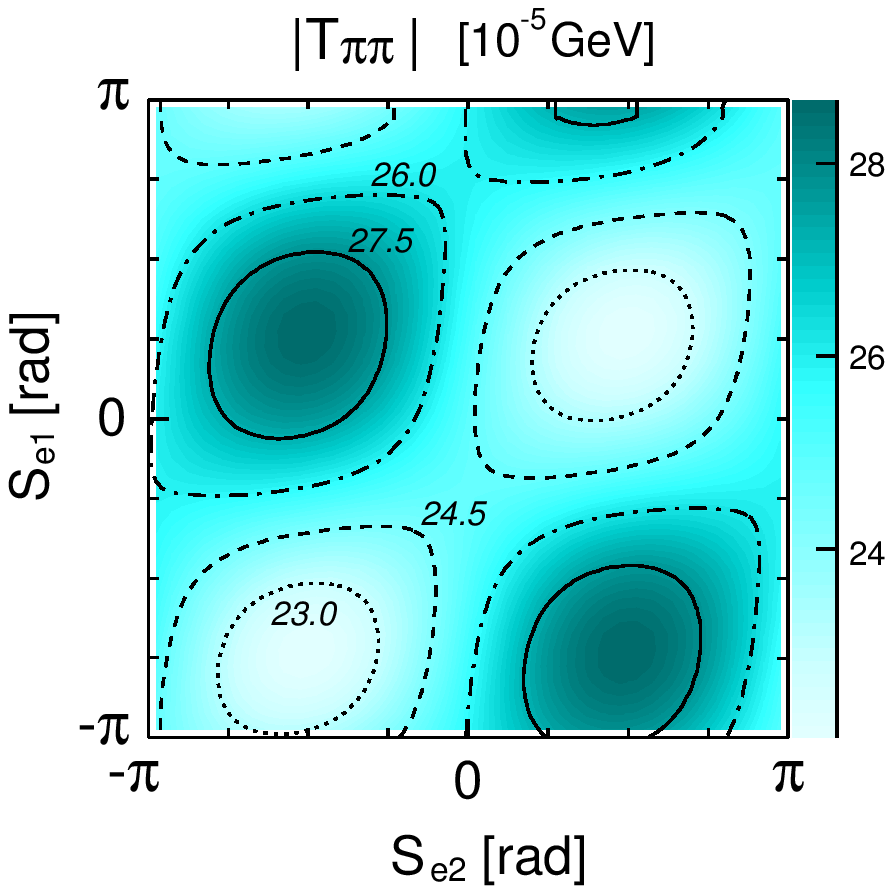}
\includegraphics[height=4.5cm]{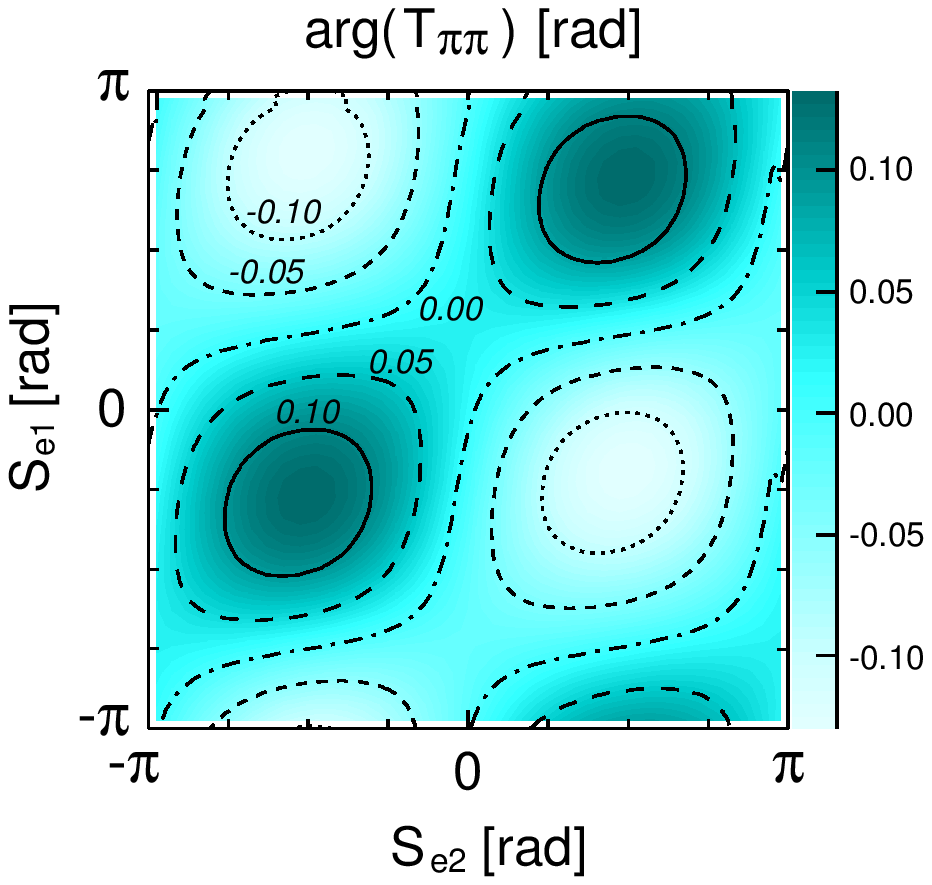}
\includegraphics[height=4.5cm]{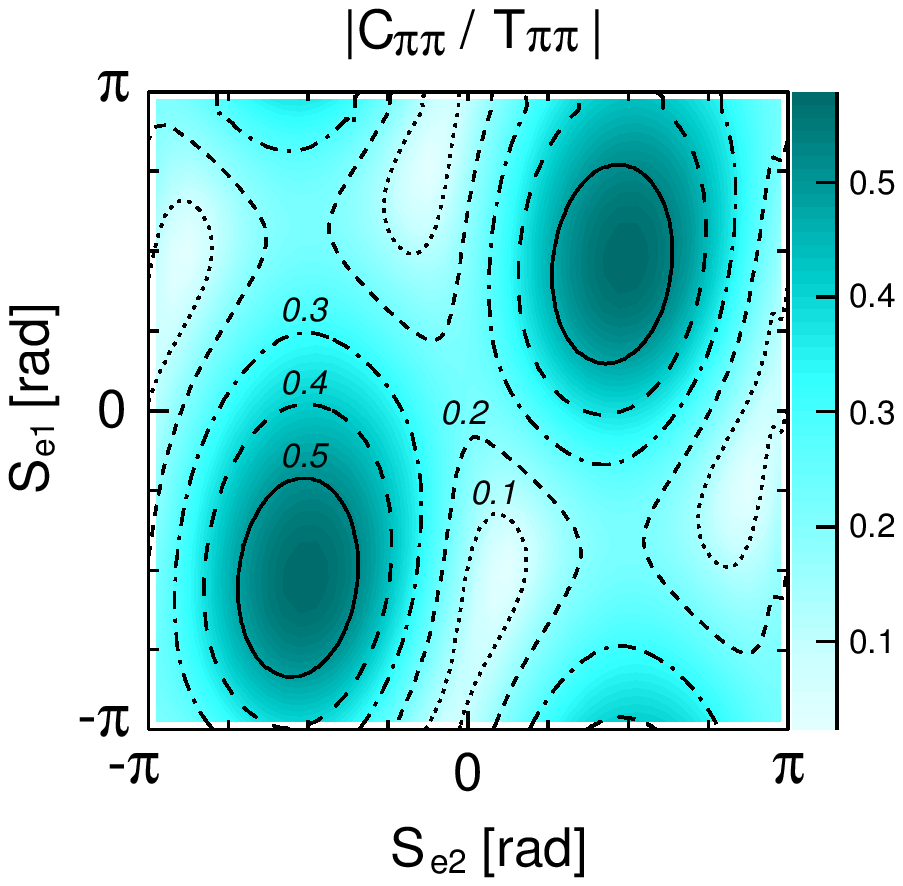}
\includegraphics[height=4.5cm]{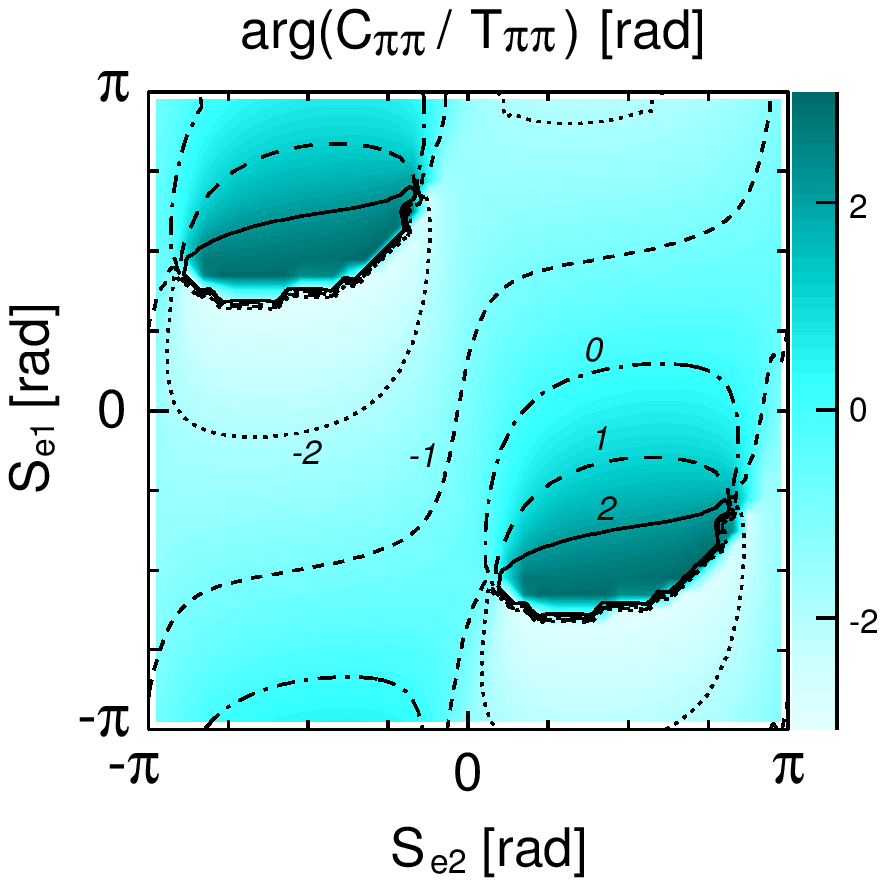}
\caption{$S_{e1}$ and $S_{e2}$ dependencies of the amplitudes $C$ and
$T$, and their ratio $C/T$ for the $B\to\pi\pi$ decays.}
\label{fig3}
\end{center}
\end{figure*}

The $S_{e1}$ and $S_{e2}$ dependencies of the
color-suppressed tree amplitude $C$, the color-allowed tree
amplitude $T$, and their ratio for the $B\to\pi\pi$ decays are displayed in
Fig.~\ref{fig3}, where the definitions of $C$ and $T$ are the same
as in \cite{LMS05}. As argued before, the destructive interference
between Figs.~\ref{fig1}(a) and \ref{fig1}(b) is moderated by the
Glauber factor, so their net contribution increases for
nonvanishing $S_{e2}$. It is observed in Fig.~\ref{fig3} that the
magnitude of $C$ reaches maximum as $S_{e2}\approx -\pi/2$. On the
other hand, Figs.~\ref{fig1}(a) and \ref{fig1}(b) acquire the same
phase factor $\exp(-iS_{e1})$ from the Glauber gluons in the $M_1$
meson. Despite of being an overall factor, it changes the
relative phase between the spectator amplitude and the factorizable
emission amplitude, which includes the important vertex corrections
at NLO \cite{LMS05}. Therefore, $C$ also depends on $S_{e1}$, whose
magnitude reaches maximum for $S_{e1}\approx S_{e2}\approx -\pi/2$.
Because $T$ receives contributions from both the factorizable and spectator
diagrams, the Glauber factors affect its magnitude and argument.
Due to the dominance of the former contribution, the Glauber
effect is minor on $T$, compared to that on $C$.
Figure~\ref{fig3} exhibits that the magnitude of the amplitude
ratio $C/T$ is enhanced by factor 3 as $S_{e1}\approx S_{e2}\approx
-\pi/2$, relative to the value at $S_{e1}= S_{e2}=0$. The result
$C/T=0.58e^{-0.9i}$ at $S_{e1}= S_{e2}= -\pi/2$ for the $B\to\pi\pi$
decays is close to the extraction in \cite{Charng2}.

\begin{figure*}[t]
\begin{center}
\includegraphics[height=4.5cm]{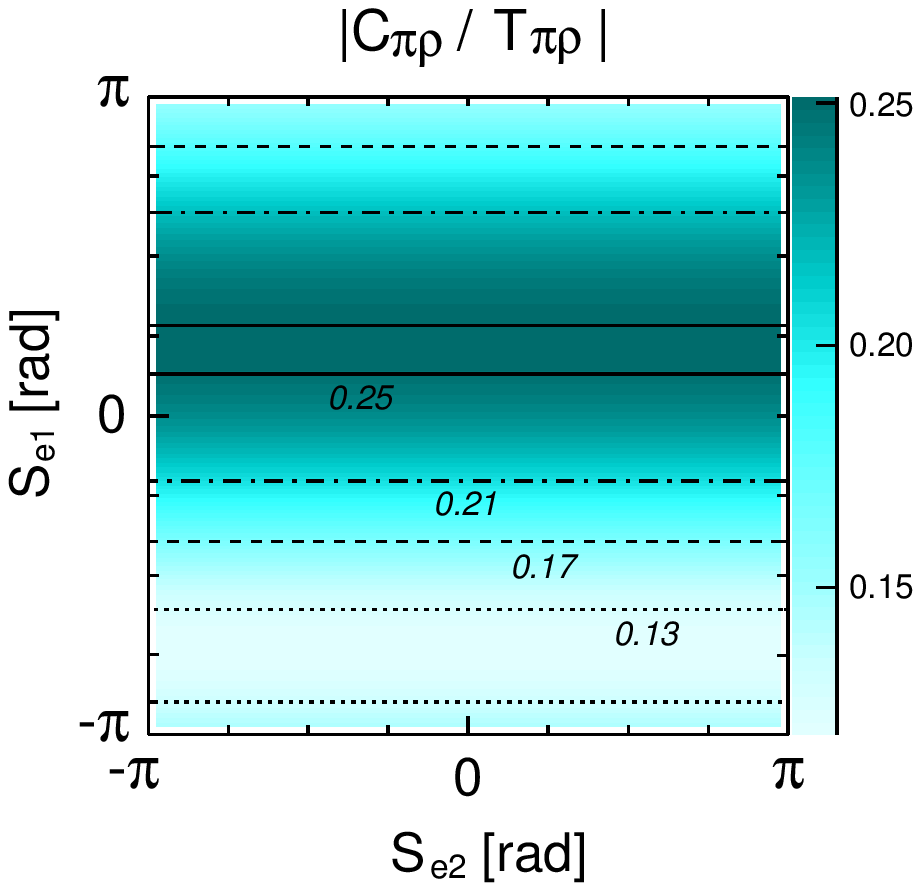}
\includegraphics[height=4.5cm]{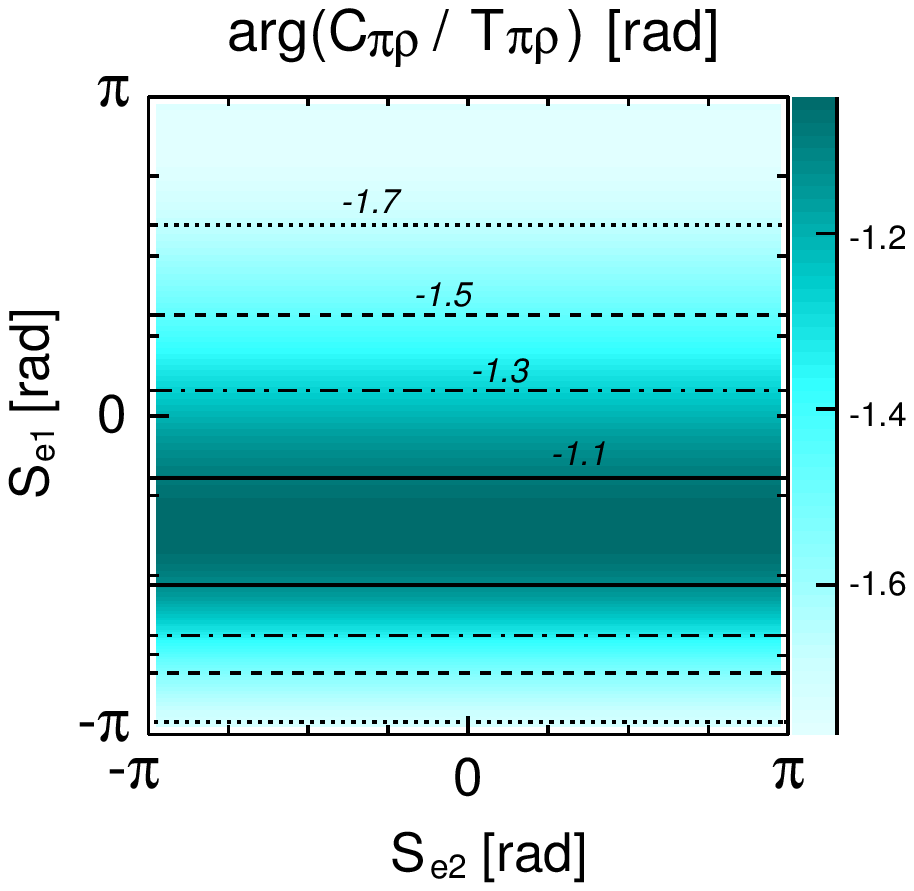}
\includegraphics[height=4.5cm]{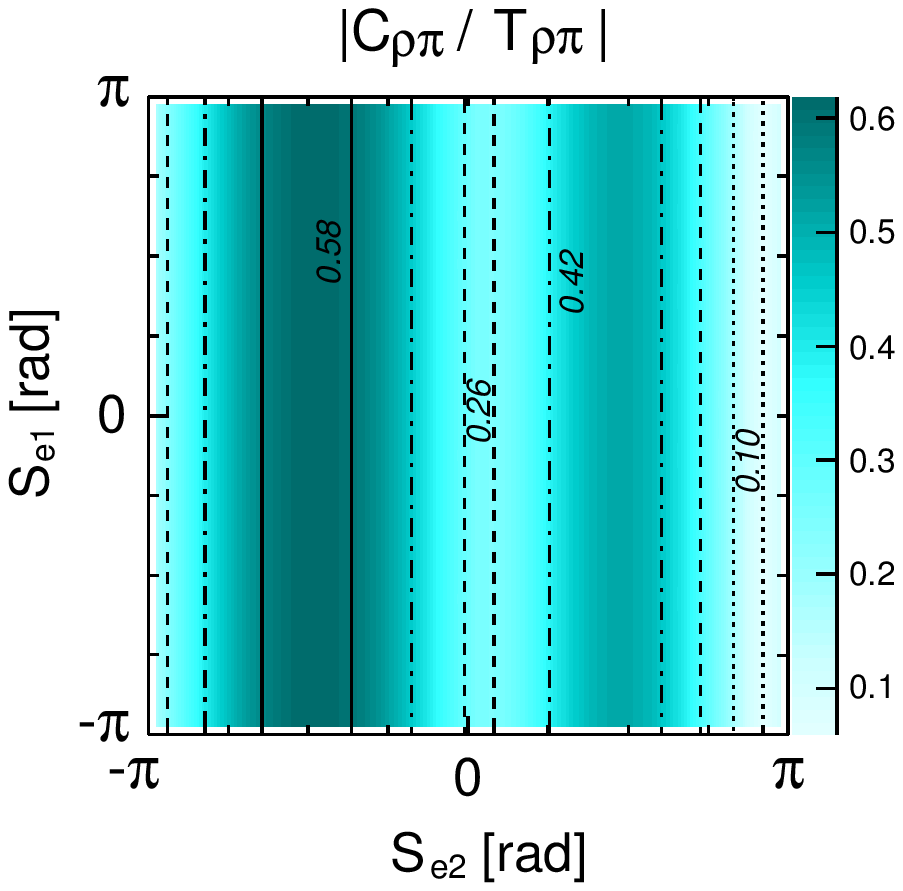}
\includegraphics[height=4.5cm]{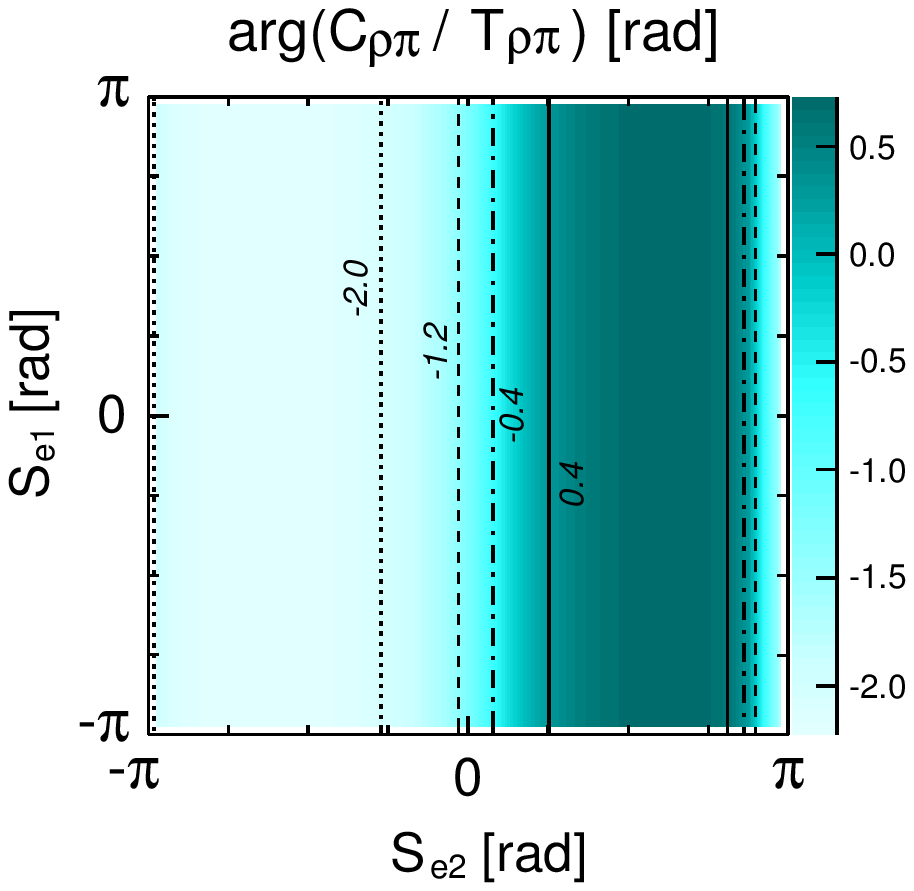}
\includegraphics[height=4.5cm]{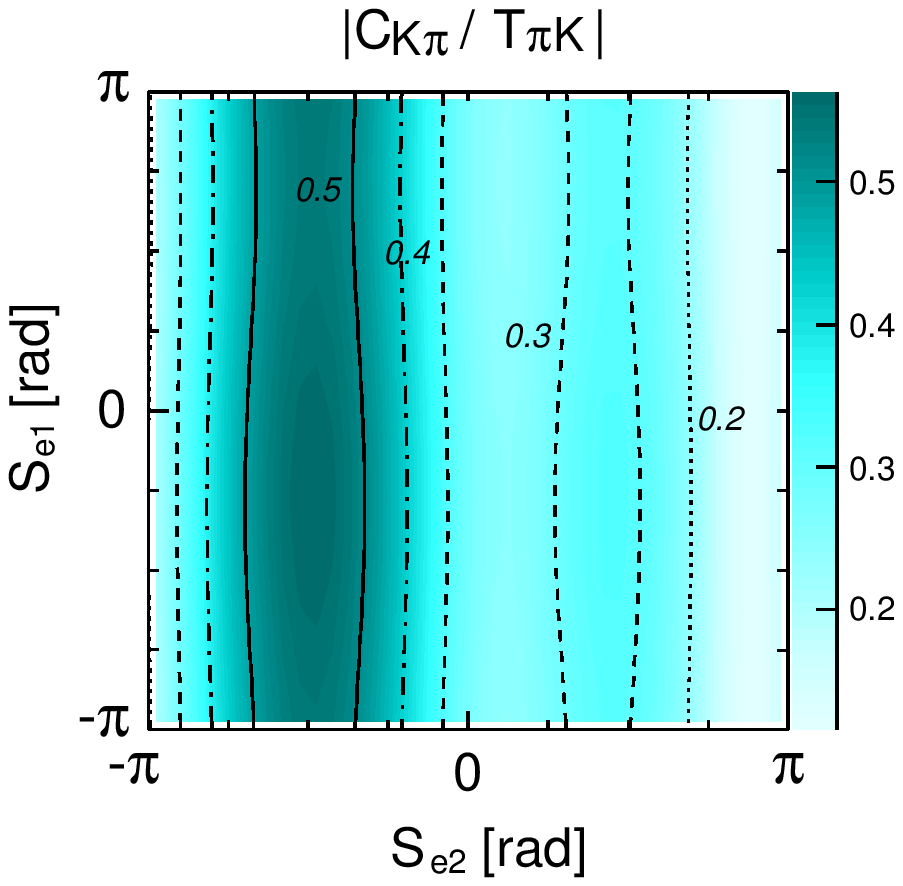}
\includegraphics[height=4.5cm]{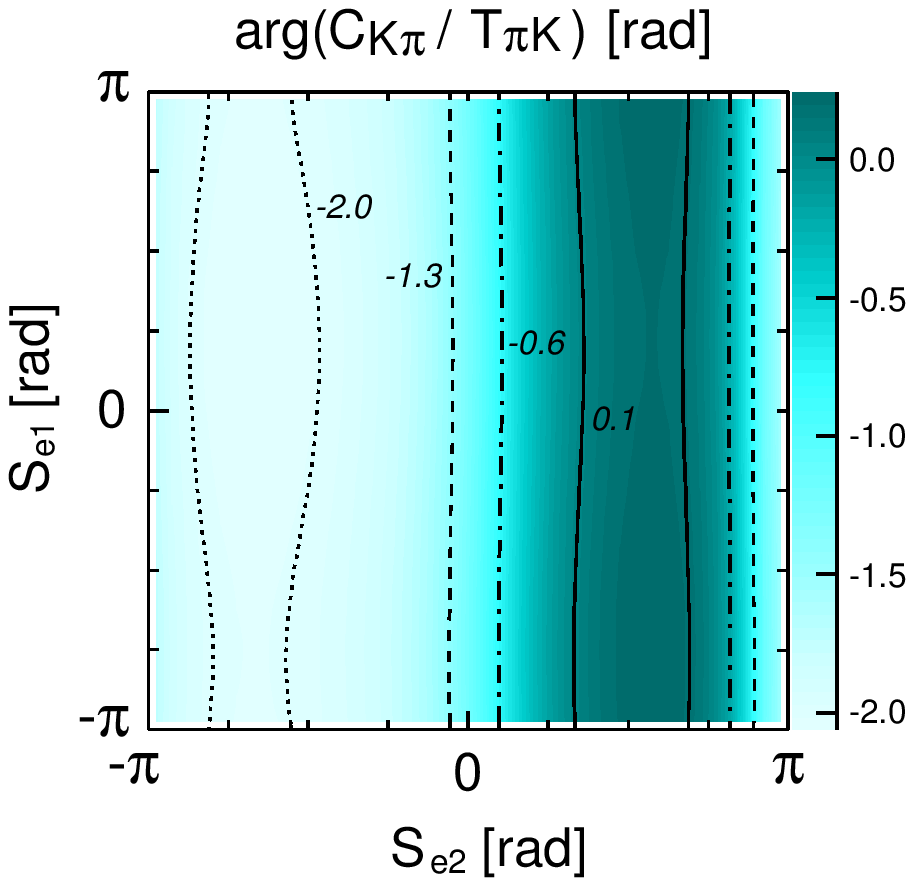}
\caption{$S_{e1}$ and $S_{e2}$ dependencies of the amplitudes $C$
and $T$, and their ratio $C/T$ for the $B\to\pi\rho$ and $\pi K$
decays.}\label{fig35}
\end{center}
\end{figure*}

Similar plots for the $B\to\pi\rho$ and $\pi K$ decays are displayed in
Fig.~\ref{fig35}. The plots for the $B\to\pi\omega$ decays, similar
to those for the $B\to\pi\rho$ ones, are not presented here. Since
only a single pion is involved in each mode, either the Glauber
phase $S_{e1}$ or $S_{e2}$ appears in the modified PQCD factorization
formula. For those modes containing the $B\to\pi$ transition, the
corresponding amplitude ratio $C_{\pi\rho}/T_{\pi\rho}$ depends on
$S_{e1}$ only: the magnitude of $C_{\pi\rho}/T_{\pi\rho}$ decreases
by about 40\%, and the argument decreases by about 10\% as $S_{e1}$
varies from zero to $-\pi/2$. For those modes with $M_2=\pi$, the
corresponding amplitude ratios $C_{\rho\pi}/T_{\rho\pi}$ and
$C_{K\pi}/T_{\pi K}$ mainly depend on $S_{e2}$: both the magnitude
and argument increase by a factor 2, as $S_{e2}$ varies from zero to
$-\pi/2$. As explained before, the variation of $S_{e2}$
modifies the interference pattern between the two spectator diagrams
in Fig.~\ref{fig1}, such that the corresponding Glauber effect
always enhances the magnitude of $C/T$. Compared to the $B\to\pi\pi$
case, the Glauber effects are minor in the $B\to\pi\rho$,
$\pi\omega$, and $\pi K$ decays as expected.

\begin{figure*}[t]
\begin{center}
\includegraphics[height=4.5cm]{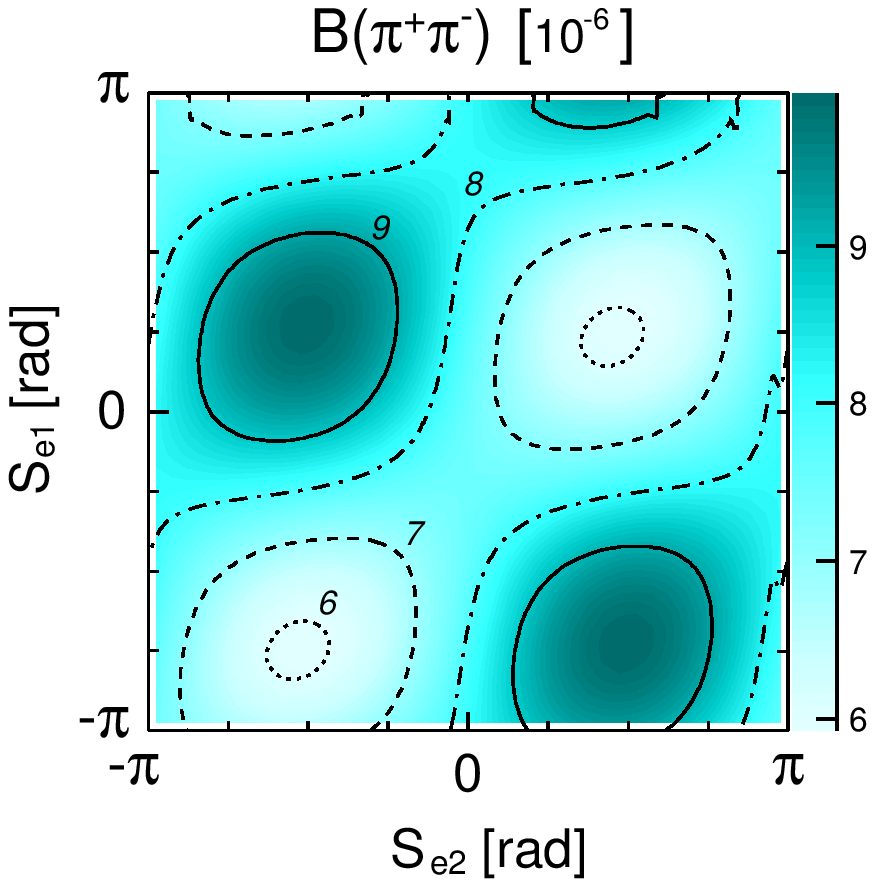}
\includegraphics[height=4.5cm]{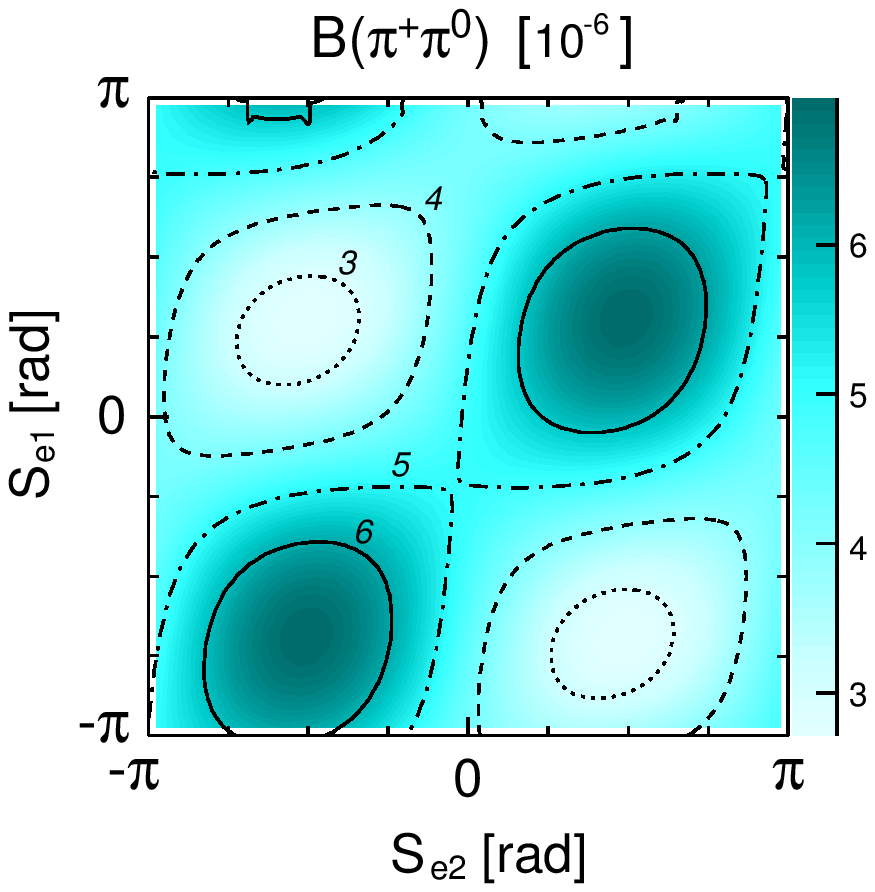}
\includegraphics[height=4.5cm]{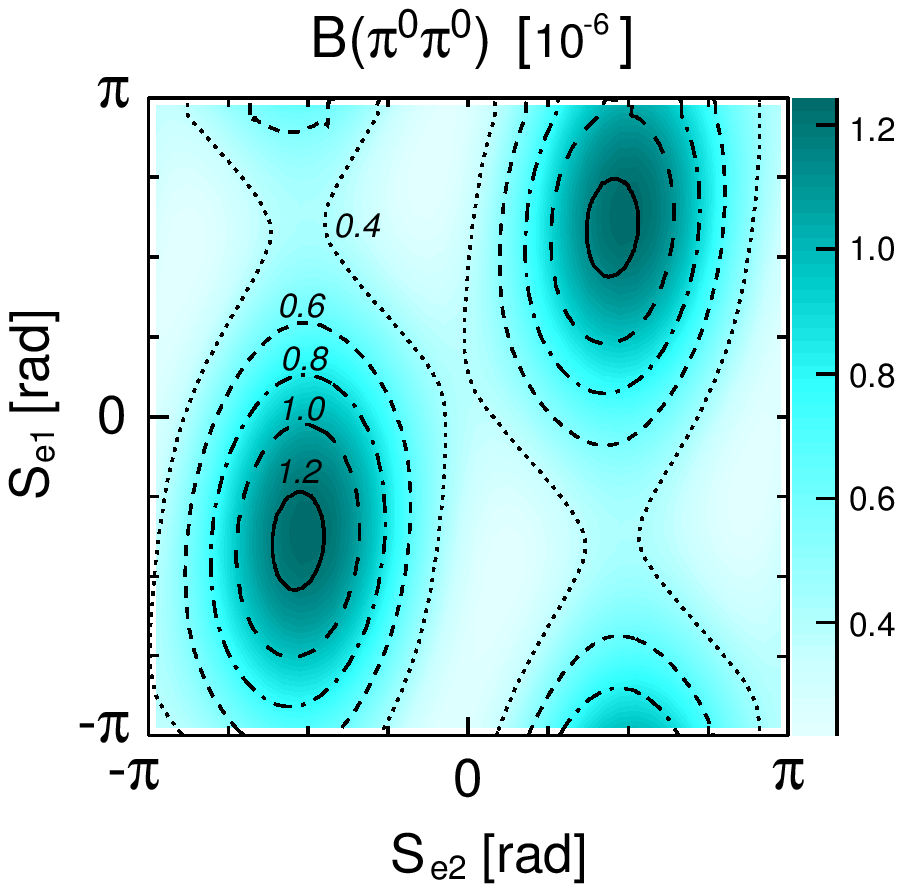}
\includegraphics[height=4.5cm]{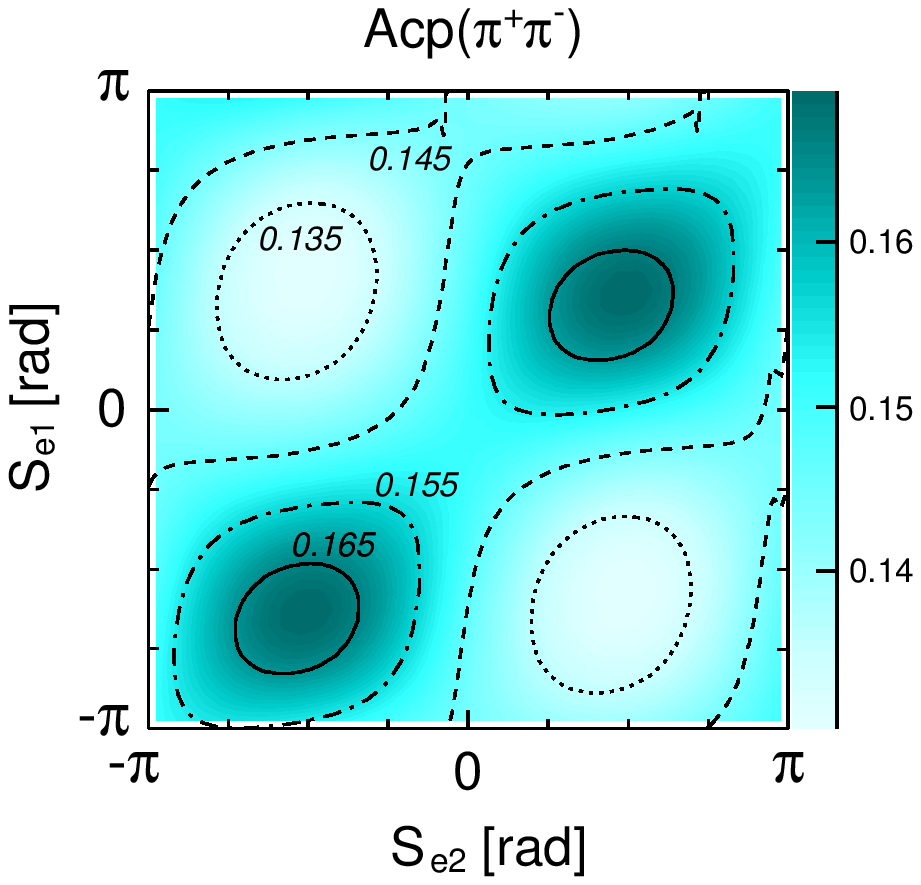}
\includegraphics[height=4.5cm]{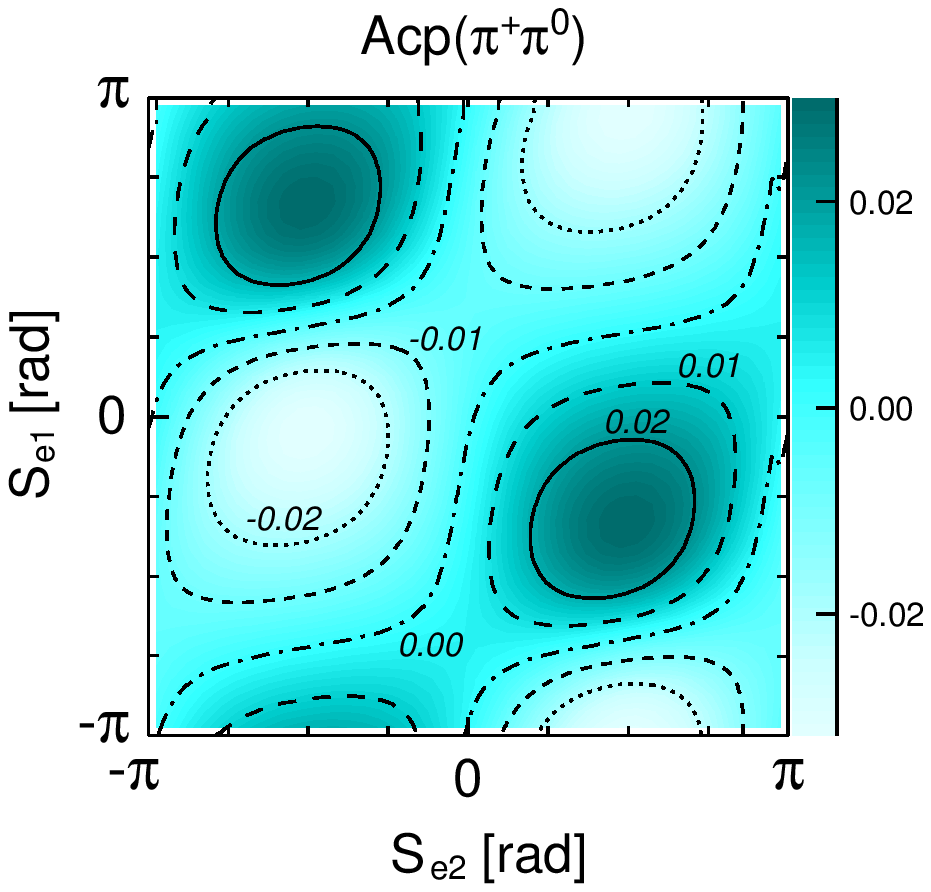}
\includegraphics[height=4.5cm]{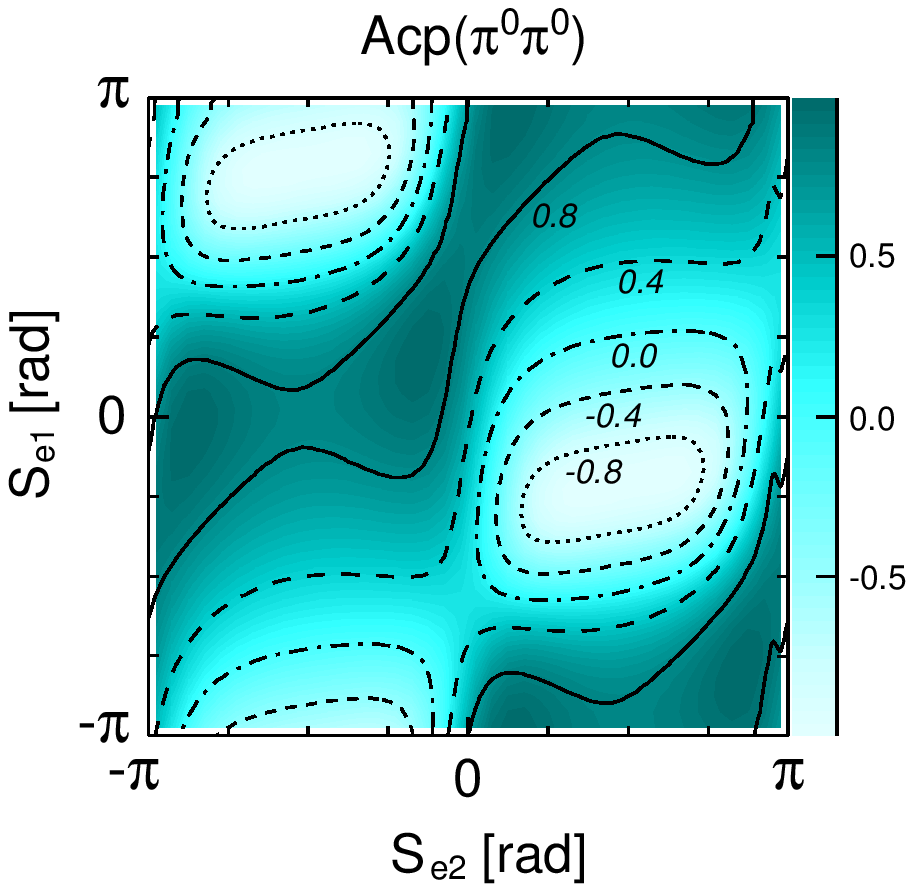}
\caption{$S_{e1}$ and $S_{e2}$ dependencies of the $B\to\pi\pi$ branching
ratios (in units of $10^{-6}$) and direct CP asymmetries.} \label{fig4}
\end{center}
\end{figure*}

The $S_{e1}$ and $S_{e2}$ dependencies of the $B\to\pi\pi$ branching
ratios (in units of $10^{-6}$) and direct CP asymmetries are shown in
Fig.~\ref{fig4}. It is found that the combined effect from the two
Glauber factors decreases the $B^0\to\pi^+\pi^-$ branching ratio
from $7.5\times 10^{-6}$ (corresponding to $S_{e1}=S_{e2}=0$) to
$6.4\times 10^{-6}$ (corresponding to $S_{e1}=S_{e2}=-\pi/2$). On
the contrary, the $B^+\to\pi^+\pi^0$ branching ratio increases from
$5.0\times 10^{-6}$ to $6.6\times 10^{-6}$. That is, the ratio of the above
two predictions becomes consistent with the data. The Glauber effect
is not dramatic, because these two modes are dominated by the
color-allowed tree amplitude $T$. The
enhancement of the $B^0\to\pi^0\pi^0$ branching ratio from about
$0.38\times 10^{-6}$ to $1.2\times 10^{-6}$ is significant, rendering
the NLO PQCD prediction agree well with the data
$(1.17 \pm 0.13) \times 10^{-6}$. Note that the above data have been updated by
combining the BaBar ones in \cite{Amhis:2012bh} with those recently
reported by Belle \cite{update}.
The improved consistency of the three predicted branching ratios with the data
is highly nontrivial, which requires the simultaneous
adjustment of the relative phases between the spectator diagrams, and
between the spectator amplitude and other emission amplitudes.
It is seen that the Glauber factor does not change much the direct CP
asymmetries in the $B^0\to\pi^+\pi^-$ and $B^+\to\pi^+\pi^0$ decays,
which contain the amplitude $T$. The impact on the
$B^0\to\pi^0\pi^0$ direct CP asymmetry is obvious in
Fig.~\ref{fig3}: the predicted $A_{CP}(\pi^0\pi^0)$ decreases from
0.59 to 0.36, closer to the central value of the data $0.03 \pm 0.17$,
when one varies the phases from $S_{e1}=S_{e2}=0$
to $S_{e1}=S_{e2}=-\pi/2$. The above data have been also updated by
combining the BaBar ones in \cite{Amhis:2012bh} with those recently
reported by Belle \cite{update}.

\begin{figure*}[t]
\begin{center}
\includegraphics[height=4.5cm]{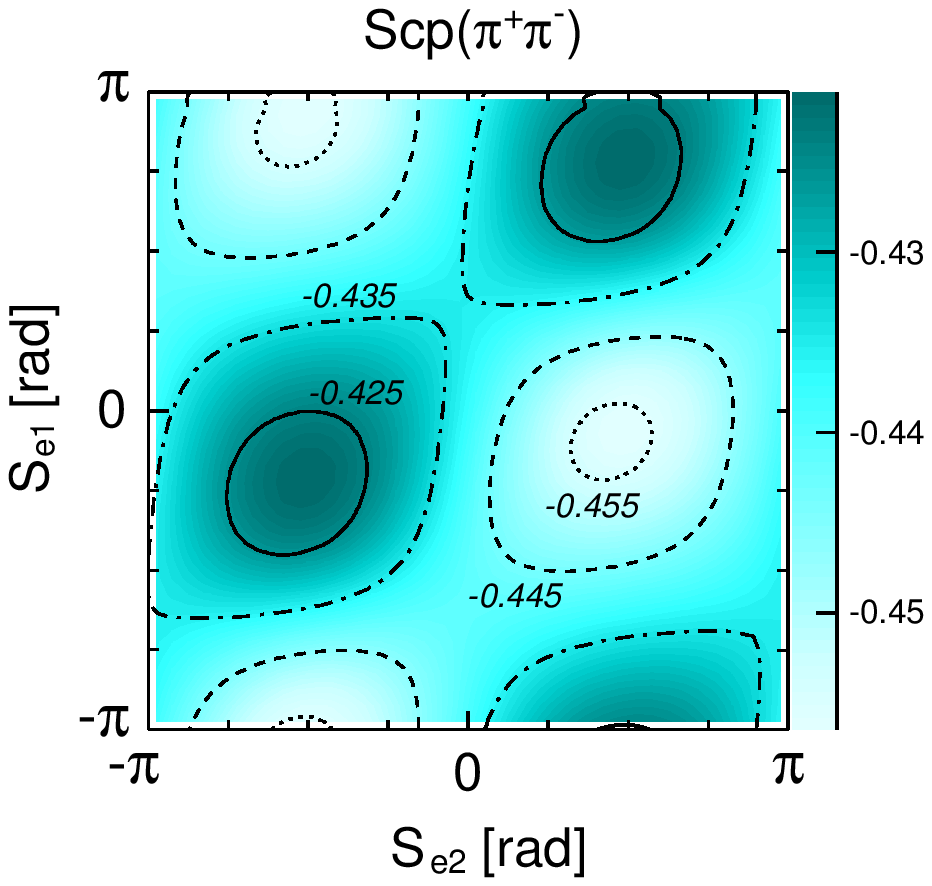}
\includegraphics[height=4.5cm]{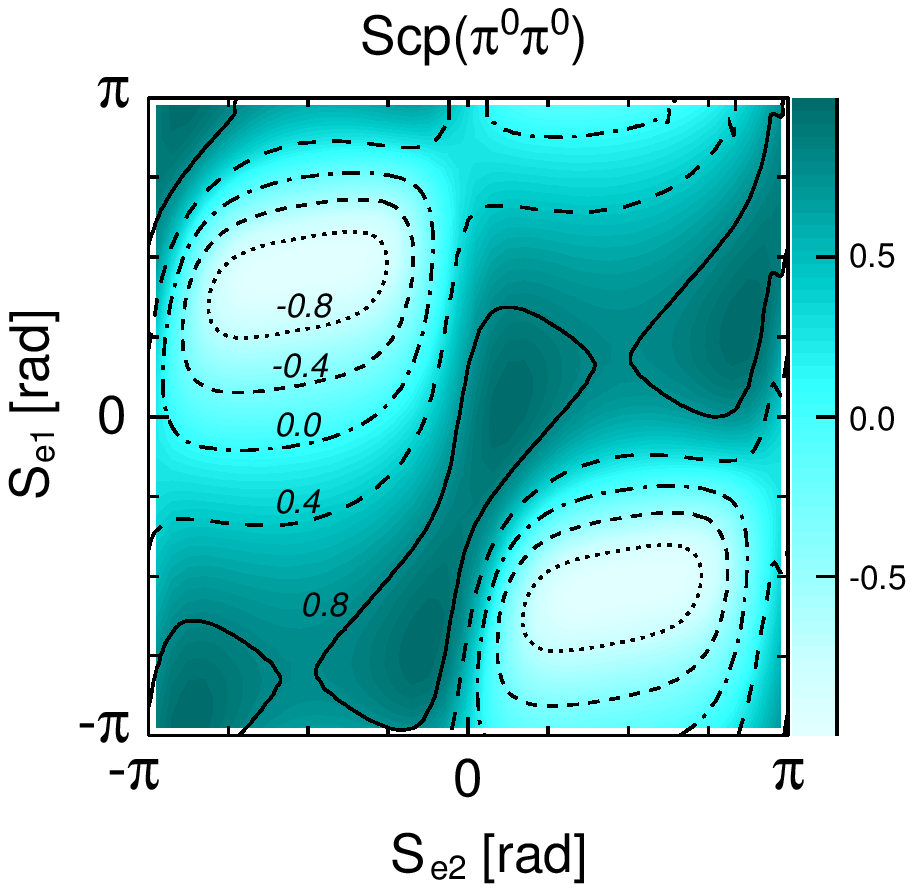}
\includegraphics[height=4.5cm]{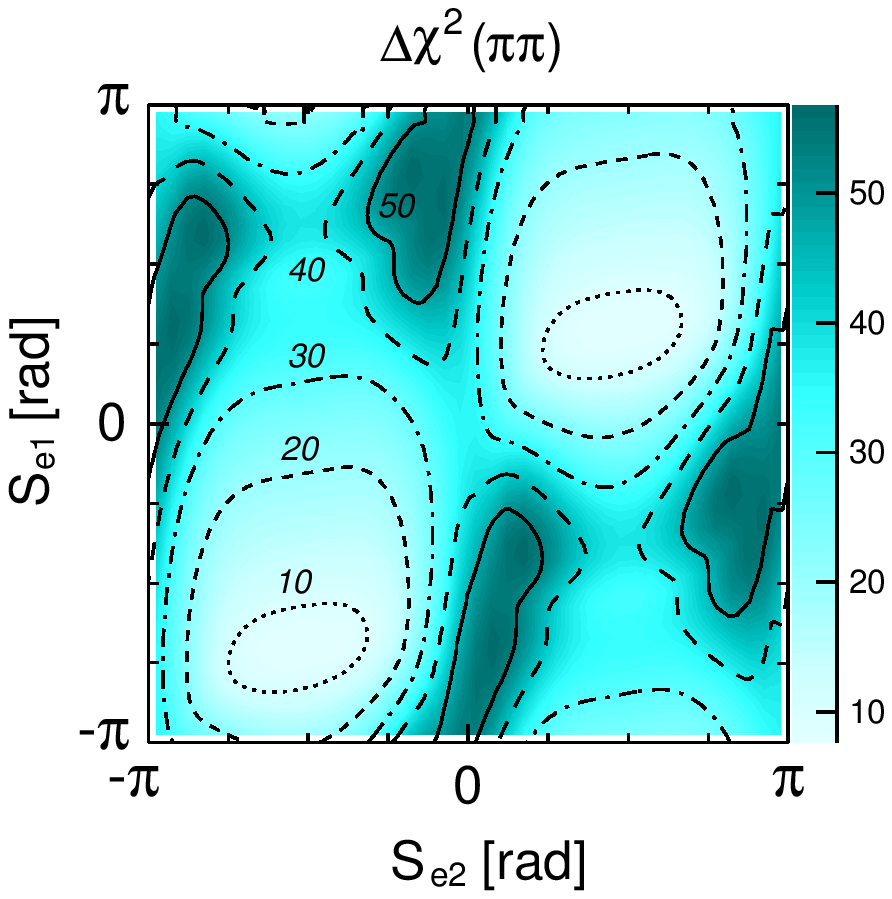}
\caption{$S_{e1}$ and $S_{e2}$ dependencies of the $B\to\pi\pi$ mixing-induced
CP asymmetries, and $\Delta\chi^2$.} \label{fig5}
\end{center}
\end{figure*}

The NLO PQCD predictions for the mixing-induced CP asymmetries in the
$B\to\pi\pi$ decays with the variation of
$S_{e1}$ and $S_{e2}$ are exhibited in Fig.~\ref{fig5}.
The prediction for $S_{CP}(\pi^0\pi^0)$ is more sensitive to the
Glauber phases compared to that for $S_{CP}(\pi^+\pi^-)$, since the
$B^0\to\pi^+\pi^-$ mode is dominated by the color-allowed tree
amplitude. The latter remains around $-0.43$ under the variation of
$S_{e1}$ and $S_{e2}$, which is lower than the data $-0.66\pm 0.06$
\cite{Amhis:2012bh}. The former reduces from $0.80$ to $0.63$, as one tunes
the phases from $S_{e1}=S_{e2}=0$ to $S_{e1}=S_{e2}=-\pi/2$. 
To quantize the improvement of the consistency between the PQCD predictions
and the data attributed to the inclusion of the Glauber
phases, we define
\begin{eqnarray}
\Delta\chi^2= \frac{({\rm data}\;{\rm mean} - {\rm theory}\;{\rm value})^2}
{\sqrt{{\rm data\;error}^2 + (0.30\times{\rm theoty\;value})^2}},
\end{eqnarray}
where the unknown theoretical uncertainty is assumed to be 30\%.
We stress that we have not attempted to undertake the best fit, but
illustrate the improvement by computing $\Delta\chi^2$. The
last plot in Fig.~\ref{fig5} summarizes the reduction of
$\Delta\chi^2$ in the global fit of the PQCD predictions with the
Glauber phases to the $B\to\pi\pi$ data. As
expected, the value drops significantly from about 36 (corresponding to
$S_{e1}=S_{e2}=0$) to around 11 (corresponding to
$S_{e1}=S_{e2}=-\pi/2$). That is, the Glauber gluons indeed affect
the ratio $C/T$ toward the indication of the data.

\begin{figure*}[t]
\begin{center}
\includegraphics[height=3.8cm]{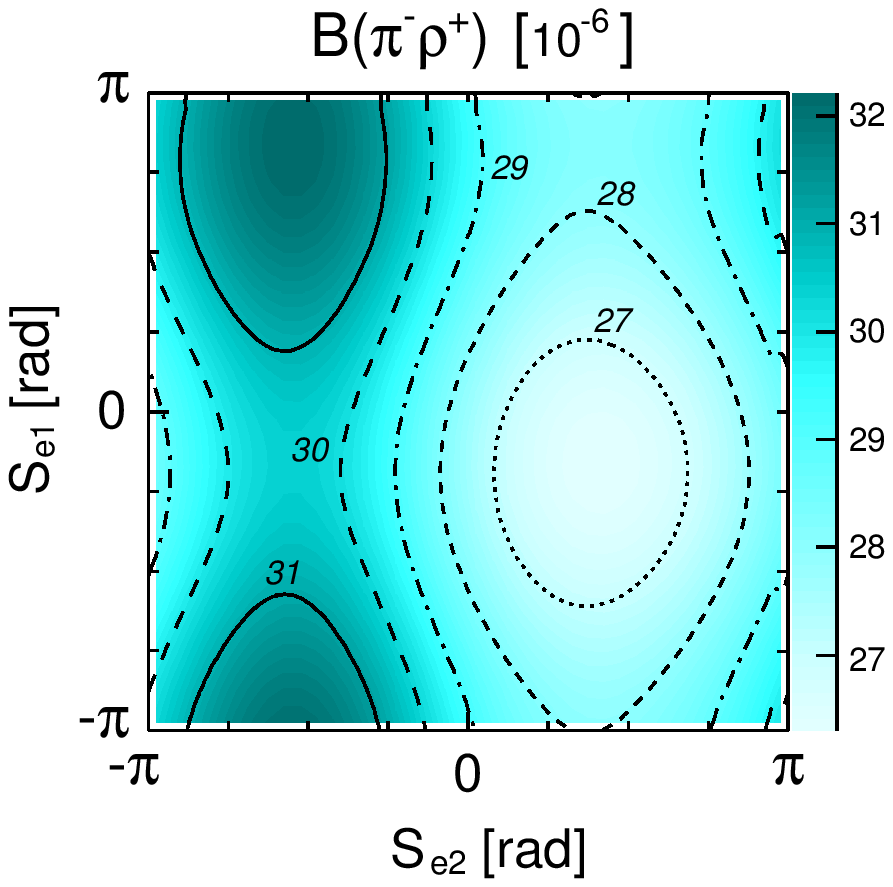}
\includegraphics[height=3.8cm]{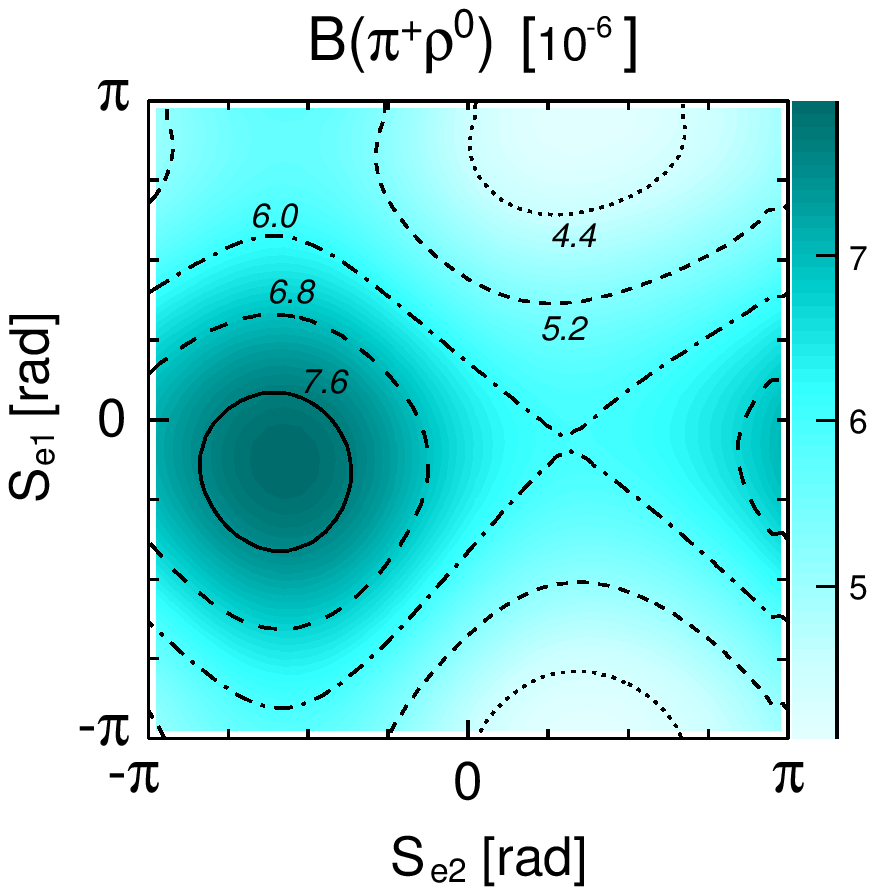}
\includegraphics[height=3.8cm]{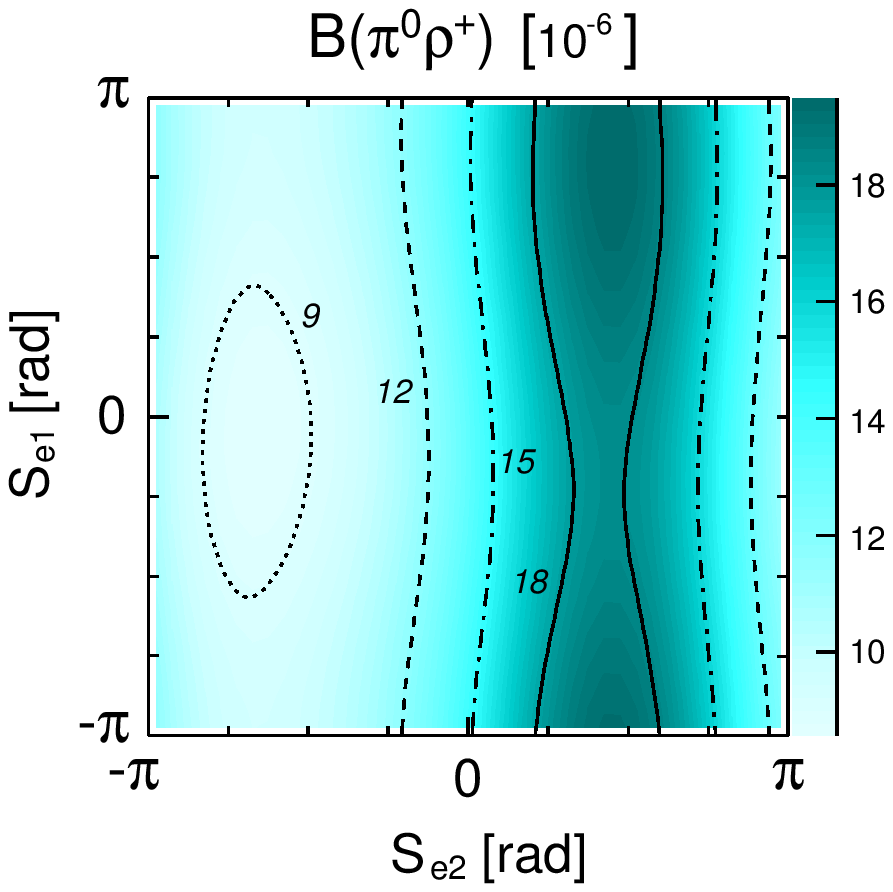}
\includegraphics[height=3.8cm]{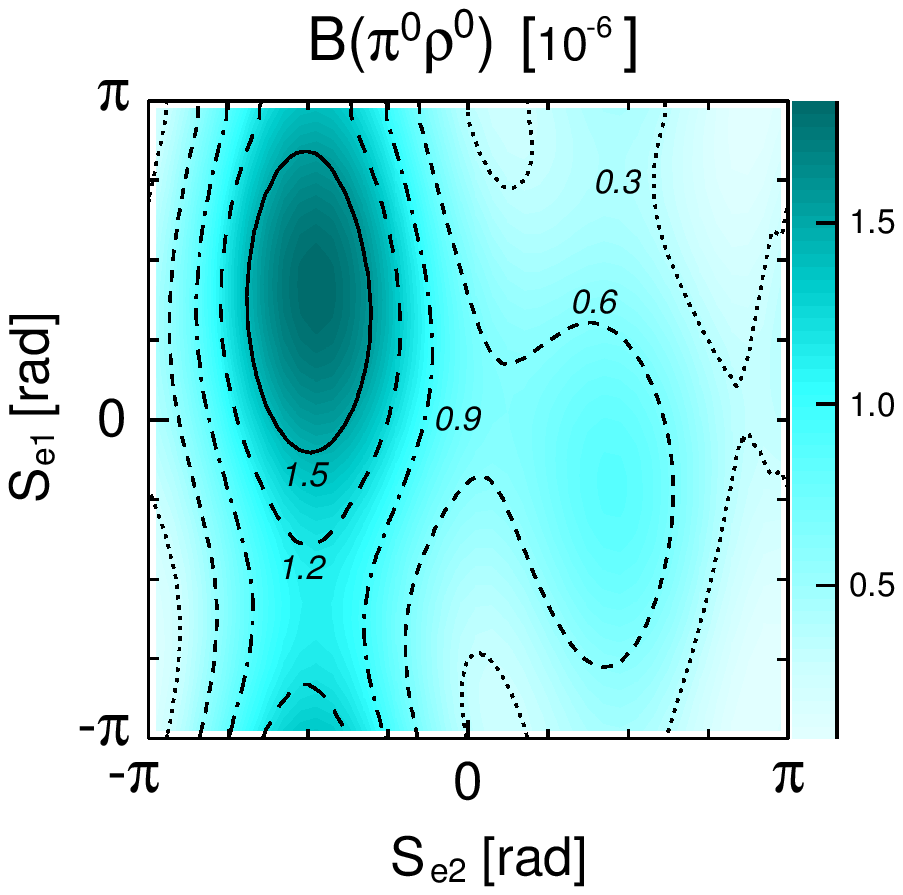}
\includegraphics[height=3.8cm]{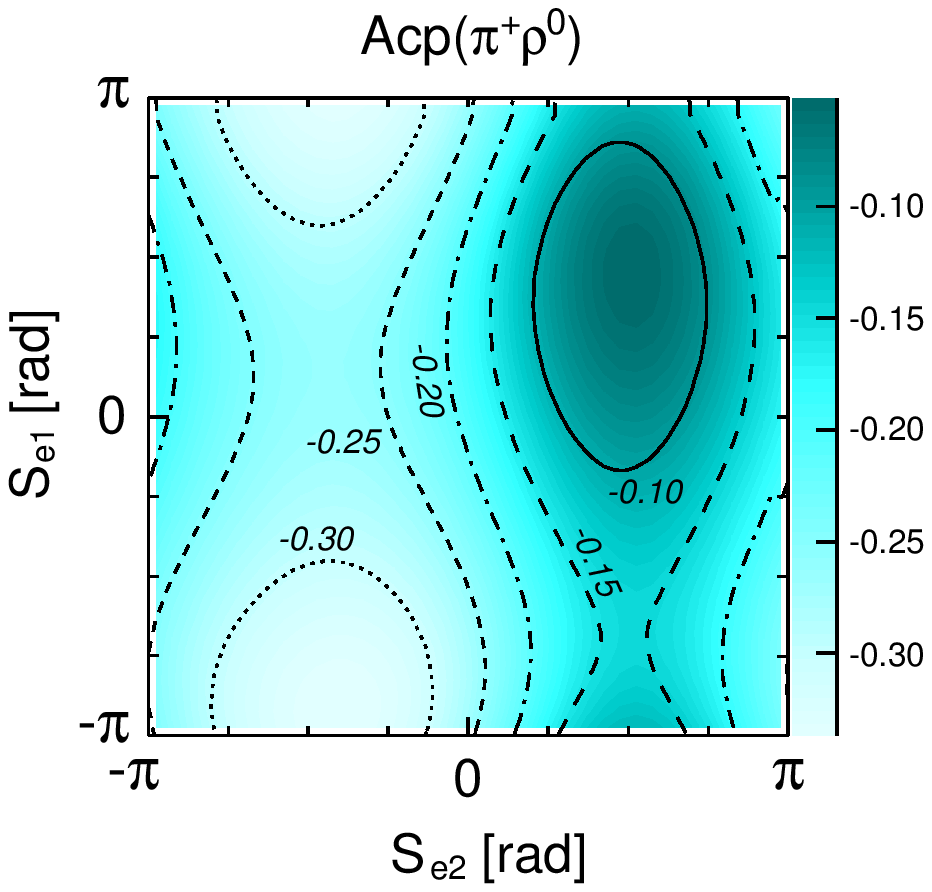}
\includegraphics[height=3.8cm]{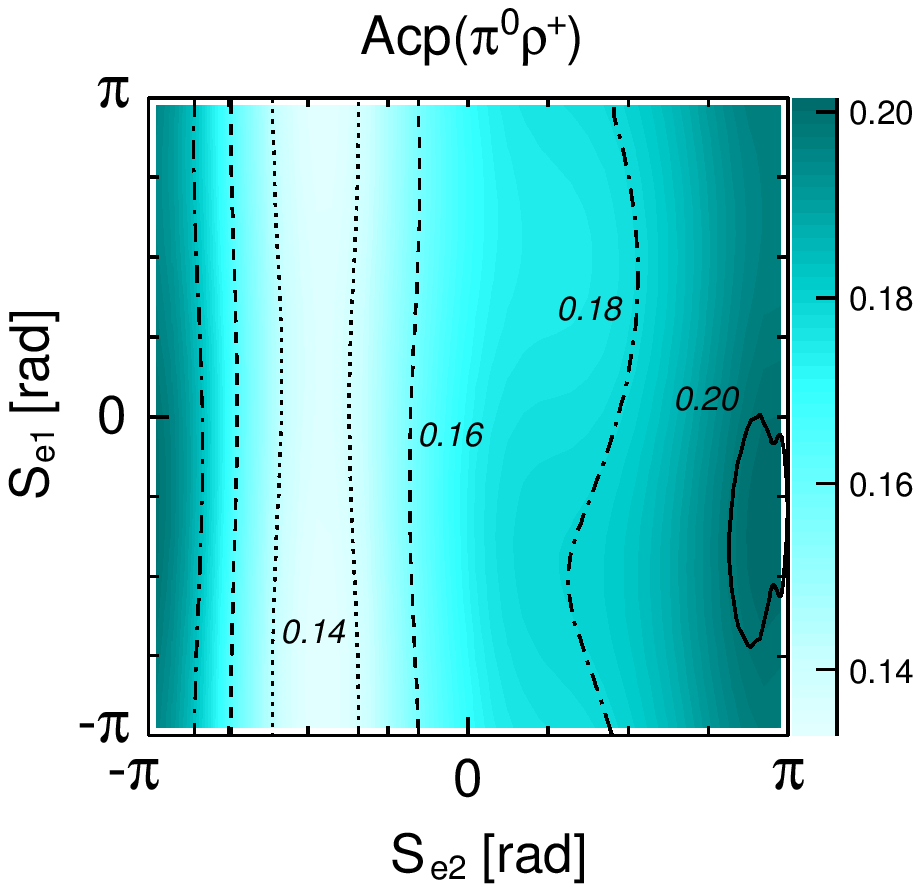}
\includegraphics[height=3.8cm]{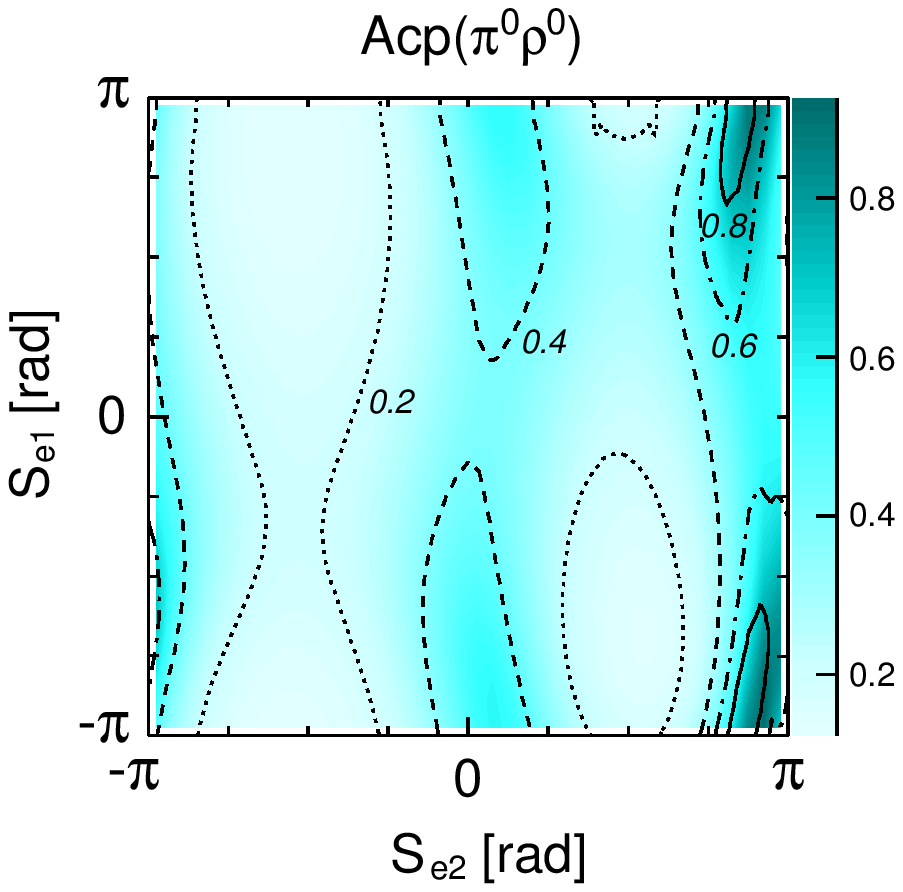}
\includegraphics[height=3.8cm]{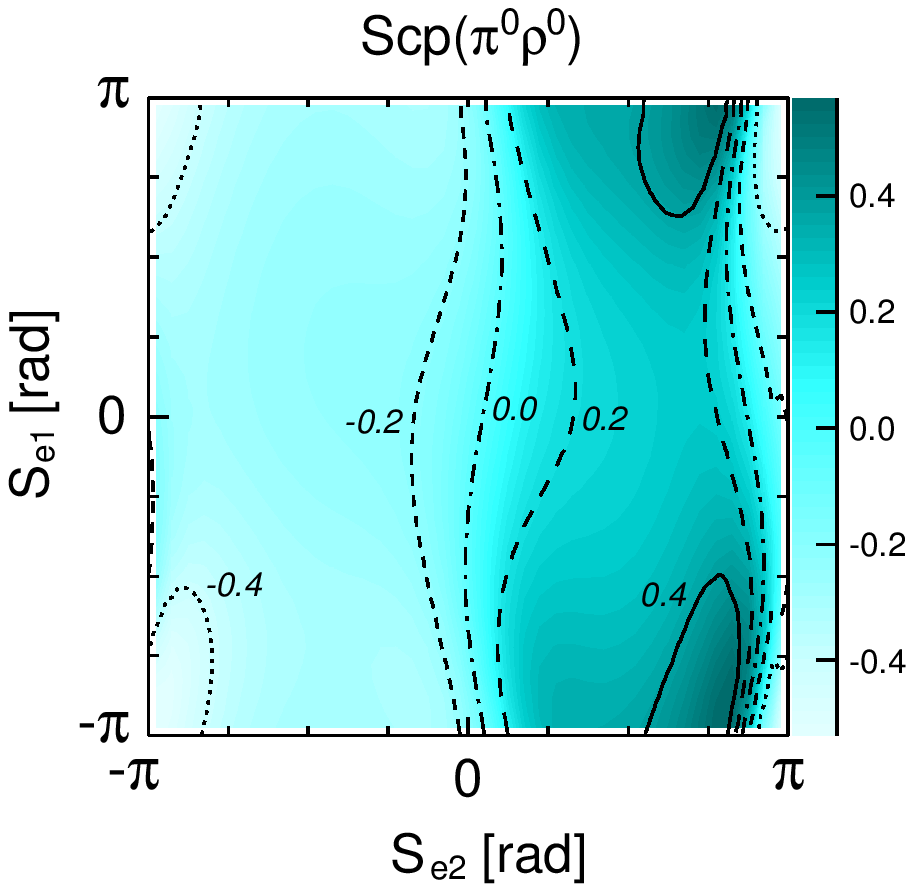}
\caption{$S_{e1}$ and $S_{e2}$ dependencies of the $B\to\pi\rho$
branching ratios (in units of $10^{-6}$), direct CP asymmetries, and
mixing-induced CP asymmetry.} \label{fig6}
\end{center}
\end{figure*}

The $S_{e1}$ and $S_{e2}$ dependencies of the $B\to\pi\rho$ branching
ratios (in units of $10^{-6}$) and direct CP asymmetries are shown in
Fig.~\ref{fig6}. Because only a single pion is involved in these
modes, the Glauber effect is minor. The NLO PQCD prediction for
the branching ratio $B(\pi^\pm\rho^\mp)$ increases a bit from $27.8\times 10^{-6}$
to $30.8\times 10^{-6}$, as one tunes the phases from
$S_{e1}=S_{e2}=0$ to $S_{e1}=S_{e2}=-\pi/2$, which slightly
overshoots the data. The predicted $B(\pi^+\rho^0)$ increases
from $6.5\times 10^{-6}$ to $7.2\times 10^{-6}$, while the predicted
$B(\pi^0\rho^+)$ decreases from $13.3\times 10^{-6}$ to
$9.3\times 10^{-6}$. The predicted $B(\pi^0\rho^0)$ changes
more dramatically under the variation of the Glauber factors, since
it is dominated by the color-suppressed tree amplitude: it is
enhanced from $0.70\times 10^{-6}$ to about $1.1\times 10^{-6}$. The
predictions for $B(\pi^+\rho^0)$ and $B(\pi^0\rho^0)$
become closer to the data. The current data for the direct CP
asymmetries in the $B\to\pi\rho$ decays and for the mixing-induced
CP asymmetry $S_{CP}(\pi^0\rho^0)$ still suffer huge uncertainties.

\begin{figure*}[t]
\begin{center}
\includegraphics[height=4.5cm]{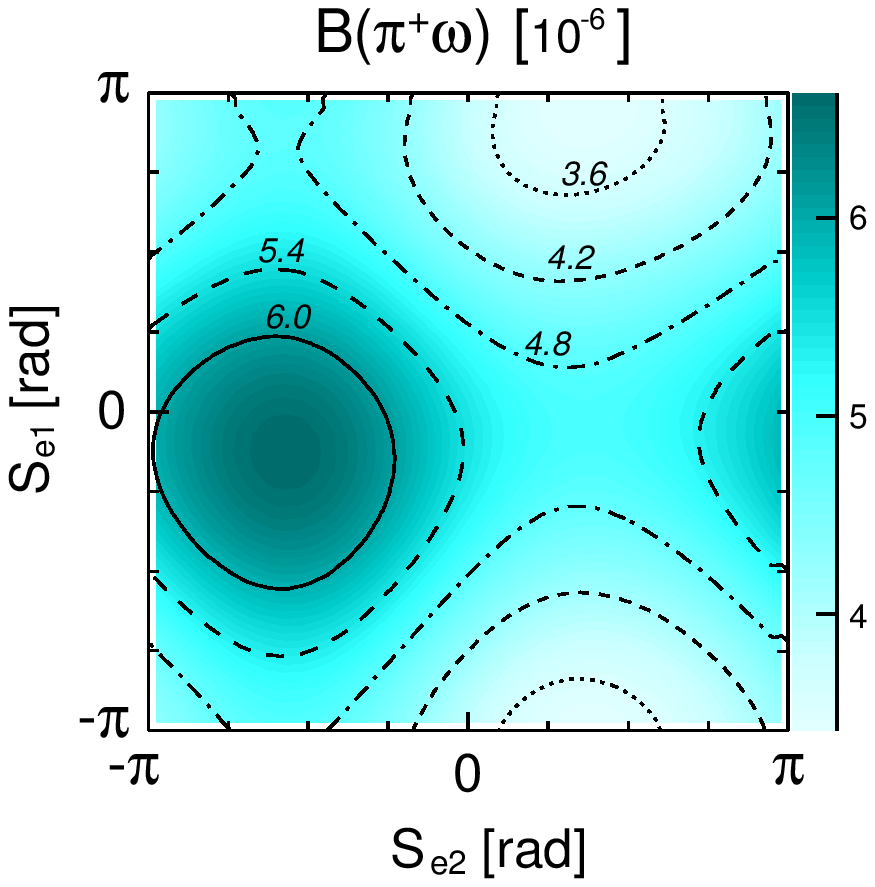}
\includegraphics[height=4.5cm]{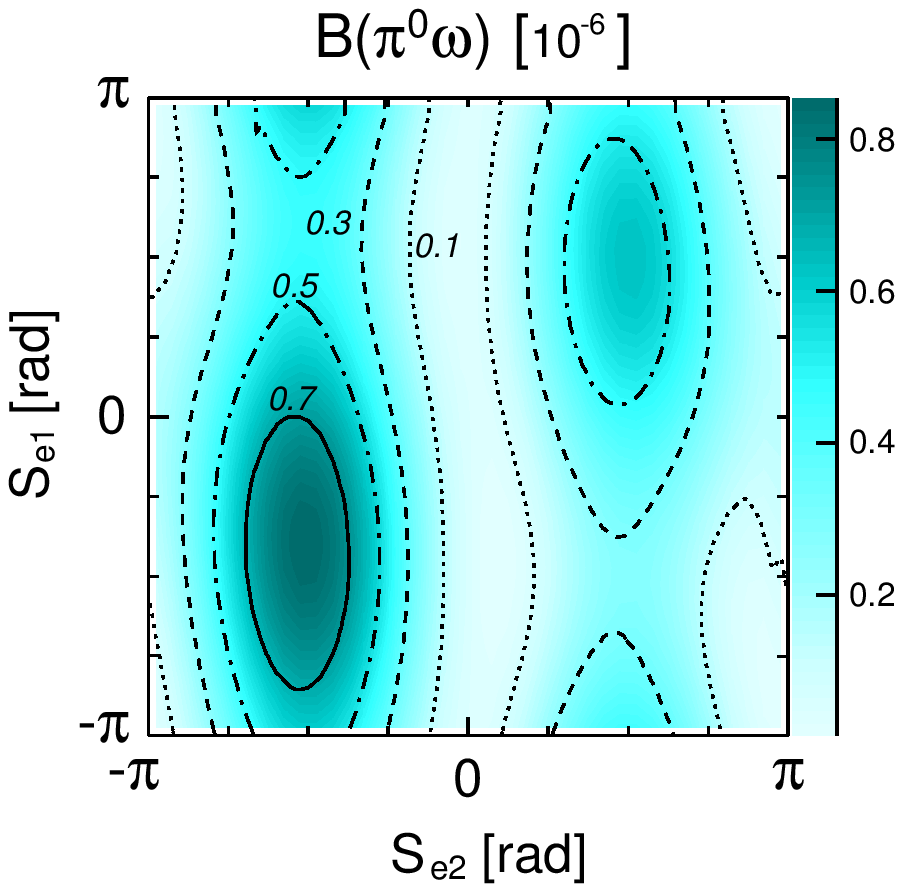}

\includegraphics[height=4.5cm]{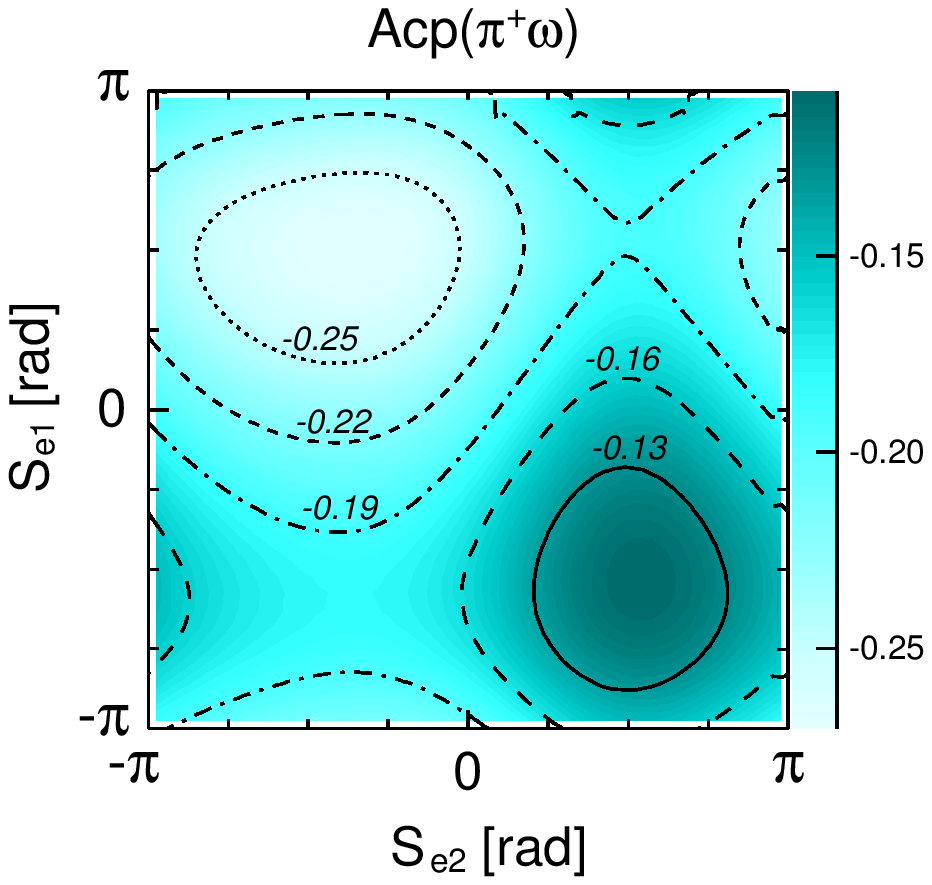}
\includegraphics[height=4.5cm]{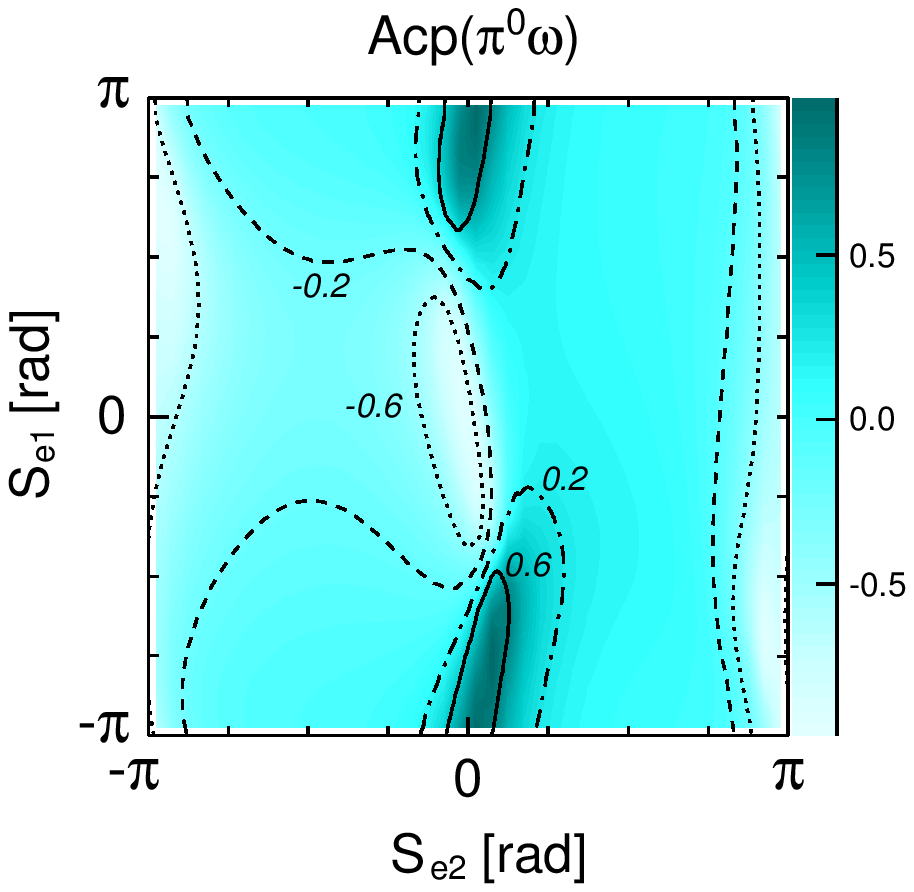}
\includegraphics[height=4.5cm]{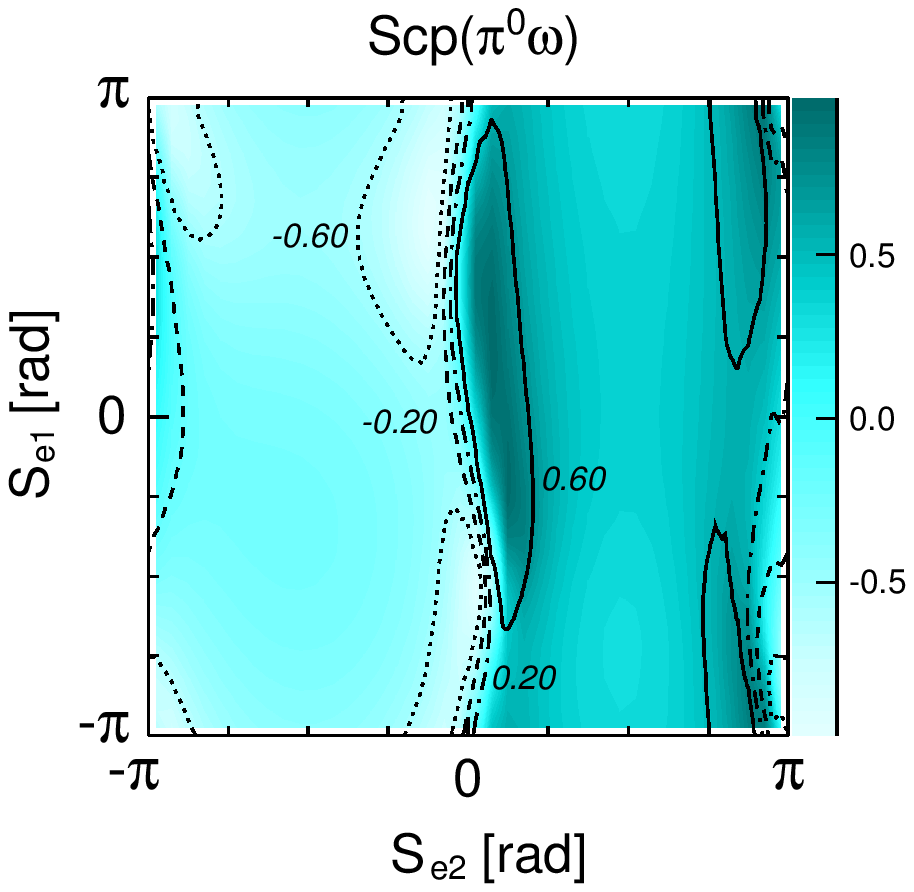}
\caption{$S_{e1}$ and $S_{e2}$ dependencies of the $B\to\pi\omega$
branching ratios (in units of $10^{-6}$), direct CP asymmetries, and
mixing-induced CP asymmetry.} \label{fig8}
\end{center}
\end{figure*}

The behavior of the $B\to\pi\omega$ modes with the Glauber phases
is similar to that of the corresponding
$B\to\pi\rho$ modes, as shown in Fig.~\ref{fig8}. The NLO PQCD
prediction for $B(\pi^+\omega)$ increases from $5.4\times
10^{-6}$ to $6.1\times 10^{-6}$, as one tunes the phases from
$S_{e1}=S_{e2}=0$ to $S_{e1}=S_{e2}=-\pi/2$. The modified result is
more consistent with the data $(6.9\pm 0.5)\times 10^{-6}$
\cite{Amhis:2012bh}. The prediction for $B(\pi^0\omega)$
increases from $0.04\times 10^{-6}$ to $0.85\times 10^{-6}$,
above the upper bound $0.5\times 10^{-6}$ \cite{Amhis:2012bh}. As remarked
before, the present formalism is a simplified one with the
convolution between the Glauber factors and the standard PQCD
factorization formulas being neglected. We shall refine our
predictions, when the data for $B(\pi^0\omega)$ become
available. As to the direct CP asymmetries, the predicted
$A_{CP}(\pi^+\omega)$ remains around $-0.2$ under the variation of
the Glauber phases. The CP asymmetries $A_{CP}(\pi^0\omega)$ and
$S_{CP}(\pi^0\omega)$ are more sensitive to the Glauber phases, and
the predicted value for the former (latter) varies from
$-0.99$ to $-0.12$ (from $-0.11$ to $-0.26$). The current data for
the direct CP asymmetries and mixing-induced CP asymmetry either
have large uncertainties, or are not yet available.

\begin{figure*}[t]
\begin{center}
\includegraphics[height=4.5cm]{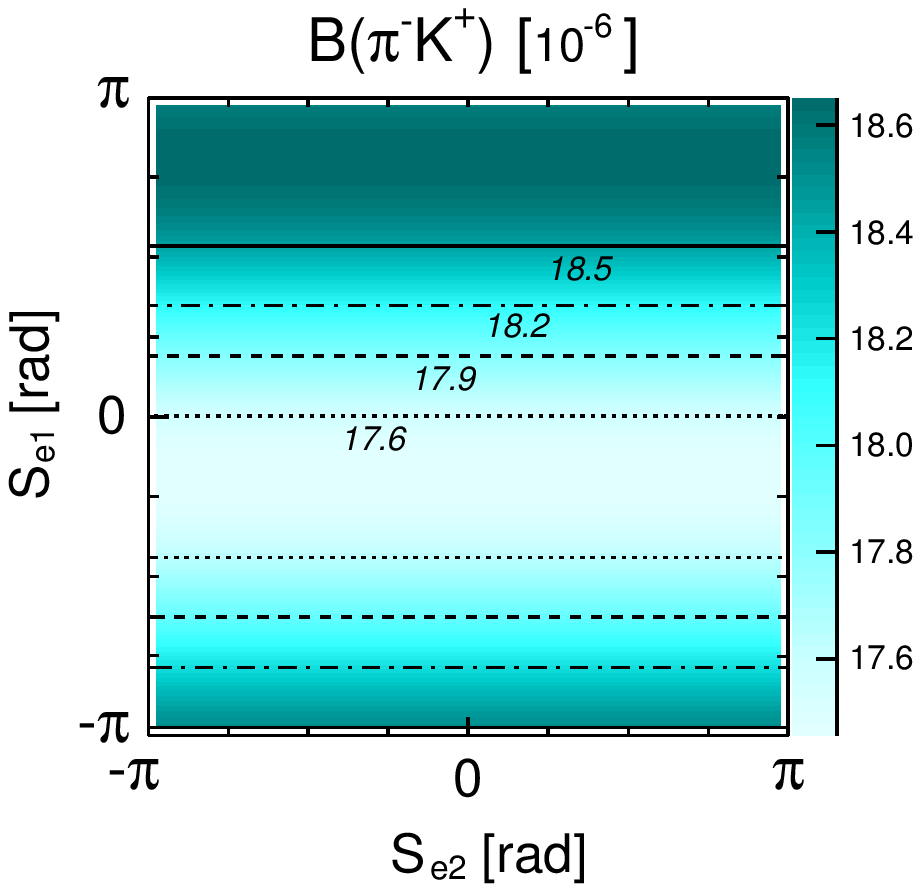}
\includegraphics[height=4.5cm]{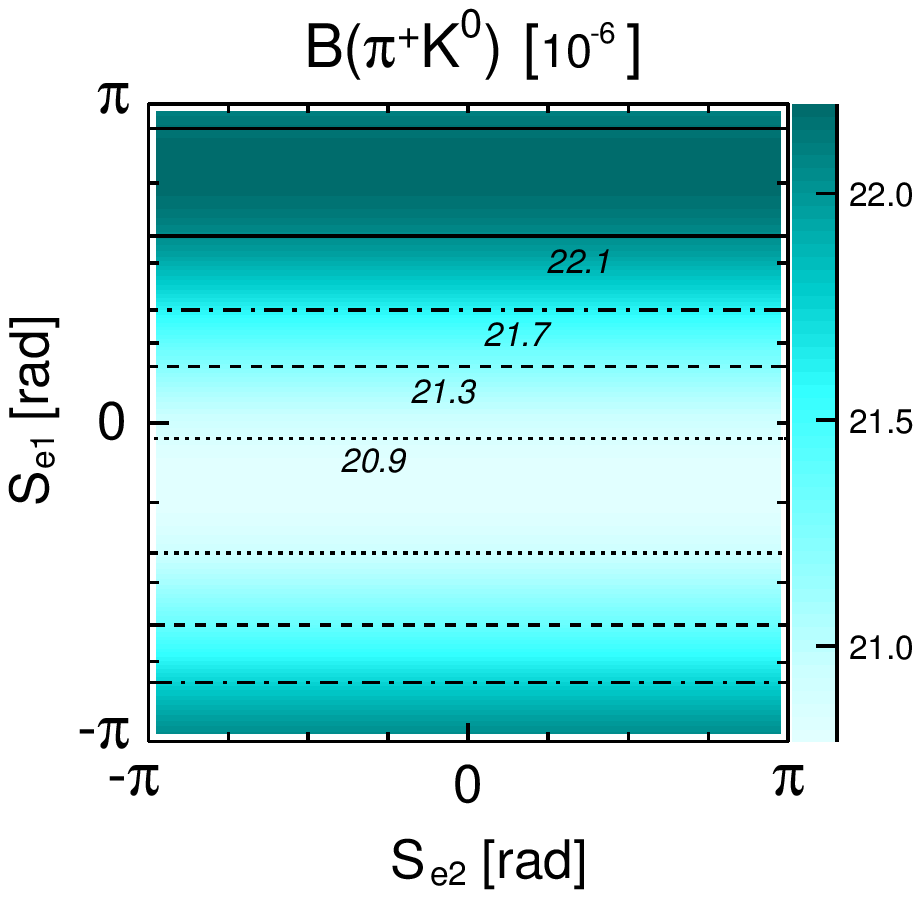}
\includegraphics[height=4.5cm]{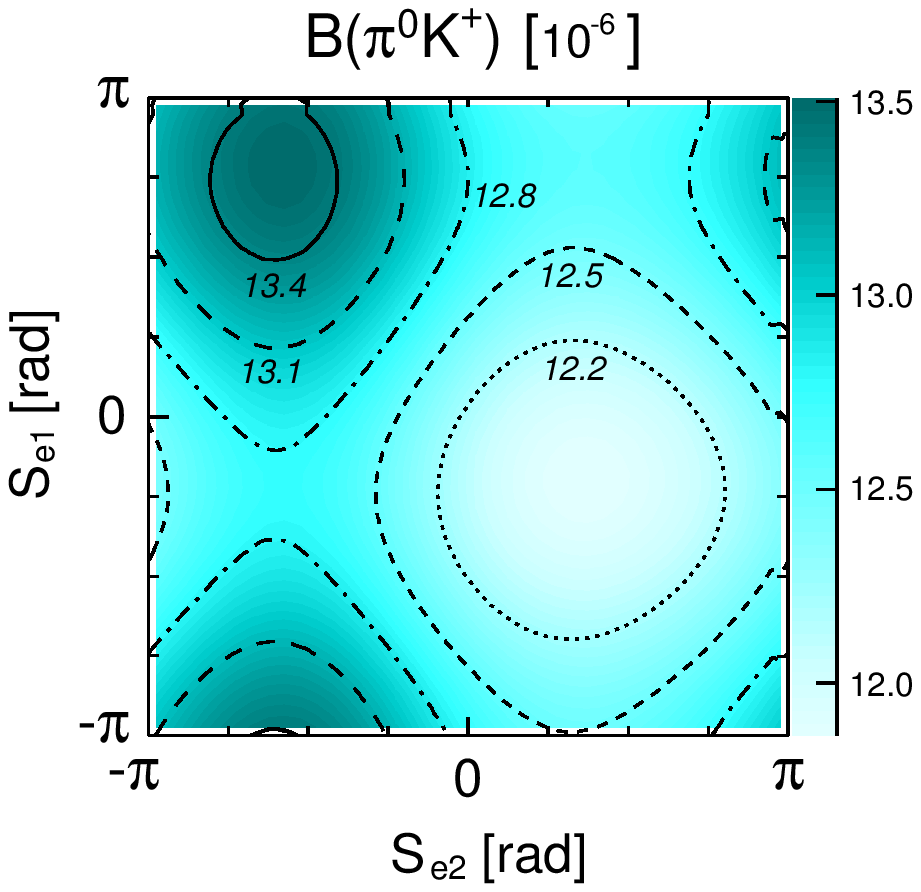}
\includegraphics[height=4.5cm]{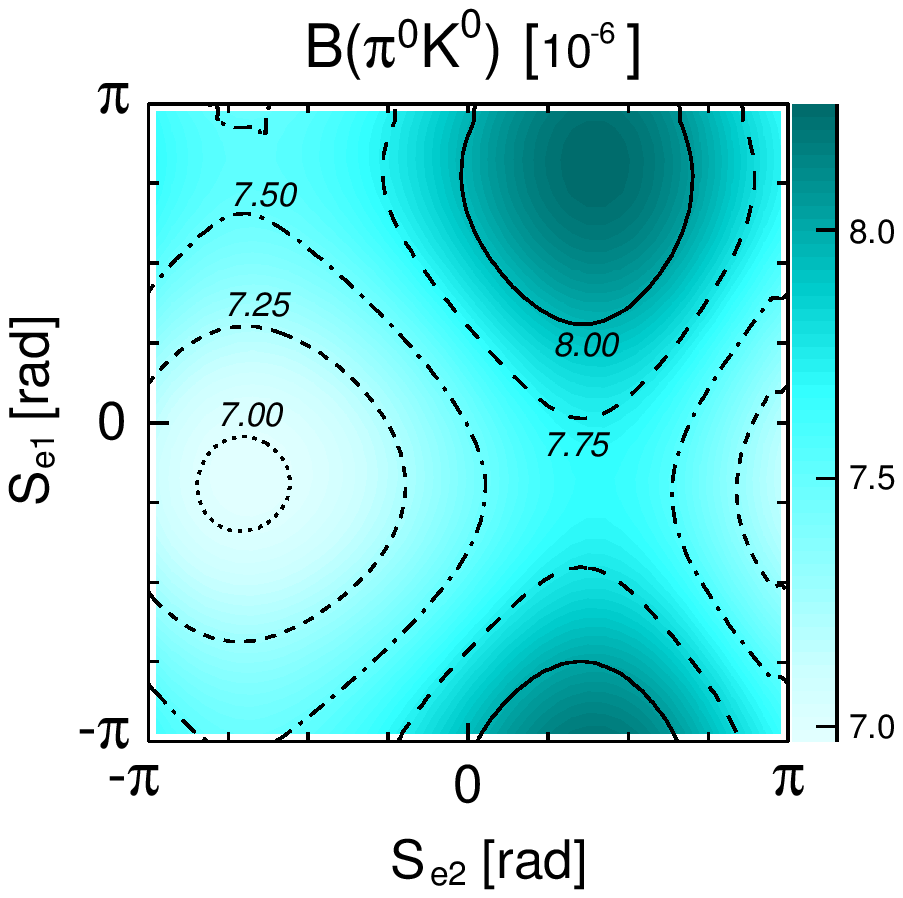}
\includegraphics[height=4.5cm]{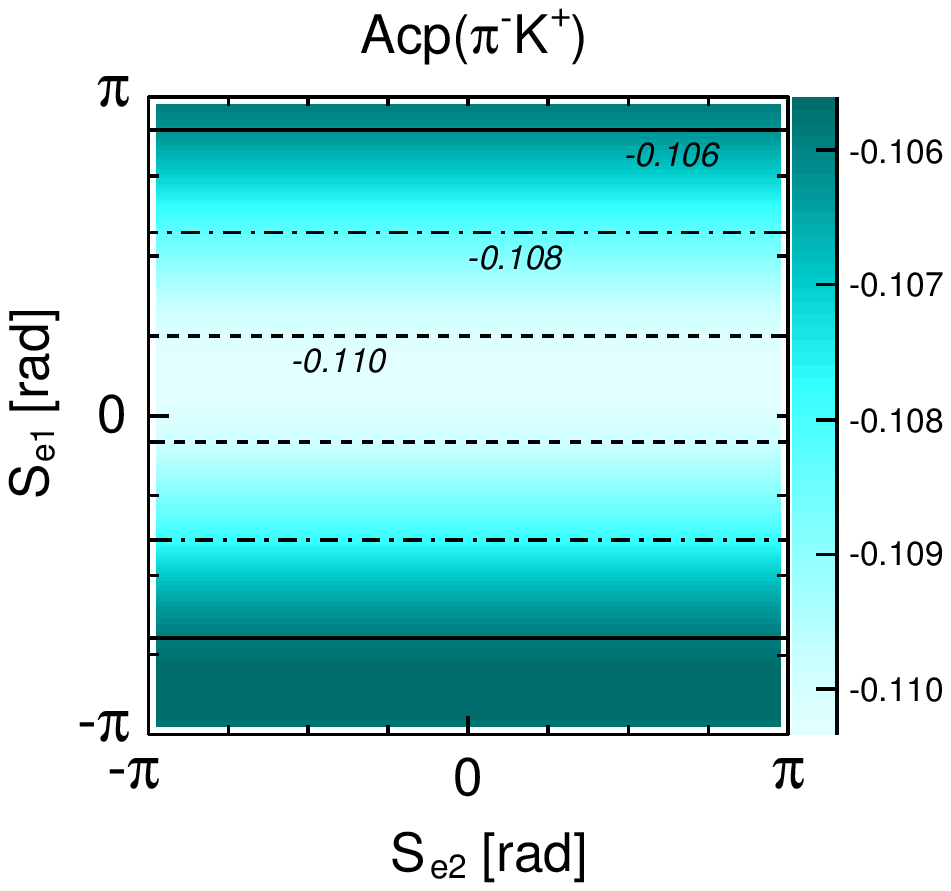}
\includegraphics[height=4.5cm]{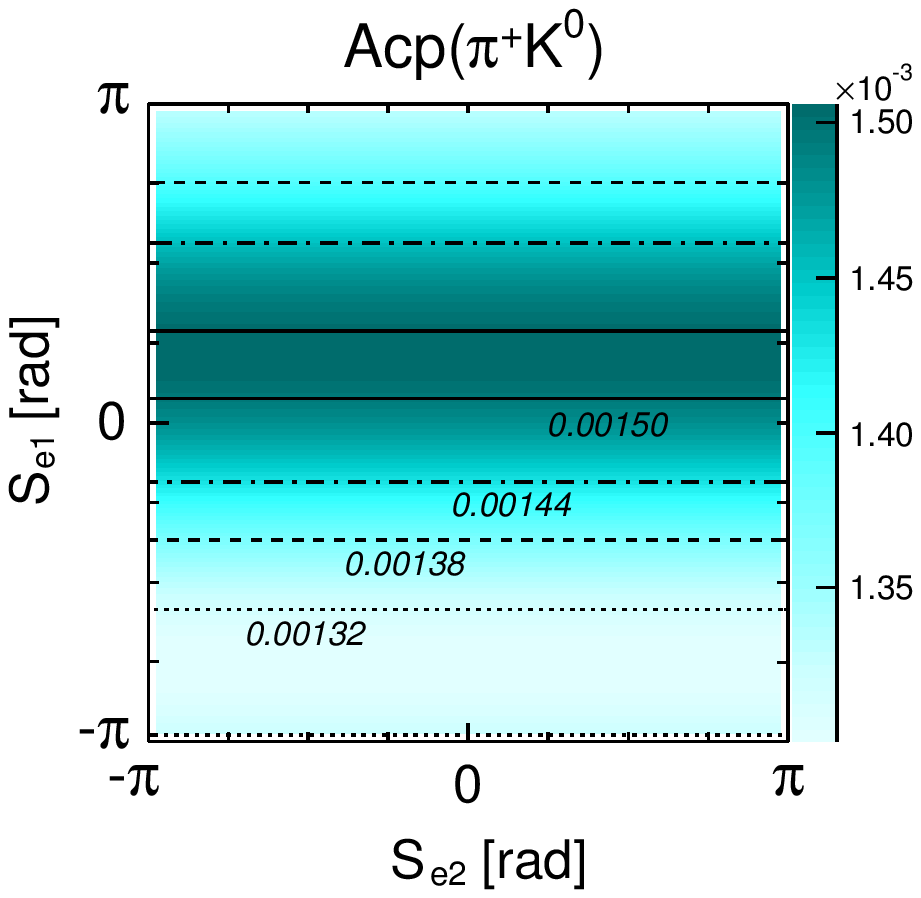}
\includegraphics[height=4.5cm]{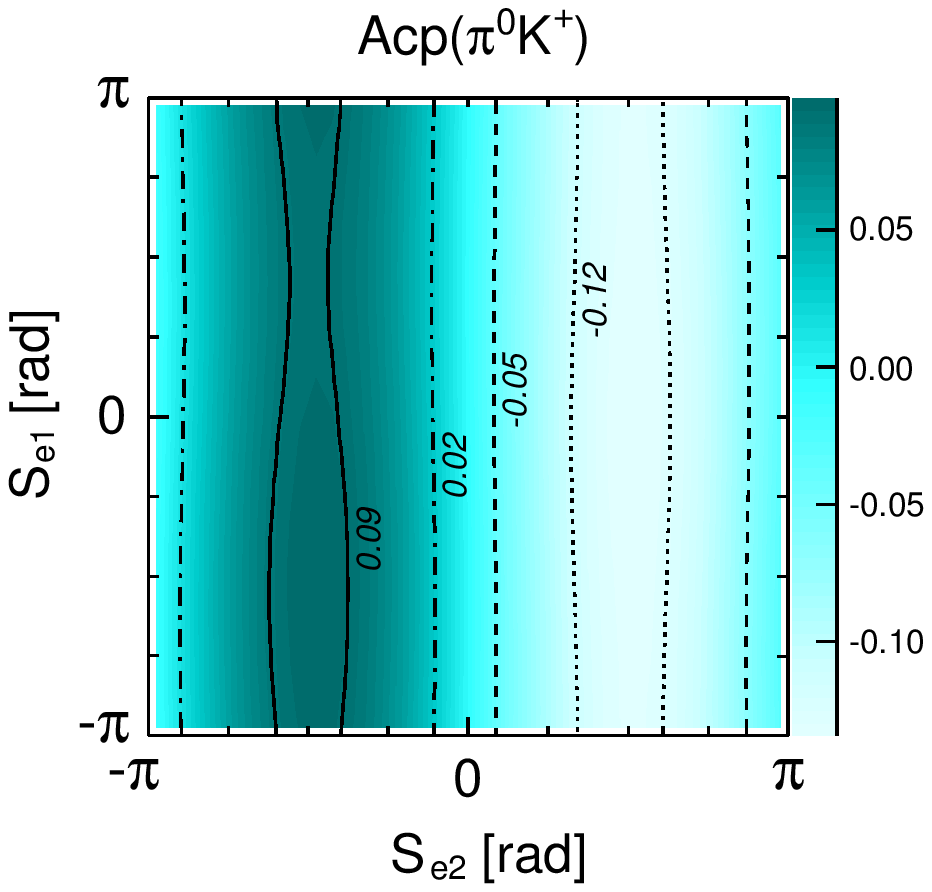}
\includegraphics[height=4.5cm]{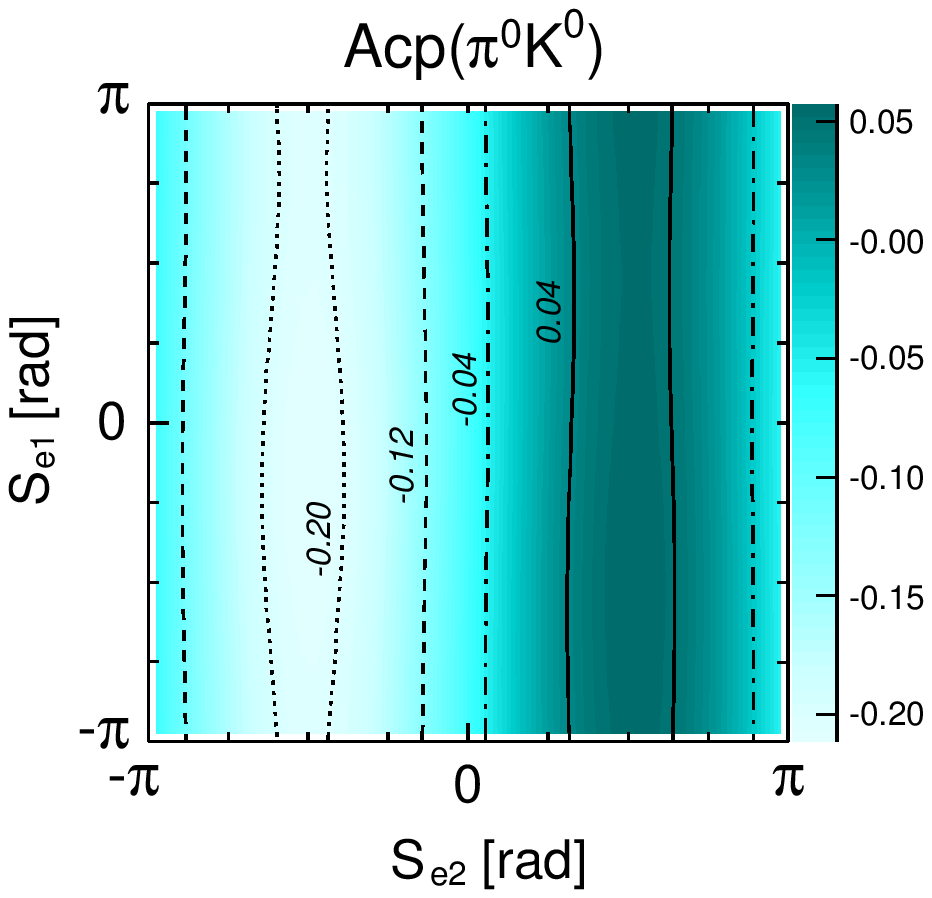}
\includegraphics[height=4.5cm]{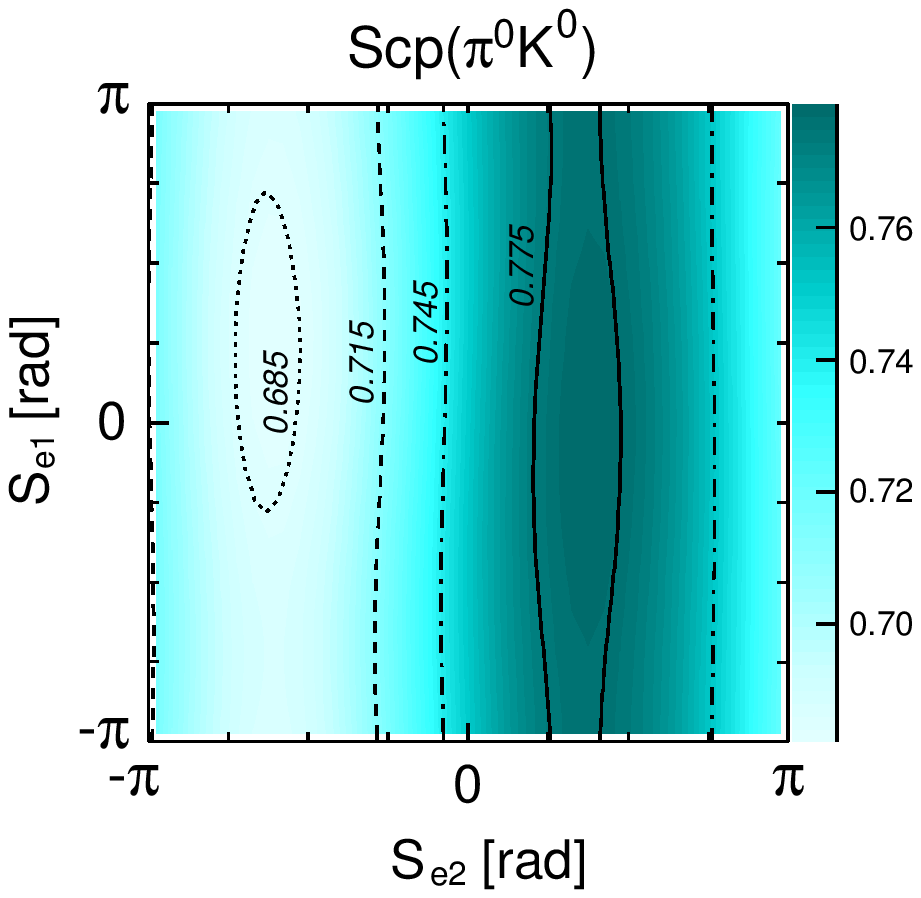}
\caption{$S_{e1}$ and $S_{e2}$ dependencies of the $B\to\pi K$
branching ratios (in units of $10^{-6}$), direct CP asymmetries, and
mixing-induced CP asymmetry.} \label{fig10}
\end{center}
\end{figure*}

The $S_{e1}$ and $S_{e2}$ dependencies of the $B\to\pi K$ branching
ratios (in units of $10^{-6}$), direct CP asymmetries, and
mixing-induced CP asymmetry are displayed in Fig.~\ref{fig10}. It is
easy to understand that the PQCD predictions for all the branching
ratios depend on the Glauber phases weakly. $B(\pi^- K^+)$ and
$B(\pi^+ K^0)$ are insensitive to the variation of $S_{e2}$, since
these two modes do not involve the color-suppressed tree amplitude.
The weak dependence on $S_{e1}$ is introduced through the
interference between the spectator diagrams and the factorizable
emission diagrams. $B(\pi^0 K^+)$ and $B(\pi^0 K^0)$ depend on both
Glauber phases, because of the involvement of the color-suppressed
tree amplitude. For a similar reason, the direct CP asymmetries
$A_{CP}(\pi^- K^+)$ and $A_{CP}(\pi^+ K^0)$ are insensitive to the
variation of $S_{e2}$, and slightly depend on $S_{e1}$. The
prediction for $A_{CP}(\pi^- K^+)$ remains as $-0.11$, as varying
$S_{e1}$, close to the data $-0.082\pm 0.006$ \cite{Amhis:2012bh}. On the
contrary, $A_{CP}(\pi^0 K^+)$ and $A_{CP}(\pi^0 K^0)$ depend on
$S_{e2}$, but are not sensitive to $S_{e1}$. Note that the amplitude $C$ contains
the $B\to K$ transition in this case, and the Glauber effect from
the kaon is assumed to be negligible. The predicted $A_{CP}(\pi^0
K^+)$ increases from $-0.01$, and becomes positive quickly, as
$S_{e2}$ approaches $-\pi/2$, a tendency in agreement with the
updated data $0.040\pm 0.021$ \cite{Amhis:2012bh}. The prediction for
$A_{CP}(\pi^0 K^0)$ decreases from $-0.08$ to $-0.21$. This
difference is attributed to the sign flip of $C$ between the above
two modes. Figure~\ref{fig10} indicates that the mixing-induced
CP asymmetry $S_{CP}(\pi^0 K^0)$ descends from $0.75$ to $0.69$.
Compared to the data $S_{CP}(\pi^0 K^0)=0.57\pm 0.17$ and $S_{CP}(c\bar
cs)=0.682\pm 0.019$ \cite{Amhis:2012bh}, the consistency has been improved.

\begin{figure*}[t]
\begin{center}
\includegraphics[height=4.5cm]{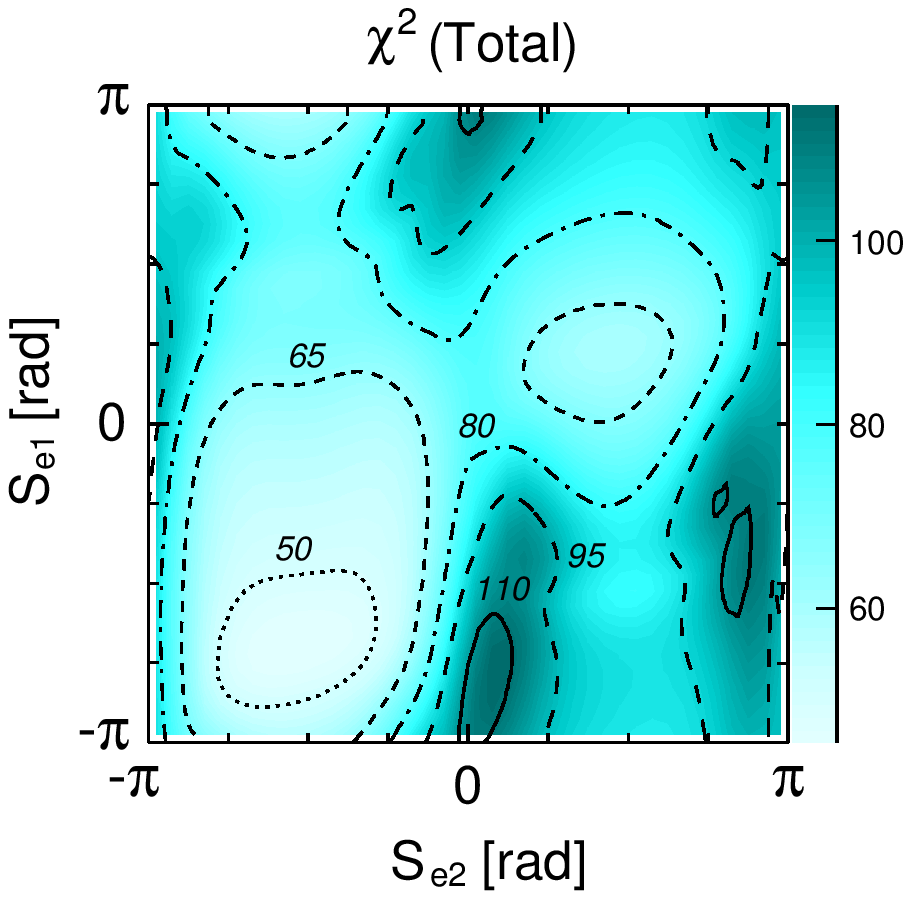}
\includegraphics[height=4.5cm]{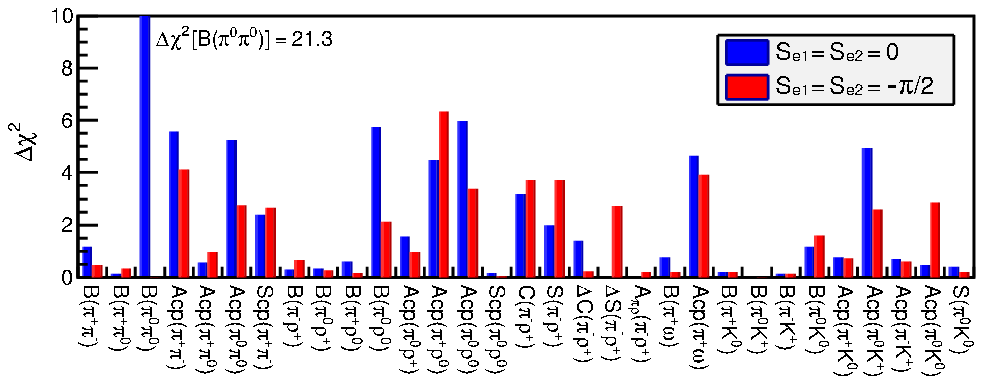}
\caption{$S_{e1}$ and $S_{e2}$ dependencies of $\Delta\chi^2$ for all
the $B\to\pi \pi$, $\pi \rho$, $\pi\omega$, and $\pi K$ decays. The difference of
$\Delta\chi^2$ for each considered quantity due to the inclusion of
the Glauber effects is also displayed.} \label{fig12}
\end{center}
\end{figure*}

At last, we display the $S_{e1}$ and $S_{e2}$ dependencies of
$\Delta\chi^2$ for the fit to all the $B\to\pi \pi$, $\pi\rho$, 
$\pi\omega$, and $\pi K$ data in Fig.~\ref{fig12}, which exhibits
significant decrease of $\Delta\chi^2$ from 76 to 49, as both
$S_{e1}$ and $S_{e2}$ change from zero to $-\pi/2$.
Figure~\ref{fig12} also shows the change of $\Delta\chi^2$ for
each mode caused by $S_{e1}=S_{e2}=-\pi/2$. 
The major reduction of $\Delta\chi^2$ arises from the
modified predictions for the $B\to\pi\pi$ decays, especially from
the $B^0\to\pi^0\pi^0$ branching ratio. The amount of reduction of
$\Delta \chi^2$ from the $B\to\pi\rho$, $\pi\omega$, and $\pi K$
modes is minor. To summarize the Glauber effects on the quantities
considered above, we present the branching ratios and direct CP
asymmetries from the data, the standard NLO PQCD predictions with
$S_e=S_{e1}=S_{e2}=0$, and the modified predictions with
$S_e=-\pi/2$ in Table~\ref{tab1}. Those for the
mixing-induced CP asymmetries are listed in Table~\ref{tab2}. At
last, we compute the CP violation parameters $C$, $\Delta C$, $S$,
$\Delta S$, and $A_{\pi\rho}$ associated with the $B^0\to
\pi^\mp\rho^\mp$ decays, which are defined in \cite{Amhis:2012bh},
and present the results in Table~\ref{tab3}.
Our predictions for these observables can be confronted with future
data.

\begin{table}[ht]
\begin{center}
\begin{tabular}{cccc|cccc}
\hline\hline
& Data~\cite{Amhis:2012bh,update} & $S_e=0$ & $S_e=-\pi/2$ &
& Data~\cite{Amhis:2012bh,update} & $S_e=0$ & $S_e=-\pi/2$
\\
\hline
$B(B^0 \to \pi^\mp \pi^\pm)$ & $5.10 \pm 0.19$ & $7.5$ & $6.4$ &
$A_{CP}(B^0 \to \pi^\mp \pi^\pm)$ & $0.31 \pm 0.05$ & $0.15$ & $0.17$
\\
$B(B^\pm \to \pi^\pm \pi^0)$ & $5.48^{+0.35}_{-0.34}$ & $5.0$ & $6.6$ &
$A_{CP}(B^\pm \to \pi^\pm \pi^0)$ & $0.026 \pm 0.039$& $-0.003$ & $-0.012$
\\
$B(B^0 \to \pi^0 \pi^0)$ & $1.17 \pm 0.13$ 
& $0.38$ & $1.2$ &
$A_{CP}(B^0 \to \pi^0 \pi^0)$ & $0.03 \pm 0.17$ 
& $0.59$ & $0.36$
\\
\hline
$B(B^0\to\pi^\mp\rho^\pm)$ & $23.0 \pm 2.3$ & $27.8$ & $30.8$ &
\\
$B(B^\pm\to\pi^0\rho^\pm)$ & $10.9^{+1.4}_{-1.5}$ & $13.3$ & $9.3$ &
$A_{CP}(B^\pm\to\pi^0\rho^\pm)$ & $0.02 \pm 0.11$ & $0.17$ & $0.13$
\\
$B(B^\pm\to\pi^\pm\rho^0)$ & $8.3^{+1.2}_{-1.3}$ & $6.5$ & $7.2$ &
$A_{CP}(B^\pm\to\pi^\pm\rho^0)$ & $0.18^{+0.09}_{-0.17}$ & $-0.20$ & $-0.31$
\\
$B(B^0\to\pi^0\rho^0)$ & $2.0 \pm 0.5$ & $0.70$ & $1.1$ &
$A_{CP}(B^0\to\pi^0\rho^0)$ & $-0.27 \pm 0.24$ & $0.38$ & $0.18$
\\
\hline
$B(B^\pm\to\pi^\pm\omega)$ & $6.9 \pm 0.5$ & $5.4$ & $6.1$ &
$A_{CP}(B^\pm\to\pi^\pm\omega)$ & $-0.02 \pm 0.06$ & $-0.20$ & $-0.18$
\\
$B(B^0\to\pi^0\omega)$ & $< 0.5$ & $0.04$ & $0.85$ &
$A_{CP}(B^0\to\pi^0\omega)$ & ----- & $-0.99$ & $-0.12$
\\
\hline
$B(B^\pm \to \pi^\pm K^0)$ & $23.79 \pm 0.75$ & $20.9$ & $21.1$ &
$A_{CP}(B^\pm \to \pi^\pm K^0)$ & $-0.015 \pm 0.019$ & $0.001$ & $0.001$
\\
$B(B^\pm \to \pi^0 K^\pm)$ & $12.94^{+0.52}_{-0.51}$ & $12.2$ & $12.9$ &
$A_{CP}(B^\pm \to \pi^0 K^\pm)$ & $0.040 \pm 0.021$ & $-0.01$ & $0.10$
\\
$B(B^0 \to \pi^\mp K^\pm)$ & $19.57^{+0.53}_{-0.52}$ & $17.6$ & $17.7$ &
$A_{CP}(B^0 \to \pi^\mp K^\pm)$ & $-0.082 \pm 0.006$ & $-0.11$ & $-0.11$
\\
$B(B^0 \to \pi^0 K^0)$ & $9.93 \pm 0.49$ & $7.5$ & $7.2$ &
$A_{CP}(B^0 \to \pi^0 K^0)$ & $-0.01 \pm 0.10$ & $-0.08$ & $-0.21$
\\
\hline\hline
\end{tabular}
\caption{Branching ratios (in units of $10^{-6}$) and direct CP
asymmetries, with the notation $S_e \equiv
S_{e1}=S_{e2}$.}\label{tab1}
\end{center}
\end{table}

\begin{table}[ht]
\begin{center}
\begin{tabular}{cccc|cccc}
\hline\hline & Data~\cite{Amhis:2012bh} & $S_e=0$ & $S_e=-\pi/2$ & &
Data~\cite{Amhis:2012bh} & $S_e=0$ & $S_e=-\pi/2$
\\
\hline
$S_{CP}(B^0 \to \pi^\mp \pi^\pm)$ & $-0.66 \pm 0.06$ & $-0.44$ & $-0.43$ &
$S_{CP}(B^0 \to \pi^0 \pi^0)$ & ----- & $0.80$ & $0.63$
\\
$S_{CP}(B^0\to\pi^0\rho^0)$ & $-0.23 \pm 0.34$ & $-0.09$ & $-0.30$ &
$S_{CP}(B^0\to\pi^0\omega)$ & ----- & $-0.11$ & $-0.26$
\\
$S_{CP}(B^0 \to \pi^0 K^0)$ & $0.57 \pm 0.17$ & $0.75$ & $0.69$ &
\\
\hline\hline
\end{tabular}
\caption{Mixing-induced CP asymmetries.}\label{tab2}
\end{center}
\end{table}

\begin{table}[ht]
\begin{center}
\begin{tabular}{cccc|cccc}
\hline\hline
& Data~\cite{Amhis:2012bh} & $S_e=0$ & $S_e=-\pi/2$ &
& Data~\cite{Amhis:2012bh} & $S_e=0$ & $S_e=-\pi/2$
\\
\hline
$C$ & $-0.03 \pm 0.06$ & $0.09$ & $0.10$ &
$\Delta C$ & $0.27 \pm 0.06$ & $0.44$ & $0.32$
\\
$S$ & $0.06 \pm 0.07$ & $-0.04$ & $-0.08$ &
$\Delta S$ & $0.01 \pm 0.08$ & $0.004$ & $-0.14$
\\
${\cal A}_{\pi\rho}$ & $-0.11 \pm 0.03$ & $-0.11$ & $-0.13$
\\
\hline\hline
\end{tabular}
\caption{CP violation parameters for the $B^0\to
\pi^\mp\rho^\mp$ decays.}\label{tab3}
\end{center}
\end{table}

\section{CONCLUSION}

In this paper we have identified the uncancelled Glauber divergences
in the $k_T$ factorization theorem for the spectator amplitudes in
the $B\to M_1M_2$ decays at NLO level. It has been shown that the
divergences are factorizable and demand the introduction of the
phase factors: those coupling the $M_1$ meson and the $B$-$M_2$
system are absorbed into the phase factor $\exp(-iS_{e1})$, and
those coupling the $M_2$ meson and the $B\to M_1$ transition are
absorbed into $\exp(\pm iS_{e2})$.
We have investigated the Glauber effects on the color-suppressed
tree amplitude $C$ and the color-allowed tree amplitude $T$ in a
simplified formalism, in which the convolution between the Glauber
factors and the standard PQCD factorization formulas is neglected.
Treating $S_{e1}$ and $S_{e2}$ as free parameters, it was observed
that the ratio $C/T$ is enhanced maximally by a factor 3, and a good fit of
the PQCD predictions to all the considered $B\to\pi\pi$, $\pi\rho$,
$\pi\omega$, and $\pi K$ data is achieved as
$S_{e1}=S_{e2}\approx -\pi/2$.

We summarize the modified NLO PQCD
predictions: $B(\pi^0\pi^0)$ and $B(\pi^0\rho^0)$ are
increased, the difference between $A_{CP}(\pi^\mp K^\pm)$ and
$A_{CP}(\pi^0 K^\pm)$ is enlarged, and $\Delta S_{\pi^0 K_S}$ is
reduced, all becoming more consistent with the data. The
major reduction of $\Delta\chi^2$ in the global fit arises from the
observables for the $B\to\pi\pi$ modes. We stress again that the
above improvement is nontrivial, since the simultaneous adjustment
of the phases between the spectator diagrams, and between the spectator
amplitude and other emission amplitudes for these modes is required.
The constraint on $C$ from the $B\to\rho\rho$ data is
evaded, because of the special role of the pion as a $q\bar q$ bound
state and as a pseudo NG boson. It seems that the implication
on new physics from the $B\to\pi K$ puzzle tends to be
weaker \cite{Ciuchini:2008eh,Baek:2009pa}.

The Glauber gluons may have the nonperturbative origin similar to
that in elastic rescattering. The correspondence
has been made explicit between the Glauber factors and the mechanism
in elastic rescattering among various $M_1M_2$ final states, including
the singlet exchange and the charge exchange \cite{CHY,Chua08}.
A derivation of the Glauber factor,
or even an evaluation of the parameters $S_{e1}$ and $S_{e2}$ by
nonperturbative methods for various mesons will lead to a deeper
understanding of the proposed mechanism.
Besides, the Glauber gluons in the nonfactorizable
annihilation amplitudes deserves a thorough investigation too, which
couple the $B$ meson and the $M_1$-$M_2$ system. The inclusion of
these additional Glauber gluons will complete the modified PQCD
formalism for nonfactorizable $B\to M_1M_2$ decay amplitudes.
The above subjects will be studied in forthcoming papers.

We expect that the Glauber effect also appears in other complicated
pion-induced processes, if it was really the mechanism responsible for
the $B\to\pi\pi$ and $\pi K$ puzzles. It has been demonstrated recently
\cite{CL13} that the existence of Glauber gluons in the $k_T$ factorization
theorem can account for the violation of the Lam-Tung relation \cite{LT78},
namely, the anomalous lepton angular distribution observed in pion-induced
Drell-Yan processes \cite{NA10-86,NA10,E615}. It was noticed that a
final-state parton is required to balance the lepton-pair transverse
momentum $q_T$, so at least three partons are involved. Since the low-$q_T$
spectra of the lepton pair are concerned, the $k_T$ factorization is an
appropriate theoretical framework. The Glauber gluons then exist
and are factorizable at low $q_T$, a kinematic region similar to the small
$x$ one for the $B\to\pi\pi$ and $\pi K$ decays. Associating the
Glauber phase factor $\exp(iS_e)$ to the $t$-channel diagrams, it has been shown
that the spin-transverse-momentum correlation between colliding partons, necessary
for the violation of the Lam-Tung relation, can be generated.
More interestingly, this resolution can be discriminated by measuring the
$p\bar p$ Drell-Yan process at GSI and J-PARC \cite{CL13}.

\begin{acknowledgments}

We thank C.K. Chua and T. Onogi for useful discussions. This work was supported in
part by Ministry of Science and Technology of R.O.C. under Grant No.
NSC-101-2112-M-001-006-MY3, by the National Center for Theoretical
Sciences of R.O.C., by ERC Ideas Advanced Grant n.~267985 ``DaMeSyFla"
and by ERC Ideas Starting Grant n.~279972 ``NPFlavour".

\end{acknowledgments}

\begin{appendix}
\section{Glauber divergences in Feynman parametrization}

In this appendix we verify the existence of the Glauber divergences
in the NLO spectator diagrams by means of the Feynman parametrization.
Starting with the integrand in Eq.~(\ref{2e}) for Fig.~\ref{fig2}(d),
we associate the Feynman parameters $x$, $t$, $z$, $1-x-y-z-t$, and $y$
with each of the denominators in sequence, obtaining a factor
$1/(q^2+2M^2)^5$, with
\begin{eqnarray}
q&=&l+x(P_2-k_2)+tk+z(k-k_1)-y(k_2-k+k_1),\nonumber\\
M^2&=&x(y+z)k_1\cdot(P_2-k_2)+y(1-y-z)k_1\cdot k_2-(1-y-z-t)(y+z)k_1\cdot k.
\label{po6}
\end{eqnarray}
Note that the Wick rotation for the variable change $q^0\to iq^0$
holds, no matter whether $M^2$ is positive or negative. 
The two poles of $q^0$ are always located in the second and
fourth quadrants. The difference is that the two poles are closer
to the imaginary axis of the $q^0$ plane, as $M^2>0$, and to the real
axis, as $M^2<0$.  After integrating out $q$, we arrive at
a power of $1/(2M^2+i\epsilon)$. To get infrared divergences,
some of the Feynman parameters need to be small, such that we have
small $M^2$. For example, the collinear divergence from the loop
momentum $l$ parallel to $P_2$ corresponds to $x\sim O(1)$, because
$(P_2-k_2+l)^2$ is small already, and $y$, $z$, and $t$ are all small,
because their associated denominators are large.
A more solid argument on the relations between the Feynman parameters
and the presence of infrared singularities can be made with the Landau equations
\cite{Landau:1959fi}.

The sign flip of $M^2$ in the last integral is required for the existence
of the Glauber divergences, such that the
principal-value prescription applies.
We first integrate out $x$ and get a power of $1/(y+z)$ as a coefficient
of the integrand. The upper bound $x=1-y-z-t$ leads to the collinear divergence
from $l$ parallel to $P_2$ as stated before. It is easy to
see that $M^2$ does not flip sign in this term,
\begin{eqnarray}
M^2_{x=1-y-z-t}&=&(1-y-z-t)(y+z)k_1\cdot(P_2-k_2)+y(1-y-z) k_1\cdot k_2-(1-y-z-t)(y+z)k_1\cdot k,\nonumber\\
&=&(1-y-z-t)(y+z)k_1\cdot(P_2-k_2-k)+y(1-y-z) k_1\cdot k_2 >0,\label{4}
\end{eqnarray}
due to the power counting $P_2^--k_2^- \gg k^-$. Hence, it does not contribute
to a Glauber divergence, and will be neglected.
We then consider another term from the lower bound $x=0$. Integrating
out $t$, we obtain the second coefficient $1/(y+z)$ for the integrand.
Similarly, the upper bound $t=1-y-z$ does not generate a Glauber divergence,
because $M^2_{x=0;t=1-y-z}=y(1-y-z)k_1\cdot k_2$
is always positive. We focus on the term from the lower bound $t=0$,
\begin{eqnarray}
M^2_{x,t=0}=(1-y-z)[ yk_1\cdot k_2-(y+z)k_1\cdot k].\label{m1e}
\end{eqnarray}
For the power counting $k_2^-\sim O(m_B)$ and $k^-\sim O(\Lambda_{\rm QCD})$,
it is obvious that the above expression can flip sign in the infrared region
$y\sim O(\lambda^2)\ll z\sim O(\lambda)$, where $\lambda\equiv \Lambda_{\rm QCD}/m_B$
denotes a small number. The above order of magnitude makes sense, viewing the
associated denominators $(k_2-k+k_1-l)^2\sim O(m_B^2)$ and
$(k-k_1+l)^2\sim O(m_B\Lambda_{\rm QCD})$. Therefore, Fig.~\ref{fig2}(d) contributes to
a Glauber divergence, as concluded in Sec.~II.

Next we investigate Fig.~\ref{fig1am}(d)
by associating the Feynman parameters $x$, $t$, $z$, and $1-x-z-t$
with each of the denominators in Eq.~(\ref{4a}) in sequence. Compared to Eq.~(\ref{2e}),
the parameter $y$ is absent, and $P_2-k_2$ in the first denominator
is replaced by $k_2$. The corresponding $M^2$ is then written as
\begin{eqnarray}
M^2=xzk_1\cdot k_2-z(1-z-t)k_1\cdot k.
\end{eqnarray}
Integrating out $x$, we find that neither terms from the upper and lower bounds,
$x=1-z-t$ and $x=0$, respectively, can flip sign:
\begin{eqnarray}
M^2_{x=1-z-t}&=&z(1-z-t)k_1\cdot (k_2- k)>0,\nonumber\\
M^2_{x=0}&=&-z(1-z-t)k_1\cdot k <0,
\end{eqnarray}
for $k_2^- \gg k^-$ in our power counting. That is, Fig.~\ref{fig1am}(d)
does not develop a Glauber divergence, as stated in Sec.~II.
Figures~\ref{fig2}(d) and
\ref{fig1am}(d) have the same amplitudes in the soft region with
$l\sim O(\Lambda_{\rm QCD})$ except a sign difference,
which is attributed to the emissions of the collinear gluon by the
valence quark and by the valence anti-quark in $M_2$. In the present analysis
based on the Feynman parametrization, Fig.~\ref{fig1am}(d) provides soft subtraction
for Fig.~\ref{fig2}(d) at $y\to 0$. A convenient way to get the sum of
Figs.~\ref{fig2}(d) and \ref{fig1am}(d) is to introduce a lower bound
$y=y_{\min}$ for Eq.~(\ref{m1e}). Obviously, Eq.~(\ref{m1e}) still develops a Glauber
divergence, as long as the hierarchy $y\ll z$ holds.

We turn to Fig.~\ref{fig2}(f), which contains the five denominators
\begin{eqnarray}
[(P_2-k_2+l)^2+i\epsilon][(k_1-l)^2+i\epsilon][(k-k_1+l)^2+i\epsilon]
(l^2+i\epsilon)[(k_2-k+k_1-l)^2+i\epsilon].\label{2f}
\end{eqnarray}
Associating the Feynman parameters $x$, $t$, $z$, $1-x-y-z-t$, and $y$
with each of the denominators in sequence, we have
\begin{eqnarray}
M^2=x(y+z+t)k_1\cdot(P_2-k_2)+y(1-y-z-t)k_1\cdot k_2-(1-y-z-t)(y+z)k_1\cdot k,
\end{eqnarray}
which is basically similar to Eq.~(\ref{po6}).
We first integrate out $x$ and get a power of $1/(y+z+t)$ as a coefficient
of the integrand. The upper bound $x=1-y-z-t$ leads to a collinear divergence
from $l$ parallel to $P_2$ meson. It is trivial to find
that $M^2$ does not flip sign in this term,
\begin{eqnarray}
M^2_{x=1-y-z-t}&=&(1-y-z-t)[(y+z+t)k_1\cdot(P_2-k_2)+y k_1\cdot k_2-(y+z)k_1\cdot k],\nonumber\\
&=&(1-y-z-t)[(y+z)k_1\cdot(P_2-k_2-k)+t k_1\cdot(P_2-k_2)+ y k_1\cdot k_2] >0,
\end{eqnarray}
due to $P_2^--k_2^- \gg k^-$. Hence, it does not contribute
to a Glauber divergence, and will be neglected.
Another term from the lower bound $x=0$ reads
\begin{eqnarray}
M^2_{x=0}=(1-y-z-t)[y k_1\cdot k_2-(y+z)k_1\cdot k],\label{m1f}
\end{eqnarray}
which can flip sign in the infrared region $y\sim O(\lambda^2)\ll z\sim O(\lambda)$,
the same as for Eq.~(\ref{m1e}). That is, Fig.~\ref{fig2}(f) contributes to
a Glauber divergence.

Correspondingly, we should investigate Fig.~\ref{fig1am}(f), which contains the
four denominators
\begin{eqnarray}
[(k_2+l)^2+i\epsilon][(k_1-l)^2+i\epsilon]
[(k-k_1+l)^2+i\epsilon](l^2+i\epsilon).\label{3f}
\end{eqnarray}
The Feynman parameters $x$, $t$, $z$, and $1-x-z-t$ are associated
with each of the denominators in sequence. Compared to Eq.~(\ref{2f}),
the parameter $y$ is absent, and $P_2-k_2$ in the first denominator
is replaced by $k_2$. $M^2$ in this case is then written as
\begin{eqnarray}
M^2=x(z+t)k_1\cdot k_2-z(1-z-t)k_1\cdot k.
\end{eqnarray}
Integrating out $x$, we observe that neither terms from the upper and lower bounds,
$x=1-z-t$ and $x=0$, respectively, can flip sign:
\begin{eqnarray}
M^2_{x=1-z-t}&=&(1-z-t)[t k_1\cdot k_2 +z k_1\cdot (k_2- k)]>0,\nonumber\\
M^2_{x=0}&=&-z(1-z-t)k_1\cdot k <0,
\end{eqnarray}
for $k_2^- \gg k^-$, and that Fig.~\ref{fig1am}(f)
does not develop a Glauber divergence.
Figure~\ref{fig1am}(f) just provides soft subtraction
for Fig.~\ref{fig2}(f) at $y\to 0$.

We then check the triple-gluon diagram in Fig.~\ref{fig2}(e), which
contains four denominators
\begin{eqnarray}
[(P_2-k_2+l)^2+i\epsilon][(k-k_1+l)^2+i\epsilon]
(l^2+i\epsilon)[(k_2-k+k_1-l)^2+i\epsilon].\label{2c}
\end{eqnarray}
Associating the Feynman parameters $x$, $z$, $1-x-y-z$, and $y$
with each of the denominators in sequence, we have
\begin{eqnarray}
M^2=x(y+z)k_1\cdot(P_2-k_2)+y(1-y-z)k_1\cdot k_2-(y+z)(1-y-z)k_1\cdot k.\label{23}
\end{eqnarray}
As integrating out $x$, the upper bound also gives a collinear
divergence relevant to the $M_2$ meson, which does not flip
sign just like Eq.~(\ref{4}). The term from the lower bound $x=0$ reads
\begin{eqnarray}
M^2_{x=0}=(1-y-z)[yk_1\cdot k_2-(y+z)k_1\cdot k],\label{a21}
\end{eqnarray}
which is the same as for Figs.~\ref{fig2}(d) and \ref{fig2}(f).

The Glauber divergence in Eq.~(\ref{a21}) can be isolated via the Ward
identity in Eq.~(\ref{split1}).
Comparing the first term in Eq.~(\ref{split1}) with Eq.~(\ref{2c}), the denominator
$(k-k_1+l)^2+i\epsilon$ is replaced by $l^2+2(k-k_1)\cdot l+i\epsilon$. Therefore,
the corresponding $M^2$ is given by
\begin{eqnarray}
M^2=x(y+z)k_1\cdot(P_2-k_2)+y(1-y-z)k_1\cdot k_2-y(1-y-z)k_1\cdot k+z(y+z)k_1\cdot k,
\label{26}
\end{eqnarray}
which can be derived simply by dropping the $-zk_1\cdot k$ term in Eq.~(\ref{23}).
The term from the lower bound $x=0$ corresponding to Eq.~(\ref{26}) is then written as
\begin{eqnarray}
M^2_{x=0}=y(1-y-z)k_1\cdot (k_2- k)+z(y+z)k_1\cdot k>0.
\end{eqnarray}
Hence, the first term in Eq.~(\ref{split1}), being free of a Glauber divergence,
is absorbed into the $M_2$ meson wave function.
It is found that the Glauber divergence in Fig.~\ref{fig2}(e) has been moved into
the second term in Eq.~(\ref{split1}), which can be combined with those in
Figs.~\ref{fig2}(d) and \ref{fig2}(f). It turns out that the Glauber divergence
associated with the $M_2$ meson has the color factor $C_F$ as claimed in \cite{LM11}.

\begin{figure}[t]
\begin{center}
\begin{tabular}{ccc}
\includegraphics[height=2.5cm]{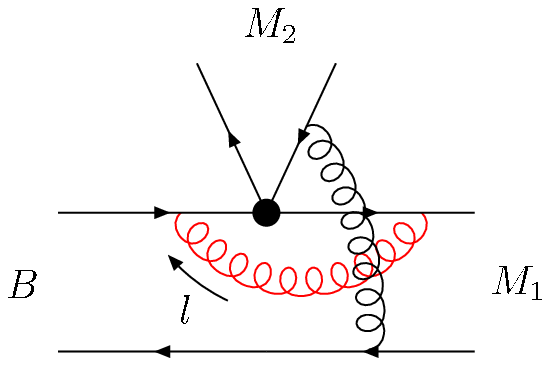}\hspace{0.3cm} &
\includegraphics[height=2.5cm]{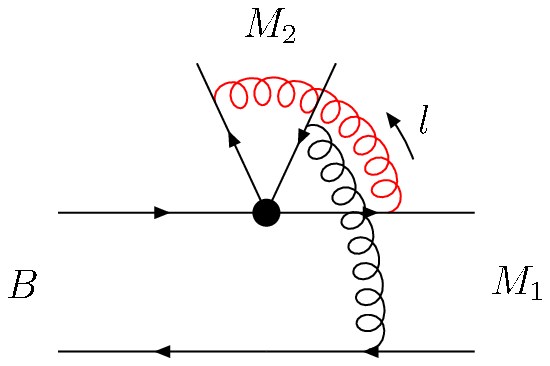}\hspace{0.3cm} &
\includegraphics[height=2.5cm]{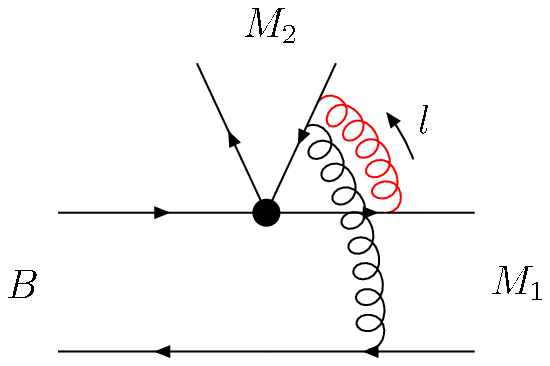}\hspace{0.3cm}  \\
\hspace{-5mm} (a) &\hspace{-5mm} (b) &\hspace{-5mm} (c) \\
\\
\includegraphics[height=2.5cm]{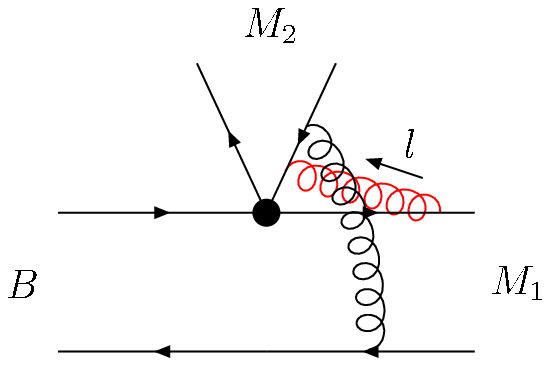}\hspace{0.3cm} &
\includegraphics[height=2.5cm]{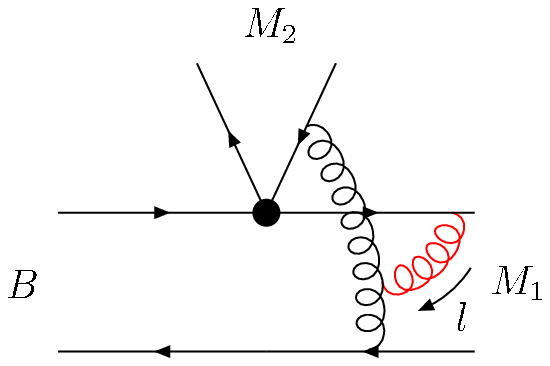}\hspace{0.3cm} &
\includegraphics[height=2.5cm]{diagrams/Ma-M1q-d.eps}\hspace{0.3cm}  \\
\hspace{-5mm} (d) &\hspace{-5mm} (e) &\hspace{-5mm} (f)
\end{tabular}
\caption{NLO diagrams for Fig.~\ref{fig1}(a) that are relevant to
the factorization of the $M_1$ meson wave function.}\label{figa}
\end{center}
\end{figure}

Consider all possible attachments of the collinear gluon emitted by
the valence quark of $M_1$ to other particle lines,
which are displayed in Fig.~\ref{figa}. Figure~\ref{figa}(c)
contains the four denominators
\begin{eqnarray}
[(k_2-l)^2+i\epsilon][(P_1-k_1+l)^2+i\epsilon]
(l^2+i\epsilon)[(k_2-k+k_1-l)^2+i\epsilon],\label{3b}
\end{eqnarray}
with which the Feynman parameters $x$, $t$, $1-x-y-t$, and $y$
are associated in sequence. It is straightforward to derive
\begin{eqnarray}
M^2=(x+y)t(P_1-k_1)\cdot k_2+y(1-x-y)k_1\cdot k_2-yt(P_1-k_1)\cdot k-y(1-y)k_1\cdot k.
\end{eqnarray}
It is appropriate to integrate out $t$ first, since its
coefficient $x(P_1-k_1)\cdot k_2+y(P_1-k_1)\cdot (k_2-k)>0$ does not flip sign
according to the power counting rules. The term from the upper bound $t=1-x-y$,
which corresponds to a collinear divergence from $l$ parallel to $P_1$, gives
\begin{eqnarray}
M^2_{t=1-x-y}=(1-x-y)[(x+y)(P_1-k_1)\cdot k_2+y k_1\cdot k_2-y(P_1-k_1)\cdot k
-y k_1\cdot k]-xy k_1\cdot k.
\end{eqnarray}
To get pinched infrared singularities, we must have
small $x,y$ due to the large denominators $(k_2-l)^2,\,(k_2-k+k_1-l)^2$.
The above expression becomes in the $x,y\to 0$ limit
\begin{eqnarray}
M^2_{t=1-x-y}=x(P_1-k_1)\cdot k_2+y P_1\cdot (k_2-k)-xy k_1\cdot k >0,
\end{eqnarray}
because the third term over the first term is of $O(\lambda^3)$ even for
$k_1^+\sim O(m_B)$ ($y$ is of $O(\lambda^2)$ then).
Another term from the lower bound $t=0$ is written as
\begin{eqnarray}
M^2_{t=0}&=&y[(1-x-y)k_1\cdot k_2-(1-y)k_1\cdot k],\nonumber\\
&\approx&y k_1\cdot (k_2- k)>0,
\end{eqnarray}
in the $x,y\to 0$ limit. The pole structures of Eq.~(\ref{3b})
can be analyzed in a way the same as in Sec.~II. It will be seen that the
interval of $l^-$ does not cover the origin, as the contour integration
over $l^+$ is performed first, or the Glauber divergences associated
with the poles of $l^-$ cancel each other at leading power in $1/m_B$,
as $l^-$ is integrated out first. In conclusion, Fig.~\ref{figa}(c)
does not contain a Glauber divergence.

The analysis of Fig.~\ref{figa}(b) is trivial. Due to the absence of $y$, it is
easy to write down
\begin{eqnarray}
M^2=xt(P_1-k_1)\cdot (P_2-k_2)>0.
\end{eqnarray}
That is, it just provides soft subtraction for Fig.~\ref{figa}(c) at $y\to 0$.

\end{appendix}

\end{document}